\pdfoutput=1

\documentclass[11pt, reqno,preprint]{article}
\pdfoutput=1
\usepackage{jheppub}
\usepackage{epsfig}
\usepackage{amssymb}
\usepackage{amsmath}
\usepackage{mathrsfs}
\usepackage{hyperref}
\usepackage{multirow}
\usepackage{bbold}
\newcommand{\eq}{\begin{equation}}
\newcommand{\eqe}{\end{equation}}
\newcommand{\eqa}{\begin{eqnarray}}
\newcommand{\eqae}{\end{eqnarray}}

\def\be{\begin{equation}}
\def\ee{\end{equation}}
\def\ba{\begin{eqnarray}}
\def\ea{\end{eqnarray}}

\title{ABJM amplitudes and the positive orthogonal Grassmannian}
\author[a]{Yu-tin Huang}
\author[b]{CongKao Wen}
\affiliation[a]{School of Natural Sciences, Institute for Advanced
Study, Princeton, NJ 08540, USA} 
\affiliation[b]{Centre for Research in String Theory, Department of Physics, Queen Mary University of London, Mile End Road, London E1 4NS, UK}

\emailAdd{yutinyt@gmail.com, c.wen@qmul.ac.uk}
\abstract{A remarkable connection between perturbative scattering amplitudes of four dimensional planar SYM, and the stratification of the positive Grassmannian, was revealed in the seminal work of Arkani-Hamed et al. Similar extension for three-dimensional ABJM theory was proposed. Here we establish a direct connection between planar scattering amplitudes of ABJM theory, and singularities thereof, to the stratification of the positive orthogonal Grassmannian. In particular, scattering processes are constructed through on-shell diagrams, which are simply iterative gluing of the fundamental four-point amplitude. Each diagram is then equivalent to the merging of fundamental OG$_2$ orthogonal Grassmannian to form a larger OG$_k$, where $2k$ is the number of external particles. The invariant information that is encoded in each diagram is precisely this stratification. This information can be easily read off via permutation paths of the on-shell diagram, which also can be used to derive a canonical representation of OG$_k$ that manifests the vanishing of consecutive minors as the singularity of all on-shell diagrams. Quite remarkably, for the BCFW recursion representation of the tree-level amplitudes, the on-shell diagram manifests the presence of all physical factorization poles, as well as the cancellation of the spurious poles. After analytically continuing the orthogonal Grassmannian to split signature, we reveal that each on-shell diagram in fact resides in the positive cell of the orthogonal Grassmannian, where all minors are positive. In this language, the amplitudes of ABJM theory is simply an integral of a product of $d\log$ forms, over the positive orthogonal Grassmannian.   }
\preprint{QMUL-PH-13-09}
\begin{document}
\maketitle

\pagebreak
\section{Introduction}

Among many others, the Grassmannian formulation of scattering amplitudes in four-dimensional maximally supersymmetric theory ($\mathcal{N}=4$ SYM)~\cite{Grassmanniandual} has been a great step forward in unearthing a host of hidden properties of the theory, in particular it is this ``dual" formulation of the S-matrix that makes both conformal and dual conformal symmetries of the theory manifest. Rational functions that are leading singularities of loop integrands, and form building blocks for tree-level amplitudes, are given by the residues of an integral over a Grassmannian manifold, which is the space of $k$-planes in $n$ dimensions. On the other hand, it is well known that many of the symmetries exposed by the Grassmannian formulation extend to the full planar loop integral in four dimensions, prior to integration. Indeed this aspect was utilized to construct the full planar integrand of $\mathcal{N}=4$ SYM~\cite{NimaAllLoop}. As the Grassmannian formulation naturally encodes the symmetry of $\mathcal{N}=4$ SYM, it begs the question of whether the planar integrand also ``lives" in the Grassmannian as well.

This was achieved in the remarkable recent work of Arkani-Hamed et. al. ~\cite{NimaBigBook}, where the residue of the Grassmannian integral can be identified with individual {\it on-shell diagrams}. The on-shell diagram representation of scattering amplitudes are built up by gluing two different fundamental three-point on-shell amplitudes together, and integrate over all the internal lines by their on-shell phase spaces. Writing the three-point amplitudes as Grassmannians, this gluing procedure is translated into the merging of smaller Grassmannians into larger ones. For on-shell diagrams that correspond to terms in the BCFW tree-level recursion~\cite{BCFWtree}, their Grassmannian representative have precisely the same dimensions as the bosonic constraints that remain after the merging procedure, thus leading to a completely localized Grassmannian configuration, i.e. rational terms. A more general class of diagrams that arises from the loop-level recursion has dimensions greater than the constraints, thus leading to left over integrals! This realizes the vision that the planar integrand indeed lives inside the Grassmannian.

The Grassmannian representation constructed via on-shell diagrams can be classified by the linear dependency of adjacent columns in the Grassmannian.  This study of linear dependency is termed stratification, and for ordered columns this is referred to as ``positroid stratification"~\cite{P, KLS} in the mathematics literature. The invariant content of the on-shell diagrams is precisely this stratification, where different on-shell diagrams that belong to the same strata, can be shown to be equivalent though a series of change of variables, whose physical interpretation simply mounts to the equivalence of distinct BCFW representations. This invariant data can be readily read off from the permutation paths associated with the on-shell diagrams. 

This new understanding leads to a direct connection between several beautiful areas of current research in mathematics and the physics of scattering amplitudes. For instance, one of the outcomes is that the Grassmannian formulation of scattering amplitudes naturally hands us to a novel ${d\log}$-form representation of loop integrands~\cite{NimaBigBook, dLog}, which manifest the relation between loop integrands and their leading singularities. Furthermore, various fundamental physical properties of scattering amplitudes, such as locality and unitarity, are tightly related to the deep mathematical structures of the {\it positive} cell of the Grassmannian.  Quite interestingly, many similar underlying mathematics have also been appeared in the study of other areas of physics, so-called Bipartite Field Theories, where a supersymmetric gauge theory is defined by bipartite graphs on a Riemann surface~\cite{BFT1, BFT2, BFT3, BFT4, BFT5, BFT6}. The physics of the theory are also captured by the positive Grassmannian as well as the zig-zag permutation paths of the graph.

In this paper we aim to explore the applications of on-shell diagrams for the study of a different but closely related theory, the three-dimensional $\mathcal{N}=6$ supersymmetric Chern-Simons matter theory (ABJM theory)~\cite{ABJM}. Before turning to the main focus of the paper, let us briefly summarise some of the known results regarding the scattering amplitudes in ABJM theory. At tree-level, there is a three-dimensional generalization~\cite{Gang} of BCFW recursion relations~\cite{BCFWtree}, based on which a recursion relation for the one-loop supercoefficients was also found~\cite{ABJMOneL}, so tree-level as well as one-loop amplitudes are in principle fully determined recursively. A few lower-point amplitudes at tree-level~\cite{Gang} and one-loop~\cite{ABJMOneG, ABJMOneI, N=6Note} have been computed explicitly. At two loops, only four and six-point amplitudes have been calculated so far~\cite{ABJMTwoL41, ABJMTwoL42, ABJMTwoL6}, with results interestingly resemble to the corresponding one-loop amplitudes in $\mathcal{N}=4$ SYM. Actually, with appropriate redefinition, four-point amplitudes in two theories match with each other exactly. This similarity was further explored in the context of Wilson and amplitude duality\cite{WilsonN=41, WilsonN=42, WilsonABJM1,WilsonABJM2, WilsonABJM3, WilsonABJM4}. Form factors and some non-planar amplitudes in ABJ(M) theories have also been studied in recent works~\cite{FF1, FF2, NP}.

More importantly, for the purpose of this paper, the scattering amplitudes in ABJM theory also enjoy both conformal and dual conformal symmetries~\cite{Bargheer, HuangLipstein}. The leading singularities can be identified with an integral over the orthogonal Grassmannian~\cite{SangminGrass}, which is simply the space of null $k$-planes in $n$ dimensions, i.e. a $k$-dimensional plane in an $n$-dimensional space spanned by $k$ vectors, $\vec{v}_i$ with $i=1,\cdots,k$, satisfying $\vec{v}_i \cdot \vec{v}_j =0$. Here we apply on-shell diagrams to build up representatives of the orthogonal Grassmannian. An important distinction between ABJM and $\mathcal{N}=4$ SYM is their fundamental building blocks. More specifically, in contrast to the three-point amplitudes in $\mathcal{N}=4$ SYM our fundamental building block, the four-point amplitude in ABJM theory, has much richer structure such as singularities and distinct branches. The branch structure is a reflection of the fact that three-dimensional massless kinematics is projectively a  circle, and thus have distinct winding numbers. This fact turns out to be intimately tied to the interplay of the two branches in the orthogonal Grassmannian.

Unlike $\mathcal{N}=4$ SYM, the BCFW bridge in ABJM theory is the fundamental vertex itself. This difference leads to the result that the on-shell diagram representation of the BCFW recursion manifest the presence of all physical singularities. As proposed in ref.~\cite{NimaBigBook}, the Grassmannian derived from the on-shell diagram of ABJM theory should also correspond to stratification of the orthogonal Grassmannian, where this invariant data is again encoded in permutation paths. Here, we verify this proposal and show that such encoding is consistent with the orthogonality of the Grassmannian. Furthermore, armed with the stratification we can easily achieve the following:
\begin{itemize}
  \item The tree contour: as each BCFW on-shell diagram gives a stratification, which would imply the vanishing of consecutive minors, this gives us sufficient information to determine the contour in the Grassmannian integral of~\cite{SangminGrass} that gives the tree amplitude. Interestingly this also gives us a straightforward way to determine terms that correspond to a composite leading singularity.
  \item A canonical representation of the orthogonal Grassmannian: using the permutation encoded in the on-shell diagram, we can build up a representative of the Grassmannian whose consecutive minors are always given by a simple product of the vertex variables. 
 \end{itemize} 
The last feature is desired since this implies that the singularities of the on-shell diagrams correspond to configurations of the Grassmannian where consecutive minors become linearly dependent. This is consistent with the identification of the on-shell diagram and particular stratification. 
For $\mathcal{N}=4$ SYM, this property is ensured by the realization that the on-shell diagrams populate the positive Grassmannian, which is defined such that all ordered minors are strictly positive. One important difficulty in proceeding with a similar analysis for ABJM theory is that due to the orthogonal constraints, the minors are forced to alternate between purely real and imaginary. Thus {\it positivity} is ill defined. Quite remarkably, this difficulty can be easily circumvented by analytically continue the Grassmannian to split signature, where all the minors are real and positivity can be defined. The positive orthogonal Grassmannian then plays the same central role as its counterpart in $\mathcal{N}=4$ SYM does.

The rest of the paper is organized as follows. In section \ref{section:OG2}, we review some basics about the scattering amplitudes in ABJM theory, in particular their description in terms of orthogonal Grassmannian. We then turn to a detailed study on the four-point amplitude, as a top-cell of the OG$_2$ Grassmannian, which turns out to have a rather rich structures of its own. After fully understanding the fundamental vertex, we proceed in section \ref{section:OGk} on the construction of more general on-shell diagrams for higher-point amplitudes by gluing four-point vertices together. Applying BCFW recursion relations, remarkably the on-shell diagram form of all tree-level amplitudes can be represented in a novel way making manifest of cyclic symmetries and physical poles, which are usually obscured in the case of four dimensions. For characterizing the invariant content of on-shell diagrams, the notion of {\it permutation} is introduced. The central focus of section \ref{section:stratification} is the applications of permutation on various important aspects of the orthogonal Grassmannian and on-shell diagrams of ABJM theory. Furthermore some intriguing structure regarding the consecutive minors of BCFW diagrams is observed, and proved generally. In section \ref{OrthogSec}, by analytically continuing to split signature, we reveal that each on-shell diagram in fact resides in the positive cell of the orthogonal Grassmannian, where all minors are real and positive orthogonal Grassmannian is well-defined. We conclude the paper in section \ref{section:conclusion} with a discussion and remarks on the BCFW recursion relation for the loop-level amplitudes.

During the completion of this work, we were made aware of the work in progress by Sangmin Lee and Joonho Kim ~\cite{Recent}, which has independently produced some results in the current paper. 
\section{Scattering amplitude of ABJM and the orthogonal Grassmannian} \label{section:OG2}
The scattering amplitudes of ABJM~\cite{ABJM} theory will be the focus of our study. It is a Chern-Simons matter theory with $\mathcal{N}=6$ supersymmetry. There are two types of Chern-Simons gauge fields, and the matter fields consist of eight scalars and eight fermions, forming complex representation of the R-symmetry group $SO(6)=SU(4)$. Due to the topological nature of the Chern-Simons term, the physical degrees of freedom consist of the 4 complex scalars $X_{\mathsf{A}}$ and 4 complex fermions $\psi^{\mathsf{A}\alpha}$ as well as their complex conjugates $\bar{X}^{\mathsf{A}}$ and $\bar{\psi}_{\mathsf{A}\alpha}$ with $\mathsf{A}=1,2,3,4$. They transform in the fundamental or anti-fundamental of $SU(4)$, and in the bi-fundamental representation under the gauge group $U(N)\times U(N)$. The index $\alpha=1,2$ denote the spinor representation in SL(2,R), the three-dimensional Lorentz group. The explicit form of the action can be found in~ \cite{ABJML, ABJML2}.

To arrange these states in  on-shell superspace, we introduce three anticommuting variables $\eta_A$ and write~\cite{Bargheer},
\be
\begin{split}
  \Phi
  ~=&~
  X_4+\eta_A\,\psi^A
  -\frac{1}{2}\epsilon^{ABC}\,\eta_A\eta_B\,X_C
  -\eta_1\eta_2\eta_3\,\psi^4\,,\\
  \bar{\Psi}
  ~=&~\bar{\psi}_4+\eta_A\bar{X}^A
  -\frac{1}{2}\epsilon^{ABC}\,\eta_A\eta_B\,\bar{\psi}_C
  -\eta_1\eta_2\eta_3\,\bar{X}^4\,.
\end{split}
\label{ABJMmap}
\ee
We have split the fields as $X_{\mathsf{A}}\rightarrow(X_4, X_{A})$ and $\psi^{\mathsf{A}}\rightarrow(\psi^4, \psi^A)$, and similarly for $\bar{X}^{\mathsf{A}}$ and $\bar{\psi}_{\mathsf{A}}$. So only an $SU(3)$ subgroup of the $SU(4)$ is manifest in this on-shell superspace formalism.

The tree-level amplitudes of ABJM can be compactly expressed as:
\eq
\mathcal{A}_n^{\rm tree}=\delta^3(\sum_{i=1}^n p_i) \delta^6(\sum_{i=1}^n q_{i})f_n(\lambda_i,\eta_i)
\label{GenAmp}
\eqe
where $p_i$ and $q_i$ are the on-shell momentum and supermomentum for each external leg:
\eq
(p_i)^{\alpha\beta}=\lambda_i^\alpha\lambda_i^{\beta},\; (q_i)^{\alpha\mathsf{A}}=\lambda_i^\alpha\eta_i^{\mathsf{A}}\,.
\eqe
Due to the bi-fundamental nature of the physical degrees of freedom, only even-multiplicity components of the S-matrix are non-trivial, $n=2k$. The delta functions in eq.(\ref{GenAmp}) are required by super Poincar\'e invariance. The function $f_n$ is given by a rational function of Lorentz invariants $\lambda_i^\alpha\lambda_{j\alpha}=\langle ij\rangle$ and contains fermionic variables $\eta_i^{\mathsf{A}}$ with degree $3(k-2)$ as required by superconformal symmetry.\footnote{Strictly speaking, it is required by the R-symmetry generator embedded in the superconformal algebra.} 

On-shell states are characterized by their little group and R-symmetry representation. In three dimensions, the little group is simply Z$_2$, under which the on-shell variables transform as:
\eq
\lambda_i^\alpha\rightarrow-\lambda_i^\alpha,\quad \eta_i^{\mathsf{A}}\rightarrow -\eta_i^{\mathsf{A}}\,.
\label{LittleG}
\eqe
For simplicity we group the on-shell variables into a $2|3$ spinor $\Lambda_i=(\lambda_i^\alpha,\eta_i^{\mathsf{A}})$. This implies that there are only two types of particle states from the point of view of the little group, those that obtain a minus sign under eq.(\ref{LittleG}), fermions, and those that do not, scalars. This is the usual statement that the physical degrees of freedom for all higher integer-spin fields are equivalent to scalars in three dimensions.\footnote{This of course does not apply to anyons, which do not have definite sign under eq.(\ref{LittleG}).} From the leading component of the superfield defined in eq.(\ref{ABJMmap}), we can deduce that the function must have the following property under little group transformations of one of its external legs:
\eq
\begin{array}{c} i\in \Phi \quad \underrightarrow{eq.(\ref{LittleG})}\;\; f_n\rightarrow f_n\\ i\in \bar{\Psi}\quad \underrightarrow{eq.(\ref{LittleG})}\;\; f_n\rightarrow -f_n \end{array}\,.
\eqe 
Thus there are two classes of amplitudes:
\eq
\mathcal{A}_n(\bar{1}2\bar{3}\cdots 2k),\quad \mathcal{A}_n(1\bar{2}3\cdots \bar{2k})\,.
\label{AmpDef}
\eqe
where we use $\bar{i}$ to represent that leg $i$ is part of the $\bar{\Psi}$ multiplet.

The connection between scattering amplitudes of ABJM and the orthogonal Grassmannian was first proposed by Sangmin Lee~\cite{SangminGrass}. As this connection is the focus of this paper, we present a brief introduction to the orthogonal Grassmannian and its properties.

Consider a $(2k)$-dimensional space $V$ equipped with a non-degenerate symmetric bi-linear form $Q^{ij}$. The orthogonal Grassmannian is then the space of $k$-planes that satisfy the orthogonal constraint $Q^{ij}v_iv_j=0$ for $v\in V$. In this paper we will consider $Q^{ij}=\eta^{ij}$, and the Grassmannian, denoted as OG($k,2k$), can be represented as a $k\times 2k$ matrix $C_{ai}$, where $a=1,\cdots,k$ and $i=1,\cdots, 2k$. Since any linear recombination of the $k$, $(2k)$-dimensional vectors represent the same $k$-plane, this description contains a GL($k$) redundancy. Thus the most general orthogonal Grassmannian at a given $k$ is $2k^2 - k^2-k(k+1)/2=k(k-1)/2$-dimensional, where $k^2$ denotes the GL(k) gauge symmetry, where as $k(k+1)/2$ correspond to the orthogonal constraint. Thus the most general configuration of the orthogonal Grassmannian, referred to as the ``\textit{top-cell}", is $k(k-1)/2$-dimensional:
\eq
Dim({\rm Top\, Cell})_{{\rm OG}(k,2k)}=k(k-1)/2\,.
\label{TopCellDim}
\eqe 

It was proposed by Sangmin Lee~\cite{SangminGrass} that: \textit{A tree-level amplitudes of ABJM theory is given by a sum of the residues of the following integral over a OG$(k,2k)$ orthogonal Grassmannian $C_{a i}$ 
\eq
\mathcal{L}_{k,2k}=\int \frac{d^{2k\times k} C_{a i}}{\rm Vol(GL(k))}\frac{1}{M_{j}M_{j+1},\cdots M_{j+k-1}}\delta^{k(k+1)/2}(C\cdot C^{T})\prod_{a=1}^k\delta^{2|3}(C_{a}\cdot \Lambda)\,,
\label{GrassInt}
\eqe
where $M_l$ represent the $l$-th consecutive minor:
\eq
M_l\equiv\epsilon(C_{l\,a_1}C_{l+1\,a_2}\cdots C_{l+k\,a_k})=(l\,l+1\,\cdots,l+k)\,.
\eqe}
Here on we will use OG$_k$ as a short hand notation for OG($k,2k$). Note that we have not yet specified the index $j$ that appears in the minor. This index will be determined by the multiplet on the external legs.

The orthogonal constraint is imposed by the degree $k(k+1)/2$ delta function:
\eq \label{orthogonal}
C\cdot C^{T}=\sum_{i=1}^{2k}C_{a i}C_{b i}=0\,.
\eqe
Note that the way that this constraint is written implies that the signature of the Grassmannian is Euclidean $\eta^{ij}=(+,+,\cdots,+)$. Later on in section~\ref{OrthogSec}, we will find it convenient to analytically continue to split signature $\eta^{ij}=(+,-,+\cdots,-)$. Note that here the signature refers to that of the Grassmannian, and not of the external data.  The orthogonal constraint implies non-trivial relationships among the minors. For example we have~\cite{SangminGrass}:
\eq
M_{i}M_{i+1}=M_{i+k}M_{i+k+1}(-1)^{k-1}\,.
\eqe
This identity exposes the cyclic-by-two-cite symmetry of eq.(\ref{GrassInt}), up to a definite sign, required by the amplitude in eq.(\ref{AmpDef}). More general identity for non-consecutive minors will be given, and discussed, in section~\ref{Sec:GenAmal}. One can count the dimension of the integral in eq.(\ref{GrassInt}), as
\eq
\frac{k(k-1)}{2}-(2k-3)= \frac{(k-3)(k-2)}{2}
\eqe 
where $(2k-3)$ are the constraints that arise from $\prod_{a=1}^k\delta^{2}(C_{a}\cdot \lambda)$ and $-3$ simply correspond to the constraints that are imposing momentum conservation on the external data $(\lambda_i)$, not on the Grassmannian. The final $(k-3)(k-2)/2$-dimensional integral is then localized by the zeroes of the minors. 

Here, we will be interested in the configuration of grassmanian manifold which is a result of this final localization. In other words, we will reverse the previous procedure and consider the top-cell \textit{first} being partially localized using the ${(k-3)(k-2)}/{2}$ number of zeroes in the minors. This leaves behind a $(2k-3)$-dimensional integral, subject to the constraints of the bosonic delta function $\delta(C\cdot \lambda)$. Thus the residues of the integral in eq.(\ref{GrassInt}), can be recast as a $(2k-3)$-dimensional submanifold of the orthogonal Grassmannian, subject to the final $(2k-3)$ bosonic degrees of freedom. As we will see, this $(2k-3)$-submanifold can be iteratively constructed.

It is convenient to use the GL($k$) symmetry to gauge fix the $k\times 2k$ matrix such that the Grassmannian takes the form 
\eq
C_{a i}=(I_{k\times k}, c),
\eqe
where $I_{k\times k}$ is the $k\times k$ identity matrix, and the matrix $c$ parametrizes the remaining $k^2$ degrees of freedom. The orthogonal constraint $Q(v,w)=0$ is now 
\eq
I_{k\times k}+c\cdot c^T=0
\label{RedConstraint}
\eqe
and thus $ic$ is simply an orthogonal matrix $O(k)$ which has two branches, $SO_+(k)$ and $SO_-(k)$. The two branches can be defined in a GL(k) invariant fashion as:
\eq
\frac{M_{\sigma}}{M_{\bar{\sigma}}}=\pm (i)^{k}\,.
\label{BranchDef}
\eqe
In the above $\sigma$ represent the set of columns entering the minor, while $\bar{\sigma}$ represent its complement. For example in G(3,6), if $\sigma=(1,2,4)$ then $\bar{\sigma}=(3,5,6)$. The fact that the orthogonal Grassmannian has two branches is directly related to the special properties of three-dimensional kinematics. To expose this connection, we take a closer look at OG$_2$.

\subsection{The branches of OG$_2$}
Let us study the top-cell of OG$_2$, which according to eq.(\ref{TopCellDim}) is one-dimensional. We begin with the following gauge fixed form:
\eq
C_{\alpha i }=\left(\begin{array}{cccc}1 & 0 &  c_{13} & c_{14} \\0 & 1 &  c_{23} & c_{24} \end{array}\right)
\label{GaugeDef}
\eqe
We will refer to such gauges where the columns that constitute unity are all adjacent as ``canonical gauge". Explicitly solving the orthogonal condition one finds:
\eqa
\nonumber&&\int dc_{13}dc_{14}dc_{23}dc_{24}\;\;\delta^{3}(CC^T)\;\star\\
&=&\int d\alpha d\beta\;\; \delta(1+\alpha^2+\beta^2)\left(\;\star\;\;\bigg|_{C=\left(\begin{array}{cccc}1 & 0 &  \alpha & \beta \\0 & 1 &  -\beta & \alpha \end{array}\right)}+\;\star\;\;\bigg|_{C=\left(\begin{array}{cccc}1 & 0 &  -\alpha & \beta \\0 & 1 &  \beta & \alpha \end{array}\right)}\right)
\label{FacGauge}
\eqae
where the two solutions correspond to the SO($2$)$_-$ and SO($2$)$_+$ part of the orthogonal grassmanian. 

Two branches of the orthogonal Grassmannian actually reflect the fact that there are two topologically distinct configurations for the external kinematics. To see this recall that:
\eq
\langle 12\rangle^2=\langle 34\rangle^2 \rightarrow \langle 12\rangle=\pm\langle 34\rangle\,.
\eqe
Thus we see that there are two inequivalent kinematic configurations. Now let us consider the first solution in eq.(\ref{FacGauge}), which through $\delta^{2}(C\cdot \lambda)$ enforces:
\eq
\lambda_1=\alpha \lambda_3 +\beta \lambda_4,\quad \lambda_2= -\beta\lambda_3 +\alpha \lambda_4
\eqe
One immediately sees that this implies
\eq
\langle 12\rangle=- \langle34\rangle.
\eqe
It is straightforward to see that the other branch implies that $\langle 12\rangle=\langle34\rangle$. Thus the different branches of OG$_2$ corresponds to the two distinct branches of the four-point momentum-space configurations!  More importantly: \textit{ the amplitude is required to live on both branches of OG$_2$}. Stating the obvious, if the amplitude is only represented on one branch of OG$_2$ it becomes non analytic as the amplitude will vanish in the other kinematic branch. Such non analyticity cannot be present for tree-level amplitudes.

At higher points, the fact that 3d kinematics has distinct branches can be understood as follows: in Minkowski space, a light-like vector can be written as $p_i^\mu=E_i(1, \cos\theta_i, \sin\theta_i)$. This means that projectively, three-dimensional massless kinematics are simply points populating the circle $S^1$. From
\eq
\langle ij\rangle=\sqrt{-2 p_i\cdot p_j}=i\sqrt{E_i\,E_j} \,\sin\left(\frac{\theta_i-\theta_j}{2}\right)
\eqe
we see that the sign of $\langle ij\rangle$  changes whenever the two points that represent $p_i$ and $p_j$ cross each other on the $S^1$:
\eq
\raisebox{-12mm}{\includegraphics[scale=0.45]{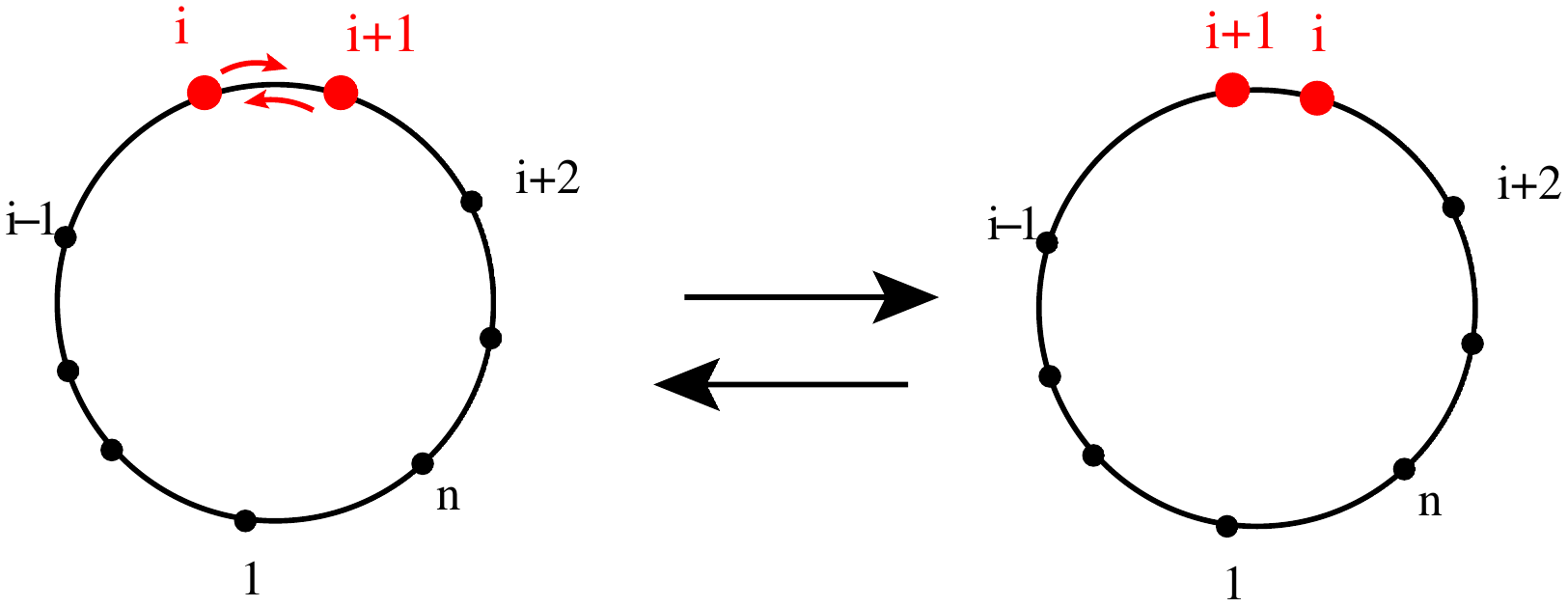}}
\label{3dcirc}
\eqe
Clearly two configurations eq.(\ref{3dcirc}) are topologically inequivalent. This can be made more precise. By judiciously adding $2\pi$ to the angles $\theta_i$, one can arrange the angles such that a given kinematics configuration has all angles strictly increasing according to their color ordering, i.e. $0<\theta_{i+1}-\theta_i<2\pi$. This gives a well-defined ``winding number" $w=(\theta_n-\theta_1)/(2\pi)$. Note that the configuration where the angles are strictly decreasing, is simply a reflection of the circle, and thus do not count as a new configuration. Now as the momenta of two points are interchanged, the winding number changes by one, indicating a distinct topological sector.

\subsection{The external states}
We now address what $j$ should be in eq.(\ref{GrassInt}). Unlike in $\mathcal{N}=4$ SYM, the on-shell degrees of freedom in ABJM theory are contained in two distinct multiplets, denoted by $\Phi$ and $\overline{\Psi}$. The scattering amplitude is then characterized by two distinct configurations: whether $\overline{\Psi}$ multiplets sit on even sites or odd sites. In terms of eq.(\ref{GrassInt}), this implies the following:
\begin{itemize}
  \item For $k=even$: $j=1$ if $\overline{\Psi}$ is on odd sites, while $j=2$ if otherwise
  \item For $k=odd$: $j=2$ if $\overline{\Psi}$ is on odd sites, while $j=1$ if otherwise
\end{itemize}
For example, at four points we have:
\eq
\mathcal{A}_4(\bar{1}2\bar{3}4)=\int \frac{dC}{(1,2)(2,3)}\delta^{3}(CC^T)\delta^{2|3}(C\cdot \Lambda),\;\quad\mathcal{A}_4(\bar{2}3\bar{4}1)= \int \frac{dC}{(2,3)(3,4)}\delta^{3}(CC^T)\delta^{2|3}(C\cdot \Lambda)
\eqe
Note use the canonical gauge in eq.(\ref{BranchDef}), we see that 
\eq
(1,2)=\alpha(3,4)
\eqe
where $\alpha=\pm$ denotes the branch. Thus $\mathcal{A}_4(\bar{1}2\bar{3}4)$ has the same measure as $\mathcal{A}_4(\bar{2}3\bar{4}1)$ does, except that instead of summing over the two branches, one now has to take the ``difference" of the two branches. To be more concrete, we begin with the canonical gauge whose ordering are now identified with the external legs. The OG$_2$ Grassmannian simply becomes:
\eq
\mathcal{A}_4(\bar{1}2\bar{3}4)=\frac{1}{2}\sum_{\alpha=\pm}\int \frac{d\theta}{s}\delta^{2|3}(C(\theta, \alpha)\Lambda),\quad \mathcal{A}_4(\bar{2}3\bar{4}1)=\frac{1}{2}\sum_{\alpha=\pm}\alpha\int \frac{d\theta}{s}\delta^{2|3}(C(\theta, \alpha)\Lambda)
\label{4pta}
\eqe
where we use the short hand notation $c_i\equiv \cos\theta_i$ and $s_i\equiv\sin\theta_i$ and 
\eq
C(\theta, a)=\left(\begin{array}{cccc}1 & 0 & i\alpha s & i\alpha c \\0 & 1 &  - ic &  is \end{array}\right)
\label{CanGauge}
\eqe
Thus we see that the two distinct four-point amplitude correspond to the  \textit{difference} or \textit{sum} of the two branches in $O(2)$. As another example consider the following ``cyclic" gauge:
\eq
C_{\alpha i }=\left(\begin{array}{cccc}1 &i\alpha s& 0  &  i\alpha c \\0 &  - ic & 1 & is \end{array}\right)
\label{CyclicGauge}
\eqe 
we see that the OG$_2$ integrand takes the form 
\eqa
\nonumber \mathcal{A}_4(\bar{1}2\bar{3}4)&=&\frac{1}{2}\sum_{\alpha=\pm}\int \frac{d\theta}{cs}\delta^{2|3}(C(\theta,\alpha)\Lambda)\\
\mathcal{A}_4(\bar{2}3\bar{4}1)&=&\frac{1}{2}\sum_{\alpha=\pm}\alpha\int \frac{d\theta}{cs}\delta^{2|3}(C(\theta,\alpha)\Lambda)\,.
\label{4ptb}
\eqae
Again the difference between the two amplitudes is the relative sign of the two branches.

The above discussion generalizes. At $(2k)$-point, if the leading singularities of $\mathcal{A}_{2k}(\bar{1}\cdots 2k)$ is given by the sum of a particular orthogonal Grassmannian configuration living on two branches, then $\mathcal{A}_{2k}(\bar{2}\cdots 1)$ is given by the difference, due to eq.(\ref{BranchDef}).

\subsection{The singularities of OG$_2$}
After solving the orthogonal constraint, we see that the OG$_2$ Grassmannian is now a one dimension integral with an integration measure that has non-trivial poles. A natural question would be: what do the the singularities in the measure correspond to? For the canonical gauge in eq.(\ref{4pta}), the singularity at $s=0$ reflects the divergence due to soft exchanges in the four-point amplitude. This can be seen from the fact that the pole on the bosonic delta functions enforce (with $s=0$,  $c=1$)
\eq
\lambda_1=-i \alpha \lambda_4,\; \lambda_2= i\lambda_3\;\;\rightarrow\;\; p_1=- p_4, \;p_2=-p_3:\;\;\vcenter{\hbox{\includegraphics[scale=0.7]{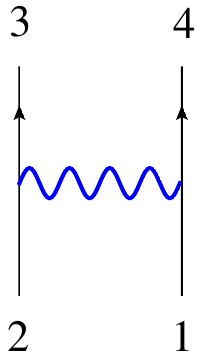}}}
\label{Sing1}
\eqe 
where the blue line indicates the soft gluon exchange. Thus we can graphically represent this as:
\eq 
{\partial}  \quad \vcenter{\hbox{\includegraphics[scale=0.9]{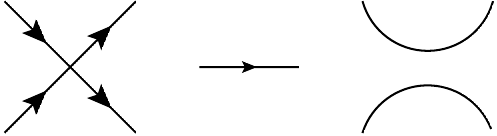}}}   \,,
\eqe
where the incoming arrows indicate the columns which are unity, and the out going arrows indicate the non-trivial entries. The operator $\partial$ denotes the singularity of the measure in a on-shell diagram. 

Let us now look at the cyclic gauge, eq.(\ref{CyclicGauge}). The measure contains two singularities, $c=0$ or $s=0$, on which the bosonic delta functions enforce $p_1+p_4=p_2+p_3=0$ or  $p_1+p_2=p_3+p_4=0$ respectively. This can be represented as:
 \eq
{\partial}\quad \vcenter{\hbox{\includegraphics[scale=0.8]{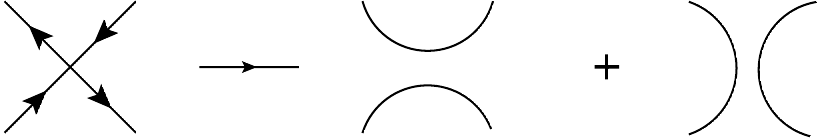}}}\,.
\eqe
Compared with the canonical gauge in eq.(\ref{Sing1}), it appears that the cyclic gauge contains one more singularity than the canonical gauge does, namely the $s$-channel soft singularity:
\eq
\vcenter{\hbox{\includegraphics[scale=0.7]{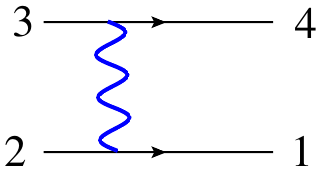}}}\,.
\eqe
This is perplexing since the two are related by a GL(2) gauge transformation. The GL(2) gauge transformation that is necessary to convert eq.(\ref{CanGauge}) to eq.(\ref{CyclicGauge}) is given by:
\eq
\left(\begin{array}{cccc}1 & i\tilde{s} & 0  &  i \tilde{c} \\0 &  i\tilde{c} & 1 & -i\tilde{s} \end{array}\right)=\left(\begin{array}{cc}1 & s/c\\0 & i/c\end{array}\right)\left(\begin{array}{cccc}1 & 0 &  i s & i c \\0 & 1 & -ic & is \end{array}\right)\,,
\eqe
where one can readily identify $\tilde{s}=-is/c$ and $\tilde{c}=1/c$. Now let's consider the singularity which was absent in the canonical gauge:  $p_1+p_2=p_3+p_4=0$. In the cyclic gauge, this corresponds to when $\tilde{c}=0$. However since $\tilde{c}=1/c$, the GL(2) gauge transformation becomes rank 1 if $\tilde{c}=0$. Thus on this singularity, the gauge transformation becomes non-invertible which explains why we did not see the singularity in the canonical gauge. 

However, we are not off the hook yet, as the missing singularity must hide in eq.(\ref{4pta}). After all, we are supposed to obtain the same four-point amplitude after using the bosonic delta functions to localize the final one-dimensional integral. The resolution is that the singularity does not appear as a singularity in the measure, but rather the degenerate limit of the bosonic delta functions. To see this, note that if $\lambda_1=i\lambda_2$, $\lambda_3=i\lambda_4$ the bosonic delta functions become 
\eq
\delta(i\lambda_2+\alpha i(is+c)\lambda_4),\;\delta( \lambda_2+(c +  is)\lambda_4)
\eqe
where for $\alpha=1$ the delta functions become degenerate. Thus the missing singularity appears as the degenerate-limit of the bosonic delta functions.\footnote{For $\lambda_1=i\lambda_2$, $\lambda_3=-i\lambda_4$, the degeneracy occurs in the other branch.}

The above discussion points out a unique aspect of ABJM amplitudes: the fundamental four-point amplitude has non-trivial singularities. This is reflected in the fact that the fundamental vertex, which is the top-cell of OG$_2$, is one-dimensional and the residue of the poles in the measure is non-trivial. From the Grassmannian point of view, the localization of the one-dimensional integral indicates that it has been localized to a special configuration. In the canonical gauge, eq.(\ref{4pta}), the configuration of $C_{ai}$ on the pole $s=0$ (and $c=1$) is :
\eq
\left(\begin{array}{cccc}1 & 0 & i\alpha s & i\alpha c \\0 & 1 &  - ic &  is \end{array}\right)\;\;\underrightarrow{s=0}\;\;\left(\begin{array}{cccc}1 & 0 & 0 & i\alpha  \\0 & 1 &  - i &  0 \end{array}\right)\,.
\eqe 
One can see that as $s=0$, column $2$ and $3$ becomes linearly dependent, and similarly for columns $1$ and $4$. Thus the singularity corresponds to special configurations of OG$_2$ for which linear-dependency develops among the columns. This special configuration can be thought of as the co-dimension one boundary of the top-cell in OG$_2$. If we restrict ourselves to the linear dependency of adjacent columns, which for $k=2$ is equivalent to the vanishing of a consecutive minor, then naively there should be 4 such boundaries. However, due to eq.(\ref{BranchDef}) which is implied by the orthogonal constraint, only two are independent. Thus there are 2 co-dimension one boundary for the top-cell of OG$_2$, and each correspond to a distinct soft-gluon exchange divergence in the amplitude! For general OG$_k$,  the linear dependency of the columns will encode even more structure that is reflected in the scattering amplitude. To expose this relation, we proceed to construct higher-point amplitudes, using the language of gluing together fundamental OG$_2$'s.

\section{OG$_k$ as on-shell diagrams:} \label{section:OGk}
In the work of Arkani-Hamed et al.~\cite{NimaBigBook}, it was demonstrated that by successively gluing together fundamental three-point G(2,3) and G(1,3) grassamannians, one builds a sub-manifold of the Grassmannian which is corresponding to the S-matrix of $\mathcal{N}=4$ SYM. Generically, it will be a sub-manifold since the dimension is lower than the dimension of the top-cell, indicating it correspond to a boundary of the top-cell. That such connection can be made is due to the identity between the kinematics of the gluing of a G(2,3) and a G(1,3), with the BCFW~\cite{BCFWtree} deformation of the external legs:
\eq
\vcenter{\hbox{ \includegraphics[scale=0.85]{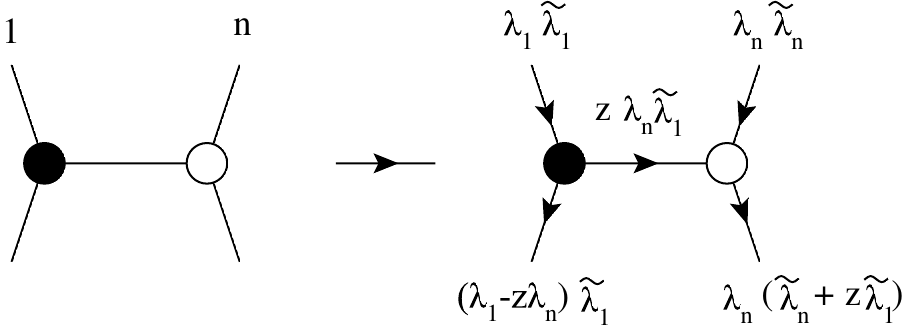} }} \,,
\eqe
where the arrows indicate the momentum flow, and we've explicitly spelled out the kinematics of each leg in the diagram implied by the constraint imposed from the bosonic delta functions in each vertex. From this point of view, the diagrams that are built from successively attaching such ``BCFW" bridges, is equivalent to the iterative construction of scattering amplitudes using lower-multiplicity, or lower-loop level, on-shell building blocks. In other words, each term in the BCFW construction to tree~\cite{BCFWtree} and planar loop~\cite{NimaAllLoop} can be recast into a particular trivalent, ``on-shell" diagram.

Through this gluing procedure, one constructs an $(n_f-1)$-dimensional G($k,n$) Grassmannian, where $n_f$ is the number of faces in the diagram. This sub manifold is then subject to $(2n-4)$ constraints that arise from the bosonic delta functions.  Note that the dimensions of the sub manifold can be greater, equal or less than the number of constraints, which correspond to multi-dimensional integrals, a rational function, or a rational function with constraints imposed on the external data beyond that of momentum conservation, respectively.  

Similar proposal was made for ABJM theory, based on the merging of fundamental OG$_2$ Grassmannians. The kinematics of the BCFW deformation is now mapped to the constraint imposed by a fundamental four-vertex:
\eq
\vcenter{\hbox{ \includegraphics[scale=0.8]{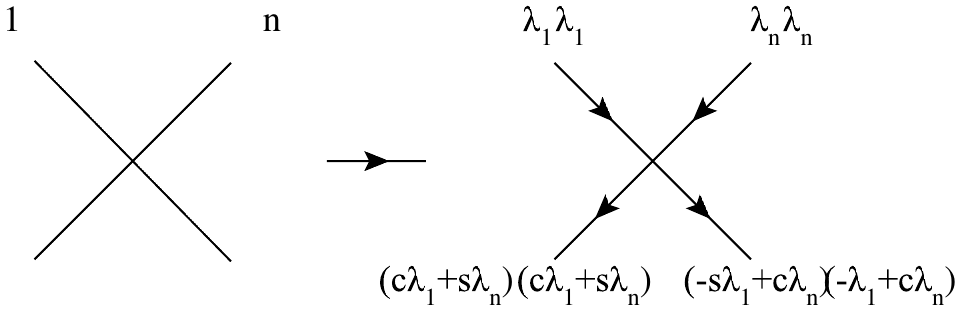} }} \,,
\eqe 
where we've used the constraints on the external data imposed in the canonical gauge. Thus it is expected that each individual term in the BCFW recursion of a ABJM amplitude~\cite{Gang}, can be represented by an on-shell diagram constructed from the gluing of four-vertices.\footnote{Note because of chirality, $ \Phi$ must be connected with $\bar{\Psi}$, so not any gluing is allowed. The ``wrong" diagrams could appear when we consider non-planar diagrams.} In this section we will study these on-shell diagrams in more detail. 

\subsection{General amalgamation of OG$_2$s\label{Sec:GenAmal}}
To illustrate the gluing procedure and the buildup of representative OG$_k$s, we begin with the merging of two OG$_2$s to form an OG$_3$. This gluing procedure is represented graphically as:
\eq
\vcenter{\hbox{ \includegraphics[scale=0.9]{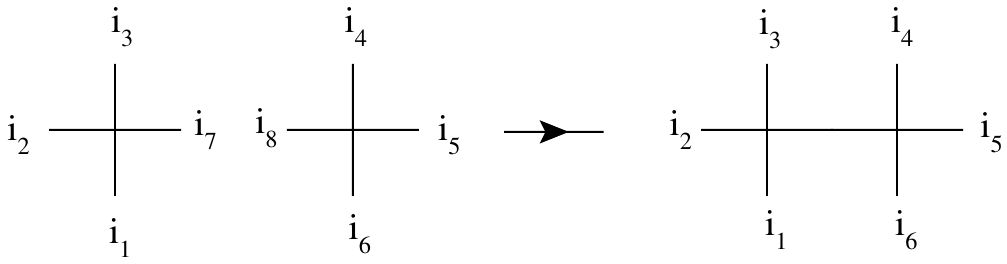} }}\,.
\eqe
What this means is that we take the two OG$_2$s, each of which is a one-dimensional integral with degree 4 bosonic delta function imposing constraints on the external data, identify one external leg from each of the two OG$_2$s and integrate away the on-shell data of the common external leg. More explicitly, in the above diagram, we identify $\Lambda_{7}=\Lambda_{I}$ and $\Lambda_{8}=\Lambda_{-I}$, where $\Lambda_{-I}=i\Lambda_{I}$ due to momentum conservation. First integrating away the bosonic part of $\Lambda_I$ we find that the degree $8$ bosonic constraints collapse to degree $3$:\footnote{As usual in the Grassmannian formulation here we treat delta-functions as contour integrals, so there is no absolute value on the Jacobian. }
\eqa
\nonumber \int d^2\lambda_I&&\prod_{a=1}^2\delta^2(C_{a 1}\lambda_{1}+C_{a 2}\lambda_{2}+C_{a 3}\lambda_{3}+C_{a I}\lambda_{I})\\
\nonumber &&\prod_{b=1}^2\delta^2(C_{b 4}\lambda_{4}+C_{b 5}\lambda_{5}+C_{b 6}\lambda_{6}+iC_{b I}\lambda_{I})\\
&=&\prod_{a=1}^3\frac{1}{C_{a I}^2}\delta^2(\sum_{j=1}^6 \tilde{C}_{a j}\lambda_{j})\,.
\eqae
where $\tilde{C}_{\alpha j}$ are the new coefficients in front of the $\lambda_{j}$s. Thus we see that in the end we've obtained a new G(3,6) Grassmannian. Combining with the fermionic part of the delta functions, the gluing procedure is written as:
\eq
\int d^{2|3}\Lambda_I \delta^{4|6}(C_{v_1}\Lambda_{v_1})\delta^{4|6}(C_{v_2}\Lambda_{v_2})=\prod_{\alpha=1}^3 C_{a I} \delta^{2|3}(\sum_{j=1}^6 \tilde{C}_{a j}\Lambda_{j})\,,
\eqe
where we've used $v_1$ and $v_2$ to label the external legs on each individual vertex. Note that there will be an extra Jacobian factor $C_{a I}$, which originated from the mismatch between bosonic and fermionic delta functions. Thus the merging procedure is nothing but a union of linear constraints on the external data, when two such external legs are identified.

We can generalize this amalgamation to merge a OG$_k$ and OG$_k'$ to form OG$_{k+k'-1}$. Starting by first combining the two Grassmannian into a OG$_{k+k'}$
\eq
C_{\alpha i}=\left(\begin{array}{cccccc}c_{11} & \cdots & c_{1,2k} & 0 & 0 & 0 \\ \vdots & \vdots & \vdots & 0 & 0 & 0 \\ c_{k1} & \cdots & c_{k,2k} & 0 & 0 & 0 \\ 0 & 0 & 0 & c_{k+1,2k+1}& \cdots &  c_{k+1,2k+2k'} \\0 & 0 & 0 & \vdots & \vdots & \vdots \\0 & 0 & 0 & c_{k+k',2k+1} & \cdots &  c_{k+k',2k+2k'}\end{array}\right)
\eqe
The non-vanishing minors of OG$_{k+k'}$ are simply given by the product of a minor in OG$_k$ and a minor in OG$_{k'}$. Now we identify the spinors of two external legs, say $2k$ and $2k+1$, integrating away the common spinor one obtains an OG$_{k+k'-1}$ Grassmannian whose entry is given by
\eq
C_{\alpha i}=\left(\begin{array}{cccccc}c_{21}-\frac{c_{11}c_{2,2k}}{c_{1,2k} } & \cdots & c_{2,2k-1}-\frac{c_{1,k-1}c_{2,2k}}{c_{1,2k} } & 0 & 0 & 0 \\ \vdots & \vdots & \vdots & 0 & 0 & 0 \\ c_{k1}-\frac{c_{11}c_{k,2k}}{c_{1,2k}} & \cdots & c_{k,2k-1}- \frac{c_{1,2k-1}c_{k,2k}}{c_{1,2k} }  & 0 & 0 & 0 \\ -i\frac{c_{11}}{c_{1,2k}}c_{k+1,2k+1} & \cdots & -i\frac{c_{1,2k-1}}{c_{1,2k}}c_{k+1,2k+1} & c_{k+1,2k+2}& \cdots &  c_{k+1,2k+2k'} \\ \vdots & \vdots & \vdots & \vdots & \vdots & \vdots \\ -i\frac{c_{11}}{c_{1,2k}}c_{k+k',2k+1} & \cdots & -i \frac{c_{1,2k-1}}{c_{1,2k}}c_{k+k',2k+1}  & c_{k+k',2k+2} & \cdots &  c_{k+k',2k+2k'}\end{array}\right)
\label{AmalResult}
\eqe 
As one can straightforwardly verify, the minor of the final OG$_{k+k'-1}$ Grassmannian is given by a linear combination of minors in the parent OG$_{k+k'}$:
\eq
(i_1,\cdots, i_{k+k'-1})=\frac{1}{c_{1,2k}}\left((i_1,\cdots,i_{k+k'-1},A)+i(-1)^{k+k'-1}(i_1,\cdots,i_{k+k'-1},B)\right)\,.
\label{AmalgaRule}
\eqe
where we've used $A,B$ to represent the identified columns $i_{2k},\,i_{2k+1}$.

A non-trivial question is whether orthogonality is preserved under amalgamation. As we combine OG$_k$ and OG$_{k'}$ into OG$_{k+k'}$, orthogonality is trivially preserved. The orthogonality of the final OG$_{k+k'-1}$ can be shown by following ref.~\cite{NimaBigBook} and rewrite the orthogonal condition as the following constraint for the minors of a OG$_{k}$ Grassmannian:
\eq
\sum_{a\in col\;{\rm OG}_{k}}(i_1,\cdots,i_{k-1},a)(j_1,\cdots, j_{k-1},a)=0\,,
\label{OrthoMinor}
\eqe 
where $\sum_{a\in col\;{\rm OG}_{k}}$ indicates a sum over all columns in OG$_k$, and $\{ i_1,\cdots,i_{k-1}\}$ and $\{j_1,\cdots,j_{k-1}\}$ can be arbitrary. A formal proof of this equivalence is given in appendix~\ref{AppendixA}. Using the amalgamation rule in eq.(\ref{AmalgaRule}) it is straightforward to show that
\eqa
\nonumber&&\sum_{a\in col\;{\rm OG}_{k+k'-1}}(i_1,\cdots,i_{k+k'-2},a)(j_1,\cdots, j_{k+k'-2},a)
=0\,,
\eqae
where one uses eq.(\ref{OrthoMinor}) to convert $\sum_{a\in col\;{\rm OG}_{k+k'-1}}$ into the sum of $a=i_{2k}$ and $a=i_{2k+1}$.

Thus starting with the fundamental four-point vertex, we can successively build up more complicated on-shell diagrams by gluing multiple four-point vertices. Each vertex contains a one-dimensional integral, the degree of freedom in the top-cell of OG$_{2}$, while subjecting to a degree $4$ bosonic delta function constraint. Each internal line that connects the two-vertex introduces a two dimensional integral $\int d^2\lambda_I$. Among the delta functions, 3 of them corresponds to overall momentum conservation, thus the total number of bosonic delta functions remaining after localizing the $\int d^2\lambda_I$ integrals is given by $4n_v-2n_I-3=2k-3$, where $n_v$ is the number of vertices $n_I$ is the number of internal lines and $n=2k$. Thus a given on-shell diagram is an $n_v$-dimensional integral subject to $(2k-3)$ constraints. Recall that the top-cell of OG$_k$ is $k(k-1)/2$-dimensional, in general the dimension of the on-shell diagrams will be less than that of the top-cell, and thus will correspond to boundaries of the top-cell.  

In practice the final form of $C_{ai}$ can be easily read off from the on-shell diagram. Here we consider two particular gauge fixing of OG$_2$ that allows one to straightforwardly read off the final answer. As a bookkeeping device for the gauge choice, one assigns two incoming and two out-going arrows for each vertex. For each internal line, the arrows must point to a definite direction. The incoming arrows will indicate the legs which correspond to the unity columns in OG$_{2}$. For each vertex, there are two-possible assignments:
\eq
\vcenter{\hbox{ \includegraphics[scale=1]{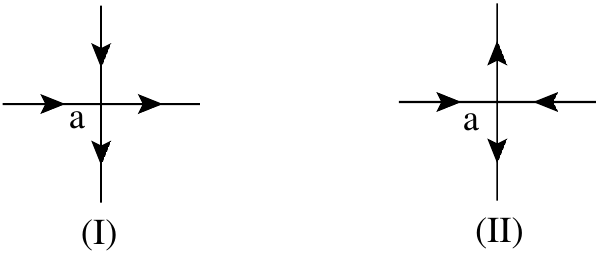} }}\,.
\label{4ptRule}
\eqe
In the above diagram (I) correspond to the canonical gauge in eq.(\ref{CanGauge}) while diagram (II) correspond to the cyclic gauge in eq.(\ref{CyclicGauge}). As one builds up higher-point diagram by gluing fundamental four-point vertices, the arrows in the diagram forms paths that connects points on the boundary through the diagram and back to the boundary:
\eq
\vcenter{\hbox{ \includegraphics[scale=0.8]{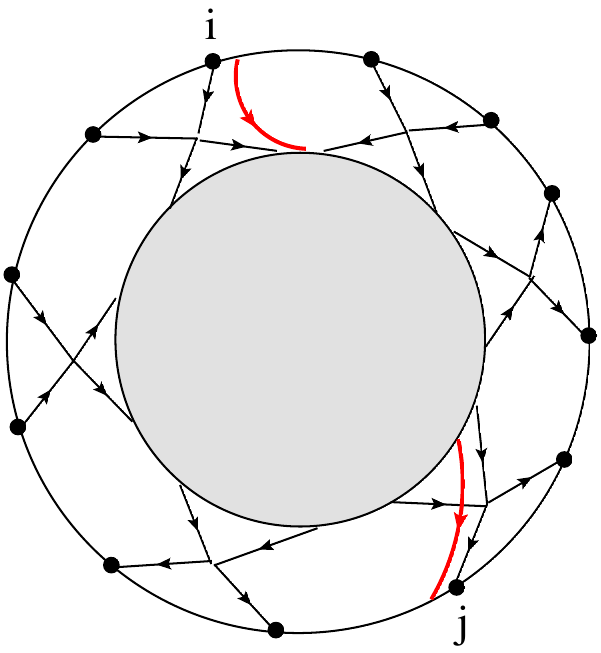} }}\,.
\eqe
Note that due to eq.(\ref{4ptRule}) there will always be $k$ ``sources" and $k$ ``sinks" on the boundary. The $c_{ij}$ in $C_{ai}$ can then be readily read off by summing over all paths that connect $i$ and $j$ in the on-shell diagram, with the appropriate $ic_{a}$ and $is_{a}$ factors assigned at each vertex, as well as an extra factor of $i$ each time an internal line is crossed:
\eq
c_{ij}=\sum_{\beta\in{\rm paths}}(i)^{n_{I\beta}} \prod_{n_{v\beta}} (c_{n_{v\beta}}\, {\rm or}\, s_{n_{v\beta}}) 
\eqe
where $\beta$ labels the paths, $n_{I\beta}$ are the number of internal lines along the path, and $n_{v\beta}$ labels the number of the vertices that are present along the path. For the particular gauges in eq.(\ref{CanGauge}) and eq.(\ref{CyclicGauge}), the extra factors that arise from crossing each vertex is: 
\eq
\vcenter{\hbox{ \includegraphics[scale=1]{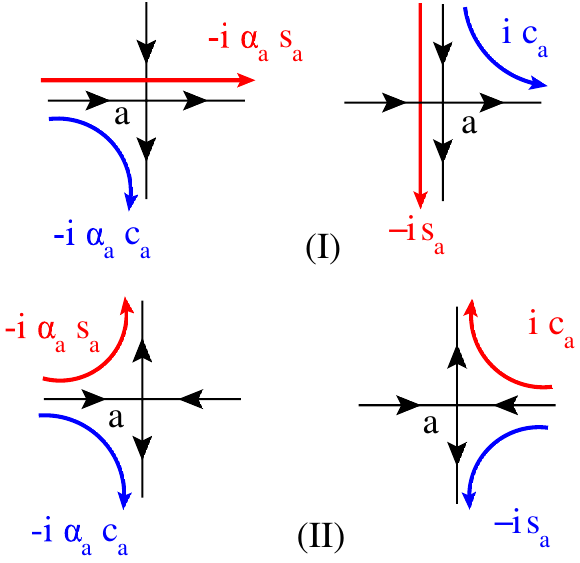} }}\,,
\label{VertexRule}
\eqe
where again $\alpha$ parametrize the branch for each individual vertex. As an example we consider the six-point factorization diagram:
$$\includegraphics[scale=1]{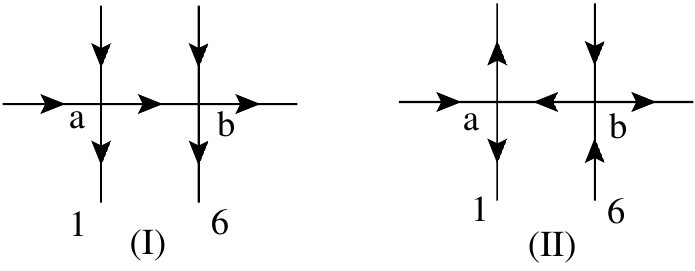}$$ 
As $n_v=2$ this diagram represents a two-dimensional integral subject to $2k-3=3$ constraints, the system is over constrained and implies non-trivial constraint on the external data beyond that of momentum conservation.  As momentum conservation and on-shellness is enforced at each vertex, the extra constraint is simply $(p_1+p_2+p_3)^2=0$. The explicit form of $C_{ai}$, both in canonical and cyclic gauge, are given as:
\eqa
\nonumber (I)\quad C_{\alpha i}&=&\left(\begin{array}{cccccc}-i\alpha_a c_a & 1 & 0 & 0 & -i\alpha_a\alpha_bs_as_b & -i\alpha_a\alpha_bs_ac_b  \\  -is_a & 0 & 1 & 0 & i\alpha_bc_as_b & i\alpha_bc_ac_b \\ 0 & 0 & 0 & 1 & ic_b & -is_b  \end{array}\right)\\
(II)\quad C_{\alpha i}&=&\left(\begin{array}{cccccc}-i\alpha_ac_a & 1 & -i\alpha_as_a & 0 & 0 & 0 \\ -i\alpha_bs_ac_b & 0 & i\alpha_bc_ac_b & 1 & -i\alpha_bs_b & 0 \\-is_as_b & 0 & ic_as_b & 0 & ic_b & 1\end{array}\right)
\eqae
If a given path goes through a closed loop, one simply obtains a geometric sum:
\eq
c_{ij}=c^{(0)}_{ij}+\frac{c^{(1)}_{ij}}{1-\Gamma_{ij}}\,,
\eqe
where $c^{(0)}_{ij}$ are paths that do not go through closed loops while $c^{(1)}_{ij}$ correspond to those that goes through one, and $\Gamma_{ij}$ are the product of variables in the given loop. Note that due to the mismatch of fermionic and bosonic delta function, the gluing procedure will generate a Jacobian factor $J=(1-\Gamma_{ij})$. 

In conclusion, by gluing the on-shell diagrams in the cyclic gauge, one obtains a $n_v$ dimensional integral:
\eqa
\nonumber \mathcal{L}_{k}&=&\sum_{\{\alpha_a\}}\int \left[\prod_{a=1}^{n_v}\frac{d\theta_a}{2g(\alpha_a)s_ac_a}\right]J\prod_{m=1}^k\delta^{2|3}\left( C_m(\theta_a,\alpha_a)\cdot \Lambda \right)\\
&=&\sum_{\{\alpha\}}\int \left[\prod_{a=1}^{n_v}\frac{1}{2g(\alpha_a)}d\log\left(\tan\theta_a\right)\right]J\prod_{m=1}^k\delta^{2|3}\left( C_m(\theta_a,\alpha_a)\cdot \Lambda \right)
\label{OnShellRep}
\eqae
where we've chosen the gauge in eq.(\ref{CyclicGauge}) for all vertices, $g(\alpha_a)=1\,{\rm or}\,\alpha_a$ depending on the chirality of the legs on the individual vertex, and $\sum_{\{\alpha_a\}}$ indicates that one is required to sum over all $2^{n_v}$ distinct configurations of $\{\alpha_a\}$. In other words, the terms in the ABJM tree-level BCFW recursion can be written as an $n_v=2k-3$-dimensional integral with a simple canonical $d\log$ measure ! The Jacobian factor $J$ may seem to be a breakdown of the $d\log$ form, it's actually the opposite as we will see later in the discussion of bubble reductions where the $d\log$ form matters more since it is a true integral there. The Jacobian factor is precisely needed to bring the integration measures after bubble reductions into a nice $d\log$ form.

\subsection{On-shell diagram as BCFW representation of tree-level amplitudes} \label{section:treelevel}
With detailed understanding of the structure of fundamental vertex, we are ready to construct on-shell diagram representation of all tree-level amplitudes in ABJM theory by the means of BCFW recursion relation. As discussed in \cite{N=6Note, NimaBigBook} the three-dimensional BCFW deformation is precisely the constraint imposed by the bosonic delta function of the fundamental four-vertex in the canonical gauge. Thus  the tree-level amplitudes of ABJM can be given by the following recursion:
\eq
\mathcal{A}_n=\sum_{i=3}^{2k-3}\quad \vcenter{\hbox{ \includegraphics[scale=0.8]{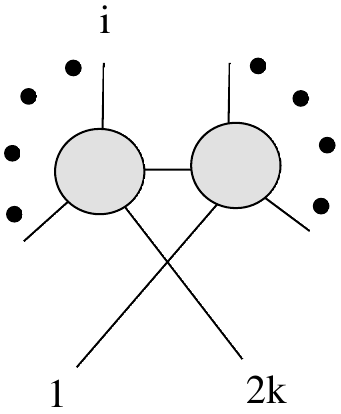} }}
\eqe
Since the four-point amplitude is given by the fundamental OG$_2$ Grassmannian, through the gluing procedure discussed previously, the  recursion formula generates a $2k-3$-dimensional representation of OG$_ {k}$. Here we would like to analyse the properties of this representation.    

\subsubsection{A six-point amplitude and its singularities}
Let us begin with the six-point BCFW on-shell diagram, 
\eq 
\vcenter{\hbox{ \includegraphics[scale=0.8]{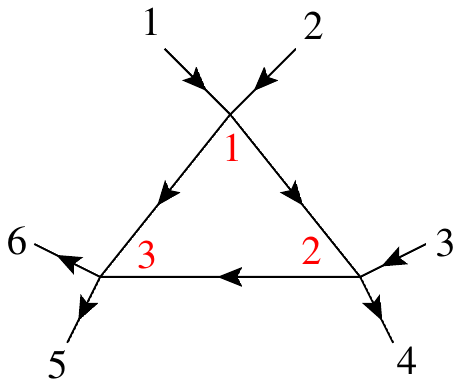} }} \, .
\label{6ptBCFW}
\eqe
It is straightforward to work out the corresponding Grassmannian according to eq.(\ref{VertexRule}), we obtain, 
\eq \label{sixptexample}
C_{ai} =\left(\begin{array}{cccccc}1 & 0 & 0 & -i c_{1, 4} & -i c_{1,5} & -i c_{1,6} \\0 & 1 & 0  & -i c_{2, 4} & -i c_{2,5} & -i c_{2,6} \\0 & 0 & 1 & -i c_{3,4} & -i c_{3,5} & -i c_{3,6}  \end{array}\right) \, ,
\eqe
where
\eqa
&& c_{1,4} = \alpha_1 s_1 \alpha_2 s_2, \, c_{1,5} = -\alpha_1\alpha_2 s_1 c_2 c_3 + \alpha_1\alpha_3 c_1  s_3, \,
c_{1,6} = \alpha_1\alpha_2 s_1  c_2 s_3 + \alpha_1\alpha_3 c_1  c_3 \nonumber \\ 
&& c_{2,4} = -\alpha_2 c_1  s_2, \, c_{2,5} = \alpha_2 c_1  c_2 c_3 +\alpha_3 s_1  s_3, \,
c_{2,6} = -c_1 \alpha_2 c_2 s_3 +\alpha_3 s_1  c_3 \nonumber \\ 
&& c_{3,4} = -c_2, \, c_{3,5} = - s_2 c_3 , \,
c_{3,6} =  s_2 s_3 \, . 
\label{6ptPa}
\eqae
One can straightforwardly test the orthogonality of the above OG$_3$, and that $M_{4}/M_1=-i\alpha_{1}\alpha_{2}\alpha_3$, as expected from eq.(\ref{BranchDef}). Not surprisingly, that the non-trivial part of the above Grassmannian, eq.~(\ref{sixptexample}), can be decomposed into a direct product of three two-dimensional rotations, 
\eqa
\left(\begin{array}{ccc}  c_{1, 4} &  c_{1,5} &  c_{1,6} \\  c_{2, 4} &  c_{2,5} &  c_{2,6} \\  c_{3,4} &  c_{3,5} &  c_{3,6}  \end{array}\right) = 
 \left(\begin{array}{ccc} \alpha_1 s_1 & \alpha_1 c_1 & 0 \\ -c_1 & s_1 & 0 \\ 0 & 0 & 1  \end{array}\right)
\cdot \left(\begin{array}{ccc} \alpha_2 s_2 & \alpha_2 c_2 & 0 \\ 0 & 0 & 1 \\ -c_2 & s_2 & 0  \end{array}\right)
\cdot \left(\begin{array}{ccc} 1 & 0  & 0 \\ 0 & -c_3 & s_3 \\ 0 & \alpha_3 s_3 & \alpha_3 c_3  \end{array}\right) \, .
\eqae
This fact is rather general, one can always decompose higher-point Grassmannian into direct product of lower-point ones, except for diagrams that involve closed loops.

The diagram in \ref{6ptBCFW} can be interpreted as a BCFW shift on legs $1$ and $2$. Thus we can identify:
\eqa
\nonumber \mathcal{A}_6(\bar{1},\cdots,6)&=&\sum_{\{\alpha_a\}}\alpha_2\int \left[\prod_{a=1}^{3}\frac{d\theta_a}{2s_a}\right]\prod_{m=1}^3\delta^{2|3}\left( C_m(\theta_a,\alpha_a)\cdot \Lambda \right)\\
\mathcal{A}_6(\bar{2},\cdots,1)&=&\sum_{\{\alpha_a\}}\alpha_1\alpha_3\int \left[\prod_{a=1}^{3}\frac{d\theta_a}{2s_a}\right]\prod_{m=1}^3\delta^{2|3}\left( C_m(\theta_a,\alpha_a)\cdot \Lambda \right)
\label{6ptonShell}
\eqae
Note that since the branch of the OG$_3$ is determined by $i\alpha_{1}\alpha_{2}\alpha_3$, we see that the integration measure of $\mathcal{A}_6(\bar{2},\cdots,1)$ has an extra minus sign compared to $\mathcal{A}_6(\bar{1},\cdots,6)$ depending on the branch, as promised.

Before moving on, let us take a brief pause and consider the singularities in the on-shell form in eq.(\ref{6ptonShell}). The singularity at $s_1=0$ correspond to the opening of the BCFW vertex, since from eq.(\ref{Sing1}), the residue of this singularity can be represented as:
\eq \label{sixptfactorization}
\vcenter{\hbox{ \includegraphics[scale=0.7]{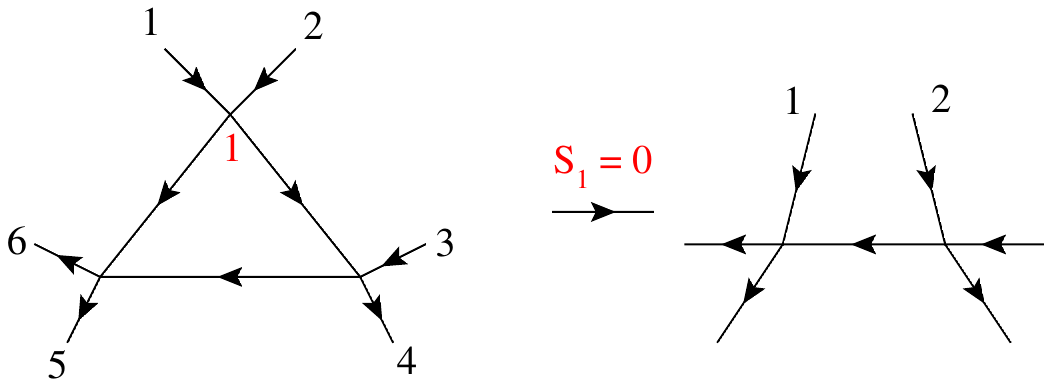} }}
\eqe
which is simply the product of two tree-level amplitudes with one additional constraint on the external data beyond momentum conservation, $(p_2+p_3+p_4)^2=0$. On the other hand, if we instead look at the singularity that correspond to $s_2=0$, one finds:
\eq \label{sixptfactorization2}
\vcenter{\hbox{ \includegraphics[scale=0.7]{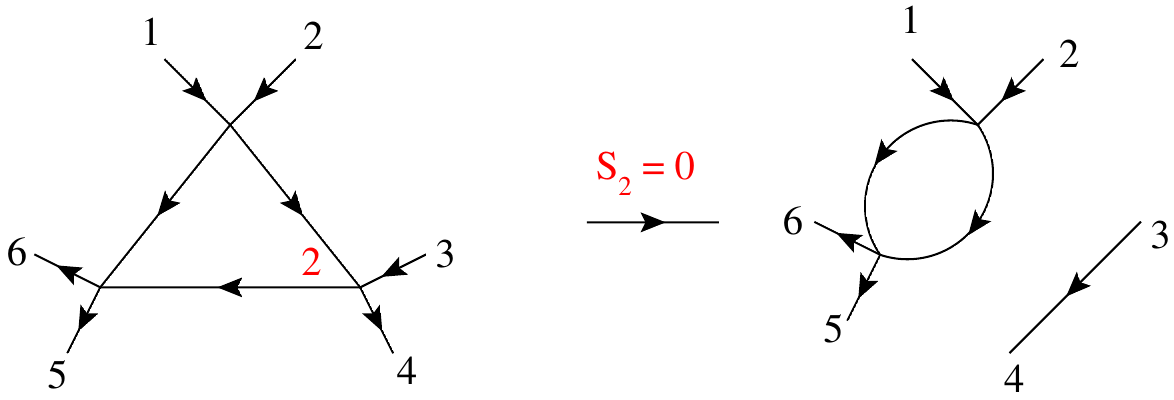}  }} \,.
\eqe
This result appears to be perplexing on two fronts. First of all, the BCFW factorization singularity similar to that of $s_1=0$ is missing. Second of all, the residue of this singularity gives a bubble diagram which correspond to a one-dimensional integral, which is rather peculiar given that this is a singularity of a rational function. The first puzzle is simply a reflection of the fact that we are using a local chart for the particular OG$_3$. The factorization singularity of vertex 2 can be explicitly seen by cyclically permuting the gauge choice in diagram \ref{6ptBCFW}.

The second puzzle is more subtle. Compared with the residue in eq.(\ref{sixptfactorization}) which is a rational function with one extra constraint on the external data, the residue in eq.(\ref{sixptfactorization2}) is a one-dimensional integral, which implies that it must impose 2 extra constraint on the external data. The degree two constraint is simply 
\eq
\lambda_{3}^a\pm i \lambda_4^a=0
\eqe 
where the $\pm$ comes from whether $c_2=\pm1$. This constraint should be familiar to us by now, it is nothing but the kinematics for a soft gluon exchange in the six-point amplitude! As we show in appendix \ref{SoftGlue}, the residue of this singularity is proportional to a four-point tree amplitude. This lead us to conclude that the bubble diagram must be proportional to the four-point tree-level amplitude! Indeed we will show in the next subsection \ref{Reduced} that through a change of variables, the bubble diagram is equivalent to a one-dimensional integral times a OG$_2$.  

Finally, note that by considering the singularity that corresponds to the BCFW factorization at each vertex in diagram, we can see the presence of all factorization channels in diagram \ref{6ptBCFW} . 
\subsubsection{The general $2k$-point amplitude}
From the above discussion, we've seen two subtleties in identifying the singularities of the on-shell diagram with the physical singularities: 
\begin{itemize}
  \item Only some singularities are manifest in any local chart. The fact that all singularities of the on-shell diagram are present is equivalent to the GL(k) gauge invariance which allows us to establish the equivalence of different local charts. 
  \item Not all singularities correspond to one constraint on the external data. Some singularities impose more than one constraint, and can be identified by simply noting that the residue is a manifold with one extra degree of freedom compared to the parent diagram.  
\end{itemize}
These features generalize to higher multiplicity BCFW diagrams.

Following the same process we can construct all tree-level amplitudes in ABJM theory by attaching BCFW bridges to the factorization channels. It is straightforward to find that the total number of on-shell diagrams for a $(2p+4)$-point tree-level amplitude is 
\eq \label{BCFWnumber}
(2p)!/(p!(p+1)!),
\eqe 
where $p=0, 1,2, \ldots $. For instance when $p=2$, namely eight-point tree-level amplitude, there are two diagrams. Here is one possible on-shell diagram representation of this amplitude,  
\eq  \label{8pttree}
\vcenter{\hbox{ \includegraphics[scale=0.5]{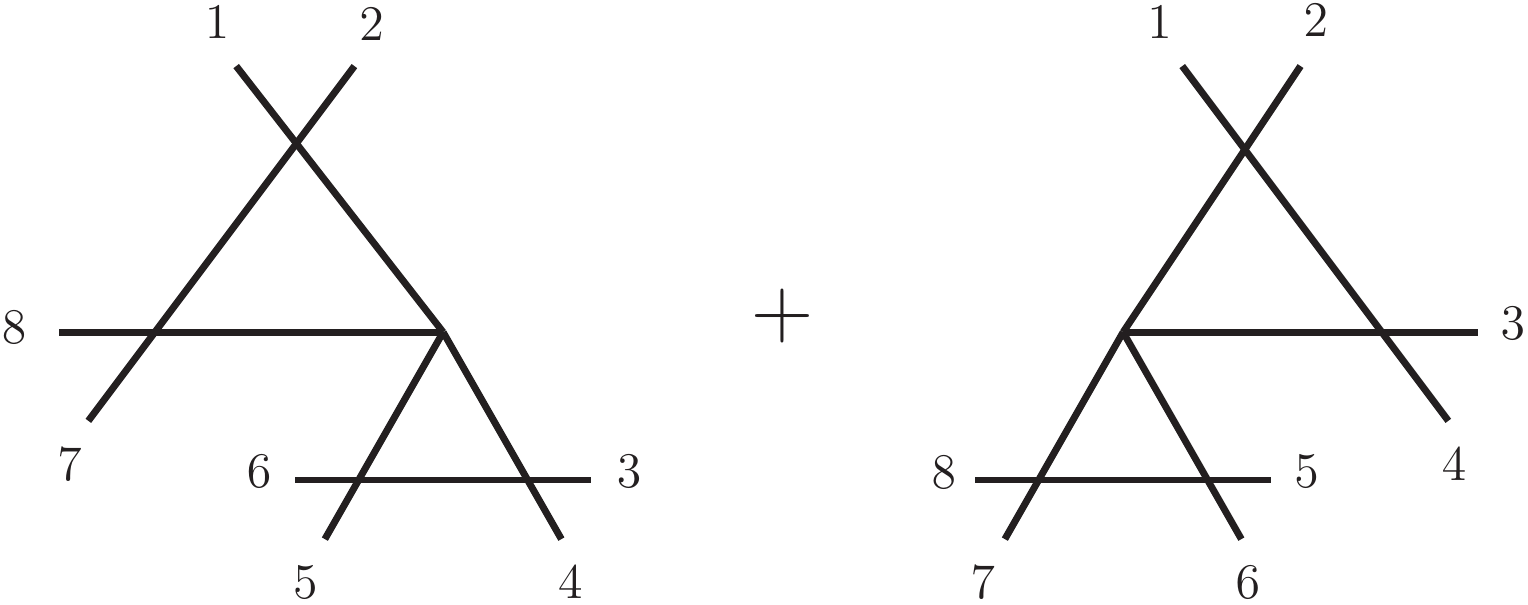} }}
\eqe
Again one can find all eight factorization channels of the eight-point amplitude manifestly by opening up four possible BCFW bridges of the on-shell diagrams. As discussed previously in the example of six-point amplitude, eq.~(\ref{sixptfactorization2}), there are two ways of opening up internal vertex $I$. One leads to a spurious pole, 
\eq  \label{8pttreespurious}
\vcenter{\hbox{ \includegraphics[scale=0.45]{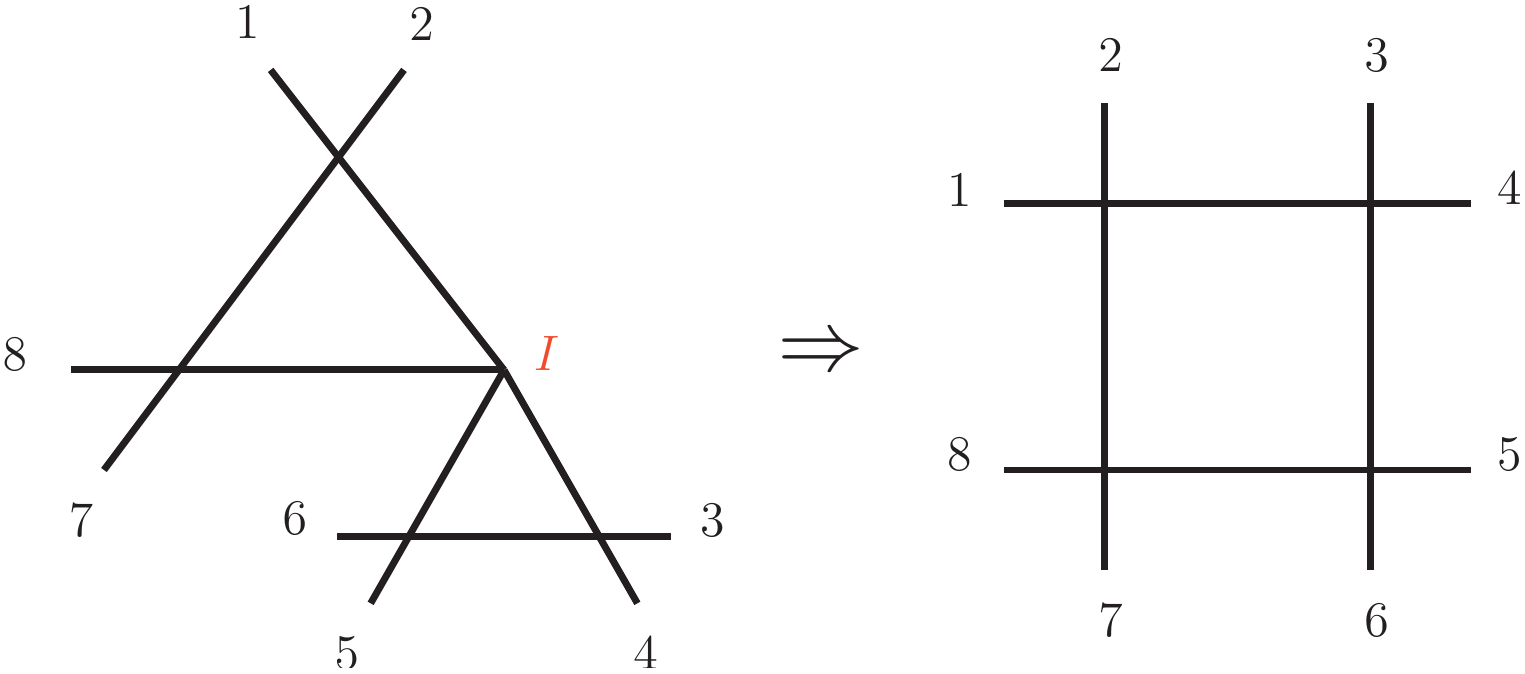} }} \, .
\eqe
It is a spurious pole of the tree-level amplitude that's because both diagrams contain the same singularities, and thus they cancel out in pair. The other singularity 
\eq  
\vcenter{\hbox{ \includegraphics[scale=0.5]{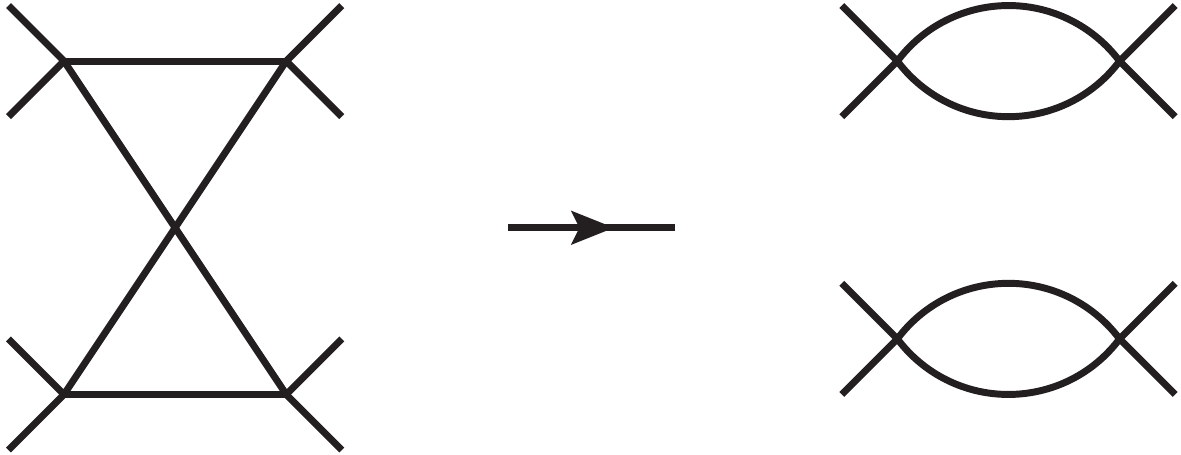} }}\, ,
\eqe
again corresponds to a soft exchange. Note that since both bubbles are one-dimensional integrals, this implies that there must be 3 extra conditions aside from momentum conservation. These 3 extra conditions can be readily identified as the momentum conservation of the four external momenta of any one of the bubbles.

By examining above simple examples, we note some nice and intriguing facts about the on-shell diagrams of tree-level amplitudes in ABJM theory. Firstly the on-shell diagrams only involve triangles, and all the triangles are connected with each other through shared vertices, not by shared lines. Secondly both the six- and eight-point amplitudes are manifestly cyclic symmetry under shift by two-site permutation, $i \rightarrow i+2$. We like to emphasize that both properties are rather surprising, in particular comparing from what we learned in $\mathcal{N}=4$ SYM: It is certainly impossible to represent all tree-level amplitudes in $\mathcal{N}=4$ SYM by boxes only, and cyclic symmetry is revealed only after a series of equivalence moves between on-shell diagrams. Here, we found in ABJM theory there is at least one representation of BCFW recursion relation which makes the required cyclic by two-site symmetry manifest, and thus all factorization channels are manifestly present.

The lesson we learned from special examples is actually a generic feature of all on-shell diagram representations of tree-level amplitudes in ABJM theory. This will be proved by induction. Let us first assume that all the lower-point amplitudes can be constructed purely by triangles, which is trivial for the six-point tree-level amplitude. Then judiciously choosing the tree-level amplitude representations in the factorization diagrams guarantee that the resulting diagram after adding a BCFW bridge can be represented by triangles only. Let us consider the following factorization diagram in detail:
\eq \label{triangleproof}
\vcenter{\hbox{ \includegraphics[scale=0.55]{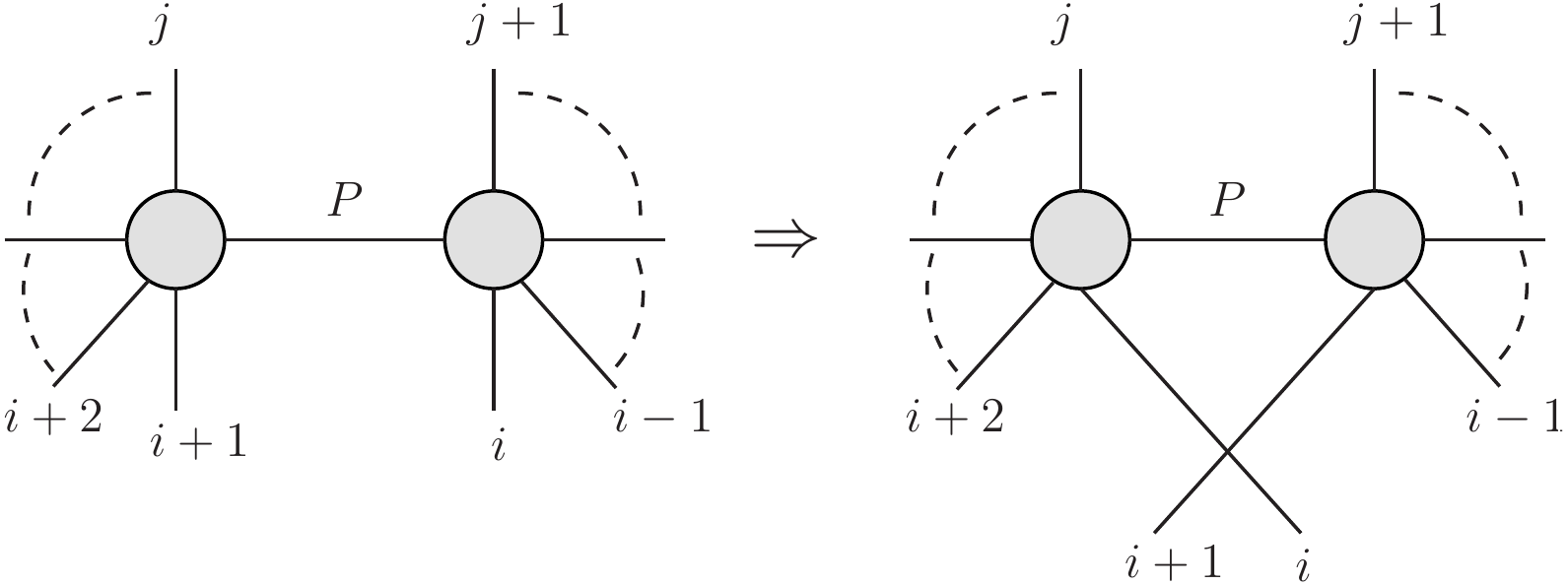} }}\,.
\eqe
By induction, the on-shell diagram representation for both left and right amplitudes are expressed solely in terms of triangles which do not share common sides. We can choose to express the left amplitude by BCFW recursion with shifts on legs $P$ and $i+1$, while the right amplitude is given by shifting legs $i$ and $P$. After erecting the BCFW bridge, we introduce a new triangle that is connected to both the left and right on-shell diagrams only through shared vertices and not sides. Thus this procedure insures that the final representation is again given by triangles which share no common sides. Furthermore, the diagrams will be one-particle irreducible.

It is straightforward to conclude that each BCFW diagram of a $(2k)$-point tree-level amplitude consists of $(k-2)$ triangles. As all triangles only have shared vertices and not shared sides, all vertices must have either 2 or 0 external legs, there are precisely $k$ external vertices and $(k-3)$ internal vertices for a BCFW diagram of $(2k)$-point tree-level amplitude. Note that as there will be a total of $k+(k-3)=2k-3$ vertices, the on-shell diagram corresponds to a $(2k-3)$-dimensional integral. This is exactly the number of the constraints, so the on-shell diagrams of tree-level amplitudes are simply rational functions as expected. 

To make this amusing geometric property more transparent it's convenient to remove the external legs, but leave triangles only. In this case, to obtain a $2k$-point tree-level amplitude, we simply take $k-2$ triangles and connect them through $k-3$ vertices in all possible topologically distinguished ways; then assign $k$ ordered numbers, $1,2, \ldots, k$, to $k$ external points. For instance, eight-point amplitude may be represented as
\eq \label{8ptriangle}
\vcenter{\hbox{ \includegraphics[scale=0.45]{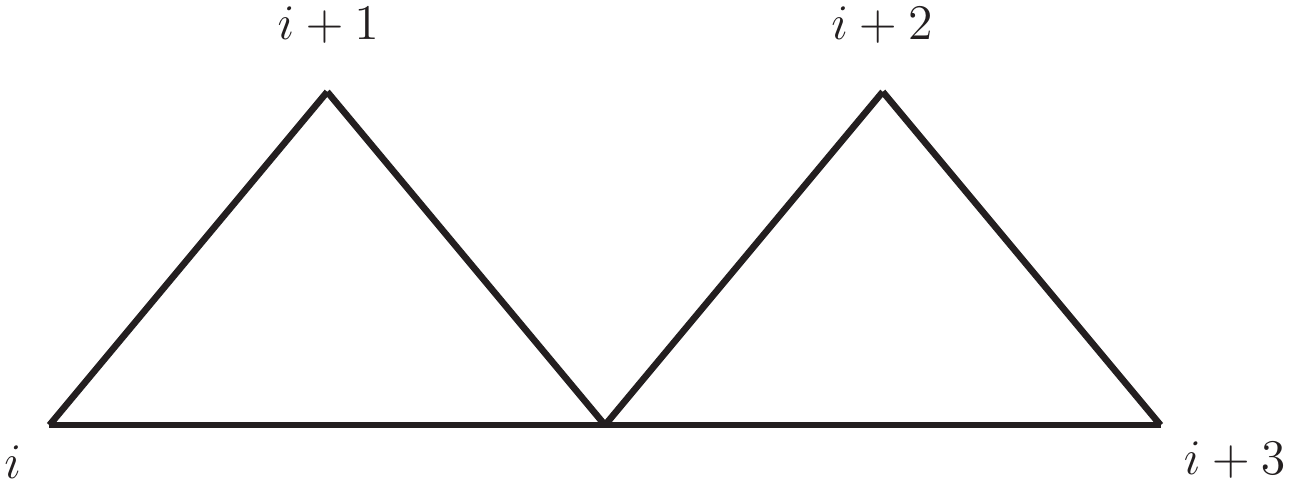} }}
\eqe
There is only one way to connect two triangles through a vertex, and two distinguishing ways of assigning external vertices, namely $i=1$ or $2$ in the above diagram. This is of course precisely the same as eq.~(\ref{8pttree}). Slightly non-trivial example would be $12$-point amplitude, where we have three different ways of putting four triangles together, 
\eq \label{12pttriangle}
\vcenter{\hbox{ \includegraphics[scale=0.7]{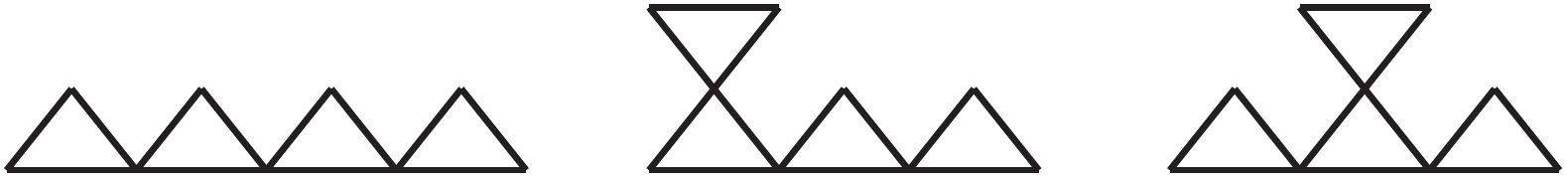} }} \, .
\eqe
Now we have six different ways of assigning $6$ numbers for the first two diagrams of eq.~(\ref{12pttriangle}), and two different ways for the last diagram. In total we have $14$ different BCFW diagrams for $12$-point amplitude, which agrees with the result of eq.~(\ref{BCFWnumber}). 

So by the construction described above, we see that the two-site cyclic symmetry of the external legs is manifest in the on-shell diagram. This property implies another remarkable fact: \textit{the on-shell diagram representation of the BCFW result manifest the presence of all physical factorization channels !} This can be understood as follows: from the outset, the BCFW recursion will manifest all factorization channels for which two chosen legs sit across the factorization channel. As the on-shell diagram representation manifests the cyclic by a two-site symmetry, all factorization channels that are related by this cyclic symmetry is manifestly present as well. One might worry about the factorization channels that are related by a cyclic rotation of one-site. Due to the special property that in ABJM only even multiplicity amplitudes are non-vanishing, the factorization channels that are related to the original BCFW channel by cyclic rotation of one-site, is in fact equivalent to a factorization channel that is cyclically rotated by even sites. For example, consider the following eight-point factorization:
\eq
\vcenter{\hbox{ \includegraphics[scale=0.8]{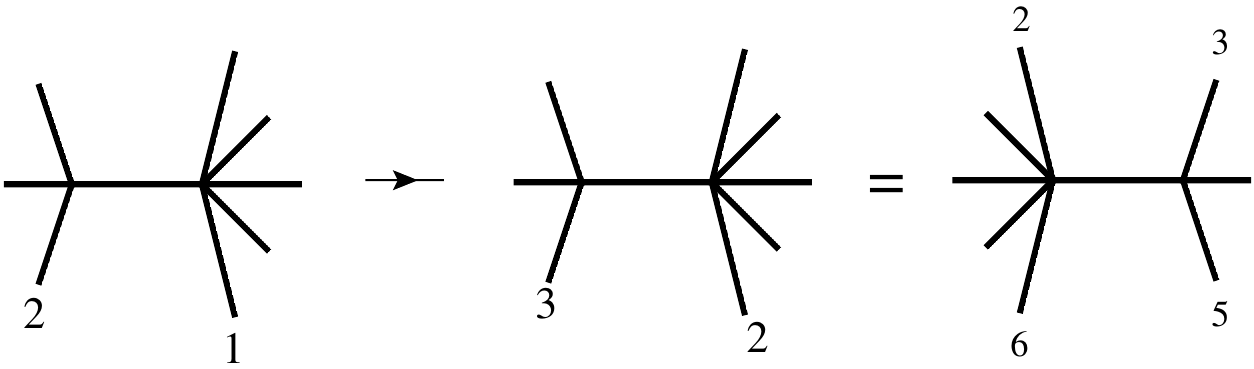} }}
\eqe
With legs $1$ and $2$ to be the chosen leg, one might worry that the factorization channel that is related by a cyclic by one-site shift, which now becomes a $(3,2)$ factorization, is absent. However the would be missing factorization channel is actually equivalent to a channel that appears in the $(6,5)$ shift, which is related by the original $(2,1)$ shift via cyclic by two-site rotation. Thus one concludes that the cyclic by two-site symmetry of the on-shell diagram guarantees that all physical poles are manifestly present. Note that spurious poles come from opening up any internal vertex, which turns two connected triangles into a box, as shown in the example of eight-point amplitude, eq.~(\ref{8pttreespurious}). It's not too difficult to see they always appear in pairs and thus cancel out each other. Another nice property regarding this representation of tree-level amplitudes is that it is manifestly inverse-soft constructible, for which we discuss in the Appendix \ref{inversesoftAppendix}. What makes this intriguing representation of tree-level amplitudes possible is the fact that BCFW bridge and the fundamental vertex is the same entity in ABJM theory.

\subsection{Equivalence moves and reducible diagrams\label{Reduced}}
Just as in $\mathcal{N}=4$ SYM, various distinct on-shell diagrams can be equivalent through a change of variables. Diagrammatically the equivalence can be established by a serious of triangle moves:
\eq \label{trianglemove}
\vcenter{\hbox{ \includegraphics[scale=0.5]{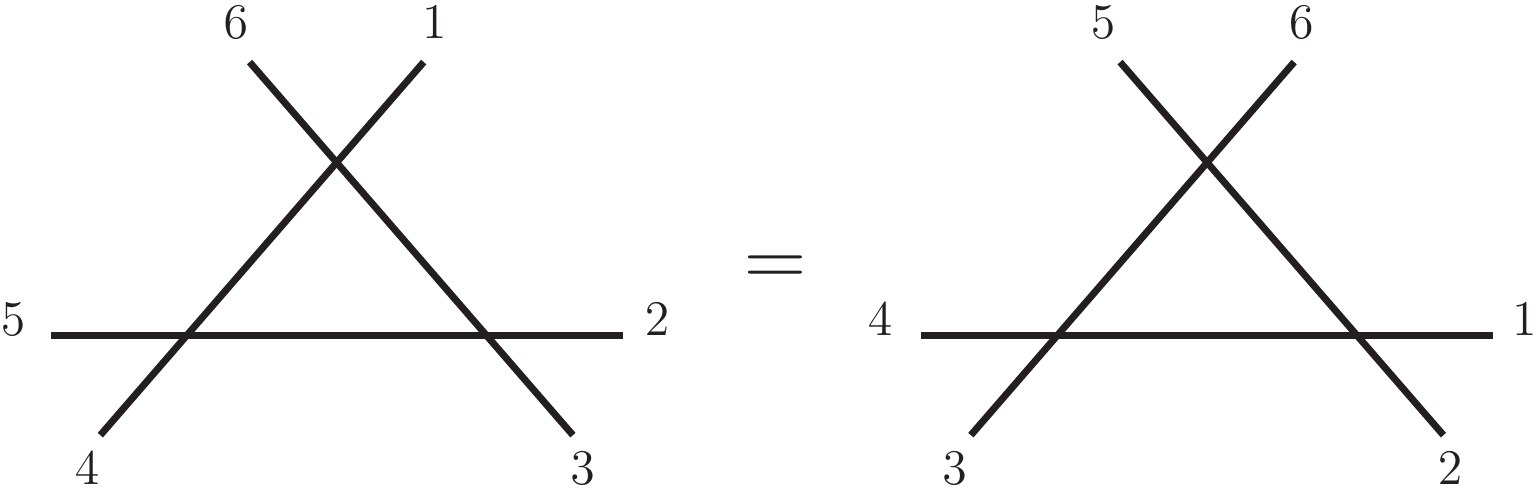} }}\,,
\eqe
which is nothing but the statement that two BCFW shifts related by cyclic permutation by one site gives equivalent result, and since there is only one diagram in the recursion, this implies a equivalence between diagrams. Note that since it is the amplitude that is equivalent, it is the combination of the two branches that is invariant. This ``triangle move" (or Yang-Baxter move) is the analogue of ``square move " in $\mathcal{N}=4$ SYM. As an example for equivalent diagrams, consider the following,
\eq \label{8pttrianglemove}
\vcenter{\hbox{ \includegraphics[scale=0.5]{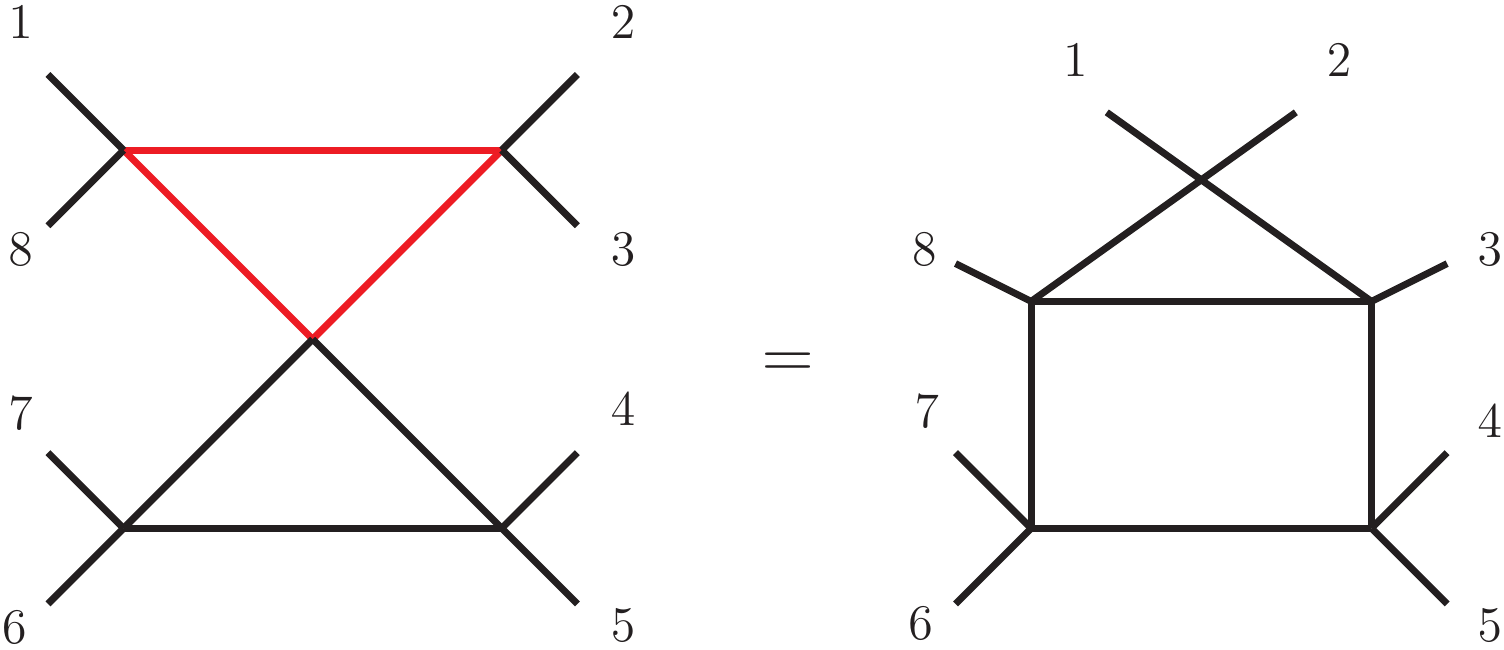} }}
\eqe
where we have applied the triangle move on the top triangle of the first diagram. For more complicated diagrams, we would like to have a way of discerning the equivalency without actively applying these moves. This invariant information is nicely captured by the permutations implied by the diagram, which we will discuss in the next section. 

The action of the triangle move on the OG$_3$ Grassmannian is rather simple, at least for certain gauge choices. For instance, for the canonical gauge in eq.~(\ref{6ptBCFW}), the Grassmannians of two diagrams in eq.~(\ref{trianglemove}), are related to each other by following similar transformation, 
\eqa
\left(\begin{array}{ccc} 0 & 0 & 1 \\ 0 & 1 & 0 \\ 1 & 0 & 0  \end{array}\right) \, ,
\eqae
and with a trivial replacement $s_i \rightarrow \alpha_i s_i$ and $c_i \rightarrow -\alpha_i c_i$.

There are also cases where through a change of variables, some degrees of freedom can be completely decoupled from the Grassmannian in the bosonic delta function. A trivial example would be the following tadpole diagram:
\eq
\vcenter{\hbox{\includegraphics[scale=0.7]{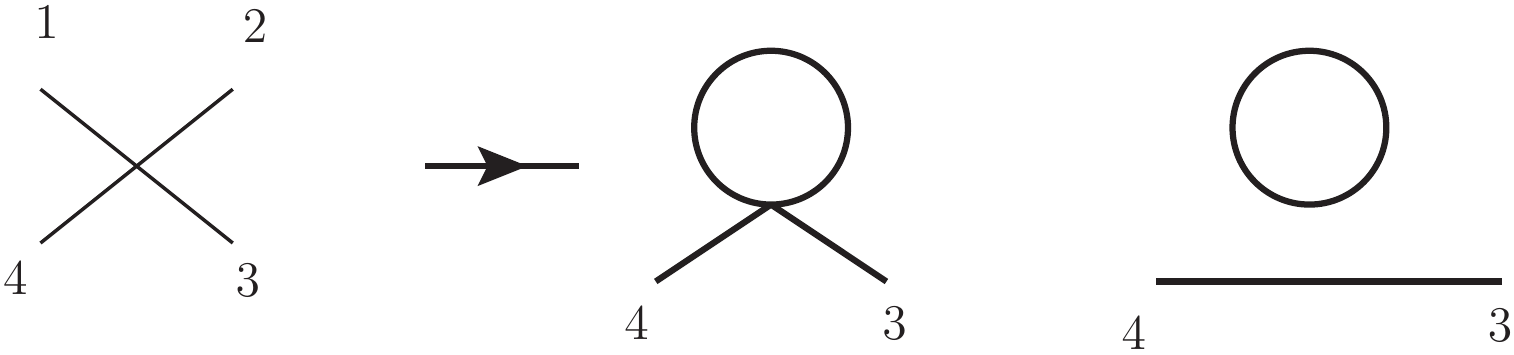}}}
\eqe
The decoupling of the integration variable in OG$_2$ can be explicitly shown as follows: starting with 
\eq
\delta^2(\lambda_1+i\alpha c\lambda_2+i\alpha s\lambda_{4}),\;\; \delta^2(\lambda_{3}-is\lambda_2+ic\lambda_{4})\,,
\eqe
The tadpole diagram is constructed by identifying $\lambda_2=i\lambda_1$, and integrating $\int d^2\lambda_1$. One may be absorbed $ \alpha =\pm 1$ in the definition of spinor $\lambda$. Furthermore we find integration measure is in a ${d\log}$ form,  
\eqa
\delta^2(\lambda_3+i\alpha\lambda_4)\int { d \theta \over c s } (1 + c) = -\delta^2(\lambda_3+i\alpha\lambda_4)\int d {\rm Log}( 1/c - 1)\, ,
\eqae
here we have chosen cyclic gauge to be precise, canonical gauge leads to a similar result. Thus as promised, the Grassmannian variables no longer appear in the bosonic delta functions and completely decouples from the remaining part of the graph.  Now consider another type of bubble,  
\eq  \label{bubblereductionB}
\vcenter{\hbox{  \includegraphics[scale=0.45]{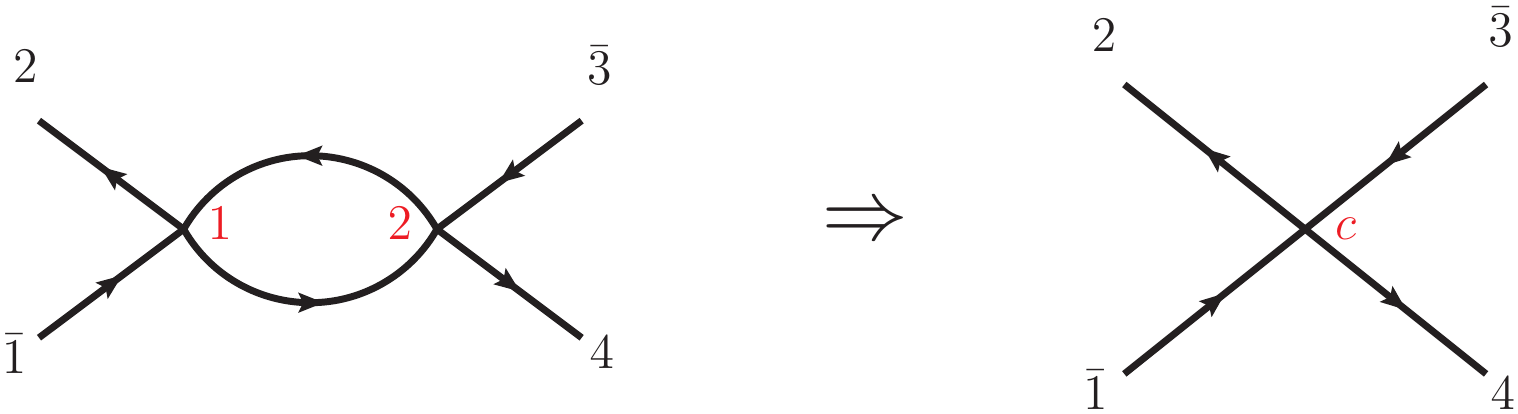} }} \, .
\eqe
Here, to illustrate the subtleties introduced by the presence of closed loops in particular gauges, we intentionally chosen a gauge such that a closed loop is formed. Reading off from the diagram the OG$_2$ Grassmannian is given as, 
\eqa
C_{\alpha i }=\left(\begin{array}{cccc}1 &  i{ \alpha_1 (   s_1 - \alpha_2 s_2  ) \over 1-\Gamma } & 0 & 
i{ \alpha_1 c_1 \alpha_2 c_2 \over  1-\Gamma } \\0 &  i { c_1 c_2 \over  1-\Gamma } & 1 & i{ s_2 - \alpha_2 s_1 \over  1-\Gamma } \end{array}\right) 
\eqae
with $\Gamma = {\alpha_2s_1  s_2}.$ For performing the reduction, we identify $C_{\alpha i }$ with the fundamental OG$_2$ Grassmannian
\eqa
C_{\alpha i } := \left(\begin{array}{cccc}1 &  i \alpha s & 0 & 
i \alpha c \\0 &  -i c & 1 & i s \end{array}\right), 
\eqae
namely we have made following identifications on the entries, 
\eqa \label{bubbletransformation}
- \alpha_1 \alpha_2 :=\alpha, \, {s_2 - \alpha_2 s_1 \over 1 -\Gamma  } := s, \, - {c_1 c_2 \over 1 -\Gamma }:=c \, .
\eqae
From these relations, the measure of the bubble now can be nicely written in terms of $c$ (and $s$), 
\eqa \label{bubblereduction2}
{d\theta_1 \over c_1 s_1} \wedge  {d\theta_2 \over c_2 s_2} (1 - \Gamma) 
= d {\rm Log}(1- {s \over s_2 }) \wedge {d \theta \over  c s } = 
d {\rm Log}(1- { \alpha s \over \alpha_1 s_1 }) \wedge {d \theta \over  c s } \, .
\eqae
Thus we see that through the change of variables in eq.(\ref{bubbletransformation}), the measure is factorized into a ${d\log}$ form multiplying tensored with the measure of the fundamental vertex. One may further redefine $(1- {s \over s_2 })$ or $(1- {s \over s_2 })$ as a new valuable, and one again finds that this extra degree of freedom decouples from the rest of on-shell diagram. Note that the Jacobian factor $(1 - \Gamma) $ plays an important role in allowing us to write the factorized measure as a $d\log$ form.

Thus we see that a reducible diagram can be rewritten as a factorized product of ${d\log}$s multiplying a reduced diagram, with the latter being independent of the arguments in the ${d\log}$s. As shown in ref~\cite{ABJMTwoL6}, the known one and two-loop amplitudes in ABJM theory can invariably be written in terms of integrals with unit leading singularities. This indicates the loop amplitudes should be written as $d\log$s multiplied by the leading singularity. Later, in section \ref{section:conclusion}, we will show indeed some loop amplitudes can be explicitly written in such a suitable form, and thus implying that it can be understood as performing a reduction on reducible graphs. However one should keep in mind here that actually any one of the $d\log$ forms in eq.(\ref{bubblereduction2}) is only well-defined in a local chart, where the other one is not valid. In fact the same issue also appears when we re-write one-loop amplitudes in a certain $d\log$ form as we will discuss in section \ref{section:conclusion}.

Unlike in four-dimensions, where the removal and adding of bubbles do not generate extra corners, in three-dimensions this does. Because of this, when combined with the triangle moves, one obtains non-trivial relations. For example: 
\eq
\vcenter{\hbox{\includegraphics[scale=0.7]{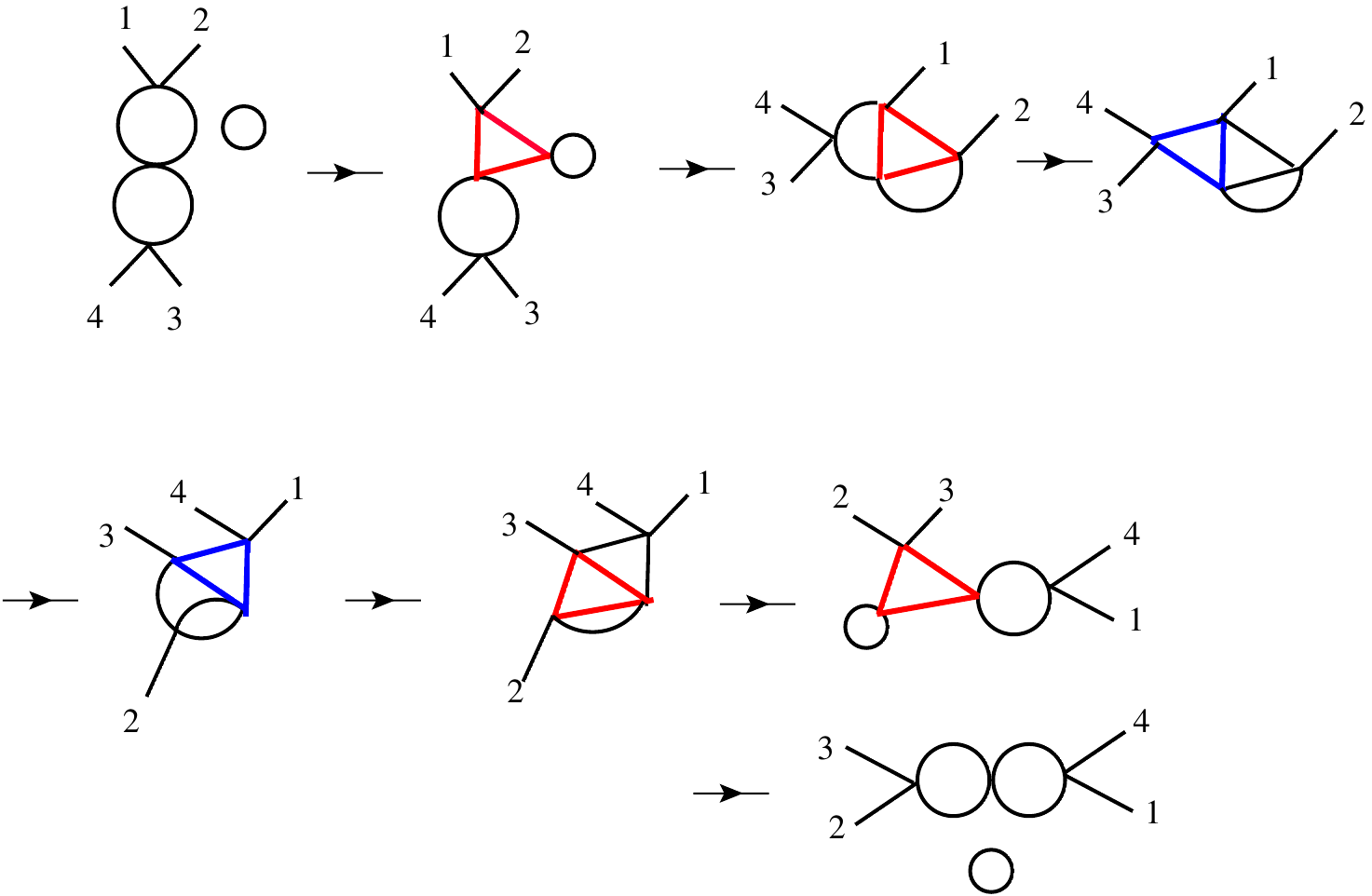}}}
\eqe
Note that the decoupled bubble in the equivalence cannot be removed. This is because there is a notion of integration contour for each bubble and a priori there is no reason why the contour for the two bubbles should be the same.

\section{The stratification of the orthogonal Grassmannian} \label{section:stratification}
In the previous section, we've seen that through BCFW construction, we end up with an OG$_k$ that is  $2k-3$-dimensional. Compared with the dimension of the top-cell, $k(k-1)/2$, one sees that beyond $k=2,3$ the on-shell diagrams constructed from the BCFW recursion will have dimensions less than the top-cell. Furthermore, various distinct diagrams can be related through equivalence moves. This raises two questions: 1. is there a GL($k$) invariant way to classify these $2k-3$ sub manifolds and 2. is there a more efficient way of identifying diagrams that are equivalent under such classification?

At a given $k$ beginning with the top-cell in  OG$_k$, we would like to identity GL($k$) invariant constraints one can impose on the top-cell, such that one lowers the dimension. Note that while the minors of the Grassmannian are only SL($k$) invariant objects, there is a GL($k$) invariant data associated with them: the rank of the minors. Thus a natural classification of the sub manifolds is the linear interdependence of the columns in the Grassmannian. We've seen this at play in our study of the fundamental OG$_2$, where the singularities of the measure correspond to linear dependency of the columns in the top-cell, their co-dimension one boundaries. The classification of all possible linear dependency of the columns is called ``\textit{matroid stratification}"~\cite{GGMS}. As shown in ref.\cite{NimaBigBook}, if one specialize to only linear dependency of \textit{consecutive columns}, hence the rank of the ordered minors, the resulting stratification, named ``\textit{positroid stratification}"~\cite{P, KLS}, remarkably characterize the sub manifolds that are built from the on-shell diagrams in $\mathcal{N}=4$ SYM. That is, each on-shell diagram corresponds to a particular stratification, which is characterized by the rank of all consecutive minors. Furthermore, the stratification is invariant under equivalence moves, and thus serve as the invariant data that is associated with equivalent on-shell diagrams. 

The relation between on-shell diagrams in ABJM theory and positroid stratification in the orthogonal Grassmannian was already discussed in ref.~\cite{NimaBigBook}. The reasoning is straightforward. Beginning with a G(2,4), whose top-cell is 4-dimensional, orthogonality is simply a way to reduce the degrees of freedom down to 1 while remaining in the top-cell. Graphically, this is simply the statement that:
\eq
\vcenter{\hbox{ \includegraphics[scale=0.7]{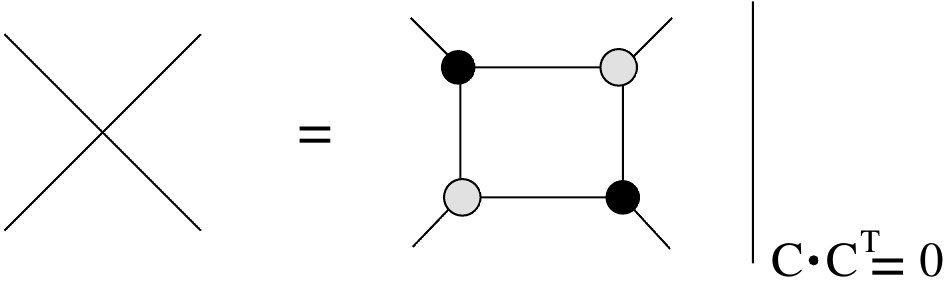} }} \,.
\eqe  
As amalgamation preserve orthogonality, in terms of positroid stratification, each ABJM on-shell diagram is completely equivalent to the stratification that is represented by the $\mathcal{N}=4$ SYM diagram which is obtained by blowing up each OG$_2$ into a top-cell in G(2,4):
\eq
\vcenter{\hbox{ \includegraphics[scale=0.5]{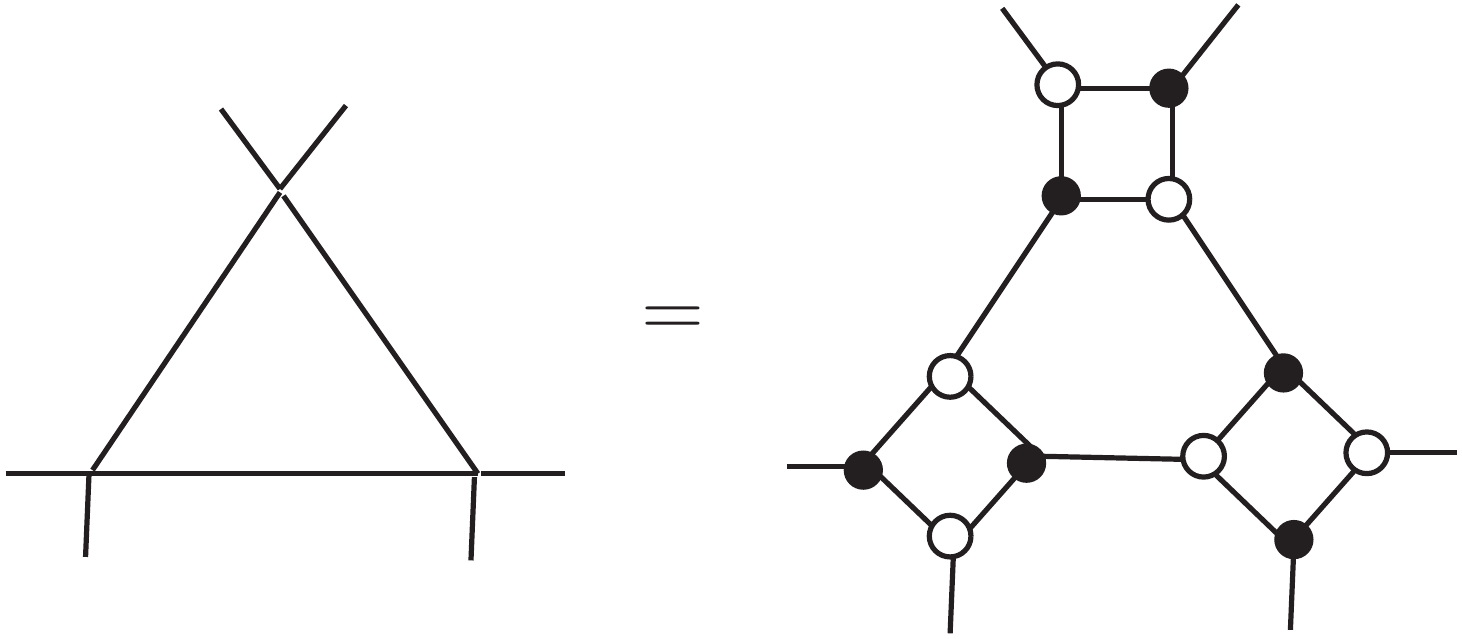} }}\,.
\eqe  

For $\mathcal{N}=4$ SYM, the corresponding stratification of an on-shell diagram is nicely encoded in the permutation path associated with the on-shell diagram~\cite{NimaBigBook}. More precisely, starting from any point on the boundary of the graph, taking a left turn whenever one encounters a black vertex, while a right turn for a white vertex, one eventually reaches the boundary. Doing the same for each point on the boundary, one obtains a set of permutations which maps the $n$-points into each other: $a\rightarrow \sigma(a)$. The remarkable property of such permutation paths is that for reduced diagrams, it encodes the stratification: given the permutation paths  $a\rightarrow \sigma(a)$, the image $\sigma(a)$ represent the closest column to $a$, such that $a$ is spanned by $a+1,a+2,\cdots \sigma(a)$. For example the permutation paths for the following G(2,4) is given as:
\eq
\vcenter{\hbox{\includegraphics[scale=0.7]{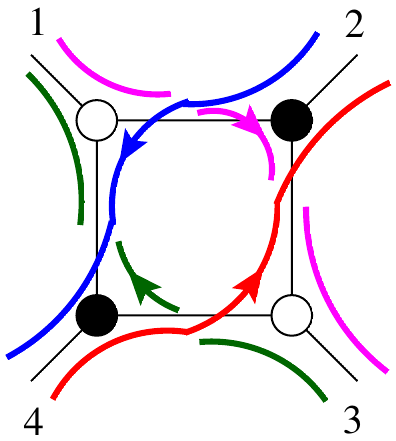}}}:\quad\quad \begin{array}{c} 1\rightarrow 3 \\2 \rightarrow4  \\3\rightarrow 1 \\ 4\rightarrow 2\end{array}\,.
\eqe
As the columns in G(2,4) are two-dimensional vectors, a column $i$ is always spanned by two distinct generic columns $i+1$ and $i+2$. Thus the permutation path tells us that there is no additional linear dependency beyond that is enforced by the dimension of the vector. This is nothing but the statement that it is in the top-cell! We refer to ref.\cite{NimaBigBook} for more details.

Now let us consider the stratification for orthogonal Grassmannian. Given the fact that the fundamental four-point vertex OG$_2$ is still a top-cell in G(2,4), the stratification is simply given by: 
\eqa \label{4ptpermutation}
\sigma_2 = [1,3][2,4] \,  ,
\eqae
where $[i,j]$ denotes a permutation path going from $i$ to $j$ and back, $i \leftrightarrow j$. Given the permutation structure of fundamental vertex, it is easy to see that the permutation of any $2k$-point on-shell diagram by gluing vertices together will be in a rather simple two-cycle form, 
\eqa \label{generalpermutation}
\sigma_k = [i_1, j_1] [i_2, j_2] \ldots [i_k, j_k] \, 
\eqae 
where we have divided $2k$ external legs into two sets, $\{i_1, \ldots, i_k \}$ and $ \{j_1, \ldots, j_k  \}$. As we've previously mentioned, the stratification is the invariant data associated with on-shell diagrams that are equivalent under triangle moves. Indeed one can find that two equivalent triangles in eq.~(\ref{trianglemove}) both have permutation, $[1,4][2,5][3,6]$. This permutation tells us that, for example, column $1$ is spanned by $2,3,4$. Since we are now in OG$_3$, any three-dimensional vector is spanned by 3 generic vectors, and therefore this implies that the six-point BCFW tree-diagram is in fact in the top-cell of OG$_3$, which is another way of seeing the Yang-Baxter move must hold.  

Note that for reducible diagrams, the permutation will change before and after the reduction, as shown in the following example:
 \eq \label{bubblepermutation}
\vcenter{\hbox{ \includegraphics[scale=0.8]{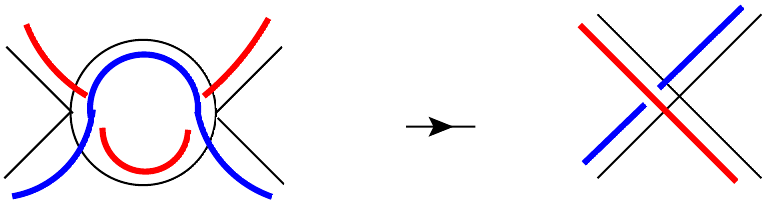} }} \, .
\eqe
Since the reduction simply corresponds to a change of variables, the stratification should not change. Thus one can conclude that permutation only reflects the stratification for reduced diagrams. However, it is not always obvious to see that, by the means of ``triangle move", whether a digram contains bubbles, in that case permutation can be very helpful: diagrams having two permutation paths forming a loop are reducible. Here is a more complicated example, 
 \eq \label{trianglebubble}
\vcenter{\hbox{ \includegraphics[scale=0.8]{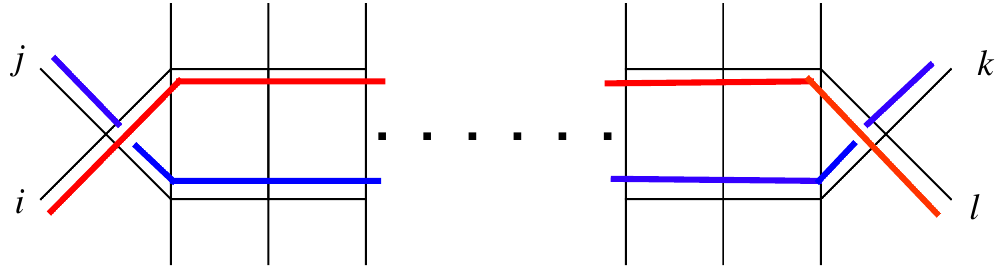} }} \, ,
\eqe
where we have permutation paths $[i,l][j,k]$, which forms a loop. One can verify by series of triangle moves that the above diagram indeed contains a bubble.

Beginning from the top-cell of OG$_k$, the codimension one boundaries are simply the vanishing of consecutive minors. Due to the identity in eq.(\ref{BranchDef}), there are precisely $k$ such boundaries. As the on-shell diagrams can be characterized by a particular positroid stratification, it should correspond to higher codimension boundaries of the top-cell. This immediately leads us to a puzzle: as we increase $k$, it is easy to see that the number of inequivalent $2k-3$-dimensional on-shell diagrams quickly out grows $k$, so how can the Grassmannian encode such diverse structure? As we will see, as some minors vanish, the others actually factorizes, revealing more poles than the number of the original minor, thus allowing for more intricate singularity structures. Before doing so, we first verify that the stratification implied by the permutations actually respects properties of the orthogonal Grassmannian.  
 
\subsection{Permutation and orthogonality}
As have been discussed in great details in reference~\cite{NimaBigBook}, the linear dependence of the columns of a G$_{k; n}$ Grassmannian,  
\eq \label{OGk}
C_{\alpha i}=\left(\begin{array}{cccc}c_{11} & c_{12} & \cdots & c_{1,n} \\ \vdots & \vdots & \vdots 
\\ c_{k1} & c_{k2} &\cdots & c_{k,n} \end{array}\right) = \left( \vec{c}_1, \vec{c}_2, \cdots, \vec{c}_n \right) \, ,
\eqe
is encoded in the permutation path of its corresponding on-shell diagram. 
Here we like to prove that the permutation assignment we had in eq.~(\ref{generalpermutation}) is consistent with the orthogonality of our OG$_{k}$ Grassmannian, namely a $k \times k$ matrix and its complement should have the same rank.

Let us review here briefly how permutation determines the linear dependency of the columns in $C_{ai}$. For instance, consider the six-point factorization diagram in eq.~(\ref{sixptfactorization}). The corresponding permutation is given as
\eqa
\sigma = [1,5][2,4][3,6]. 
\eqae  
Now, the two-cycle $[2,4]$ in $\sigma$ shows that vector $\vec{c}_2$ is spanned by vectors $\vec{c}_3$ and $\vec{c}_4$, equivalently the $3 \times 3$ matrix 
\eqa
\left( \vec{c}_2, \vec{c}_3, \vec{c}_4 \right)
\eqae
has rank $2$, and the minor $(234)$ vanishes. As for the complementary matrix $\left( \vec{c}_5, \vec{c}_6, \vec{c}_1 \right)$, exactly the same conclusion can be drawn based on the two-cycle $[1,5]$ in $\sigma$. This fact is of course required by orthogonality. 

We like to generalize this observation, and thus to prove that stratification implied by the permutations is consistent with orthogonality. Use the same notation as eq.~(\ref{generalpermutation}), we denote a general permutation as
\eqa 
\sigma_k = [i_1, j_1] [i_2, j_2] \ldots [i_k, j_k],
\eqae 
and two sets of $2k$ external legs denoted as $I = \{i_1, \ldots, i_k \}$ and $J = \{ j_1, \ldots, j_k \}$, such that $I$ and $J$ are the imagines of each other under the action of $\sigma_k$. Now consider a consecutive matrix and its complement: 
\eqa
M = \left( I_1~ J_1 \right), \quad M_c = \left( I_2~ J_2 \right) \, .
\eqae
Here $I_1 (J_1)$ is a subset of $I (J)$, $I_2 (J_2)$ is the complement. The matrix $M$ is built up by consecutive vectors of $\vec{c}_i$ and $\vec{c}_j$ with $i \in I_1$ and $j \in J_1$. Similarly for matrix $M_c$. Let us further denote $k_{I_1}, k_{I_2}$ and $k_{J_1}, k_{J_2} $ as the numbers of the elements inside the subsets $I_1, I_2$ and $J_1, J_2$ respectively. We then have following relations,
\eqa
k_{I_1} + k_{J_1} = k_{I_2} + k_{J_2} = k_{I_1} + k_{I_2} =  k_{J_1} + k_{J_2} = k \,,
\eqae
which lead to $k_{I_1} = k_{J_2}, k_{I_2} = k_{J_1} = k - k_{J_2}$. 

Now, we are ready to read off the ranks of matrices $M$ and $M_c$ from permutation. Assume that there are $n$ elements of $I_1$ are permuted into $J_1$ under $\sigma_k$, namely matrix $M$ has rank $k-n$. Then on the other hand there must be $k_{I_1}- n$ elements of $J_2$, which are permuted into $I_1$ under $\sigma_k$, put it another way, it means there $k_{J_2} - (k_{I_1}- n) = n$ elements of $J_2$ is permuted to $I_2$ in $M_c$ under $\sigma_k$. So indeed matrix $M_c$ has the same rank, $k-n$, as $M$ does. This ends the proof. Of course, permutation alone can only determine the fact that $M$ and $M_c$ have the same ranks, the information is not enough to tell whether the determinant $|M|$ is equal to $|M_c|$ when they have full rank $k$. As we will show later in section \ref{section:representative} there is a precise and concrete one-to-one map between an orthogonal Grassmannian and a permutation in the two-cycle from.

\subsection{Permutation and Integration contour}

Armed with the connection between permutation paths and stratification, one can easily make a connection between the BCFW on-shell diagrams and the contour in the Grassmannian integral eq.~(\ref{GrassInt}). We will devote this subsection to explore this connection. Let us start with simplest, but already highly non-trivial case: the on-shell diagram representation for the eight-point tree-level amplitude. For the convenience, we quote the on-shell diagrams here again, 
\eq   \label{8pttree2}
\vcenter{\hbox{ \includegraphics[scale=0.5]{8pttree} }}\,.
\eqe
From the diagram we can read off the permutation paths, 
\eqa
\sigma_1 = [1,5][2,7][3,6][4,8], \quad \sigma_2 = [1,4][2,6][3,7][5,8] \, .
\eqae
From permutation $[3,6]$ of $\sigma_1$ we can conclude that the column $\vec{c}_3$ is spanned by $\vec{c}_4, \vec{c}_5, \vec{c}_6$, which means that the rank, denote it as $R_3$, of following $4 \times 4$ matrix
\eqa
\left(  \vec{c}_3, \vec{c}_4, \vec{c}_5, \vec{c}_6   \right)
\eqae
is $3$, and so the minor $M_3=0$. The other two-cycle $[2,7]$ implies the $M_7=0$ which is expected due to eq.(\ref{BranchDef}). Note that the remaining two-cycles $[1,5]$ and $[4,8]$ do not imply any non-trivial linear dependence. Similarly from permutation $[1,4]$ of $\sigma_2$, we find $R_1 =3$ and minor $M_1 =0$. Since the Grassmannian integral in eq.~(\ref{GrassInt}) is one-dimensional for $k=4$, we find that the two eight-point on-shell diagrams are associated with contours of eq.~(\ref{GrassInt}) which circles the zero locus of $M_1=0$ and $M_3 =0$. Note eq.~(\ref{8pttree2}) can be viewed as BCFW recursion of eight-point amplitude with legs $1$ and $2$ shifted. One can of course consider diagrams with BCFW shifts on legs $2$ and $3$, which would be the same diagram as the one in eq.~(\ref{8pttree2}) but with one-side cyclic shift on external legs, namely $i \rightarrow i+1$. So then for this BCFW shift the eight-point tree-level contours are now given by the locus of $M_2=0$ and $M_4=0$, this is the same contour originally given in ref.~\cite{Gang}. Of course two contours are related to each other by residue theorem.

With the current understanding of the tree-level contours, we can further study the singularities, in particular the difference between physical poles and spurious poles. As we discussed in previous section, \ref{section:treelevel}, one can find physical (and spurious) poles by opening up external (and internal) vertices. For example for eq.~(\ref{8pttree2}) we have the following residues:
\eq   \label{8ptpoles}
\vcenter{\hbox{ \includegraphics[scale=0.8]{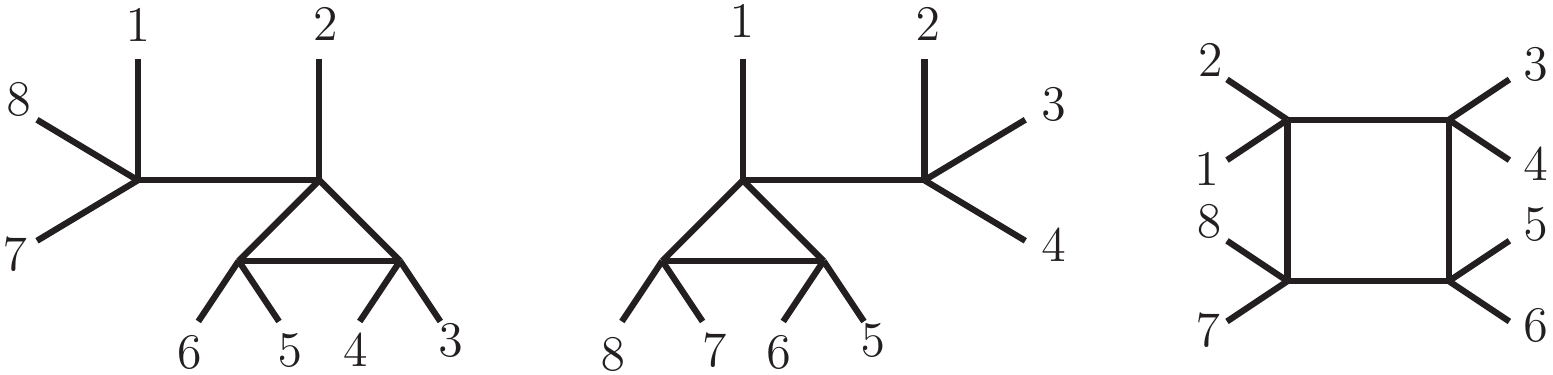} }},
\eqe
where the first two diagrams correspond to opening the vertices on which legs 1 and 2 are attached, while the last correspond to the opening of the internal vertex for each diagram. The first two diagrams correspond to physical factorization poles, whereas the last one is spurious. Here we like to understand the different nature of the physical and spurious singularities, from the viewpoint of the contour integral eq.~(\ref{GrassInt}). Again we can readily read off the permutations for each diagram,  
\eqa
{\sigma_{1}}_p = [1,7][2,5][3,6][4,8], \quad  {\sigma_{2}}_p = [1,6][2,4][3,7][5,8], 
\quad \sigma_s =  [1,4][2,7][3,6][5,8] \, ,
\eqae
where subscript $p$ stands for physical poles, and $s$ is for spurious poles. Note that each diagram in eq.(\ref{8ptpoles}) has 4 degrees of freedom, which is one less than $(2k-3)=5$. This implies that it imposes one extra constraint on the external data, which is expected for the residue of a singularity. Said in another way, as the $C_{ai}$ for the BCFW diagrams are completely determined by the external data, its boundary correspond to special configurations of the \textit{external data} that results in the development of new singularity: the vanishing of an extra minor. Stating the obvious, for generic external data two minors cannot simultaneously vanish since there is only on degree of freedom in the original grassmanian integral, and it is only for special kinematics that two minors can become identical and vanish simultaneously.

Let us see the above discussion work in details and analyse the behaviour of the minors for the diagrams in eq.(\ref{8ptpoles}). For both factorization diagrams we find ${M_{2}} = {M_{3}} =0$, whereas for the spurious diagram we find ${M_{1}} = {M_{3}} =0$. Thus indeed the singularities correspond to configurations where the minors become identical and can vanish simultaneously. If we denote the zeroes of the minors on the one-dimensional complex plane, which is determined by the external data, then the above discussion simply implies that two zeroes become degenerate. Recall that the BCFW contour correspond to the sum of residues where ${M_{1}}=0$ and ${M_{3}}=0$. Denoting the BCFW contour on the one-dimensional complex plane, one sees that the spurious singularity correspond to zeroes within the contour becoming degenerate, while physical singularities correspond to one zero inside the contour becoming degenerate with a zero outside of the contour:
\eq   \label{8ptcontour2}
\vcenter{\hbox{ \includegraphics[scale=0.7]{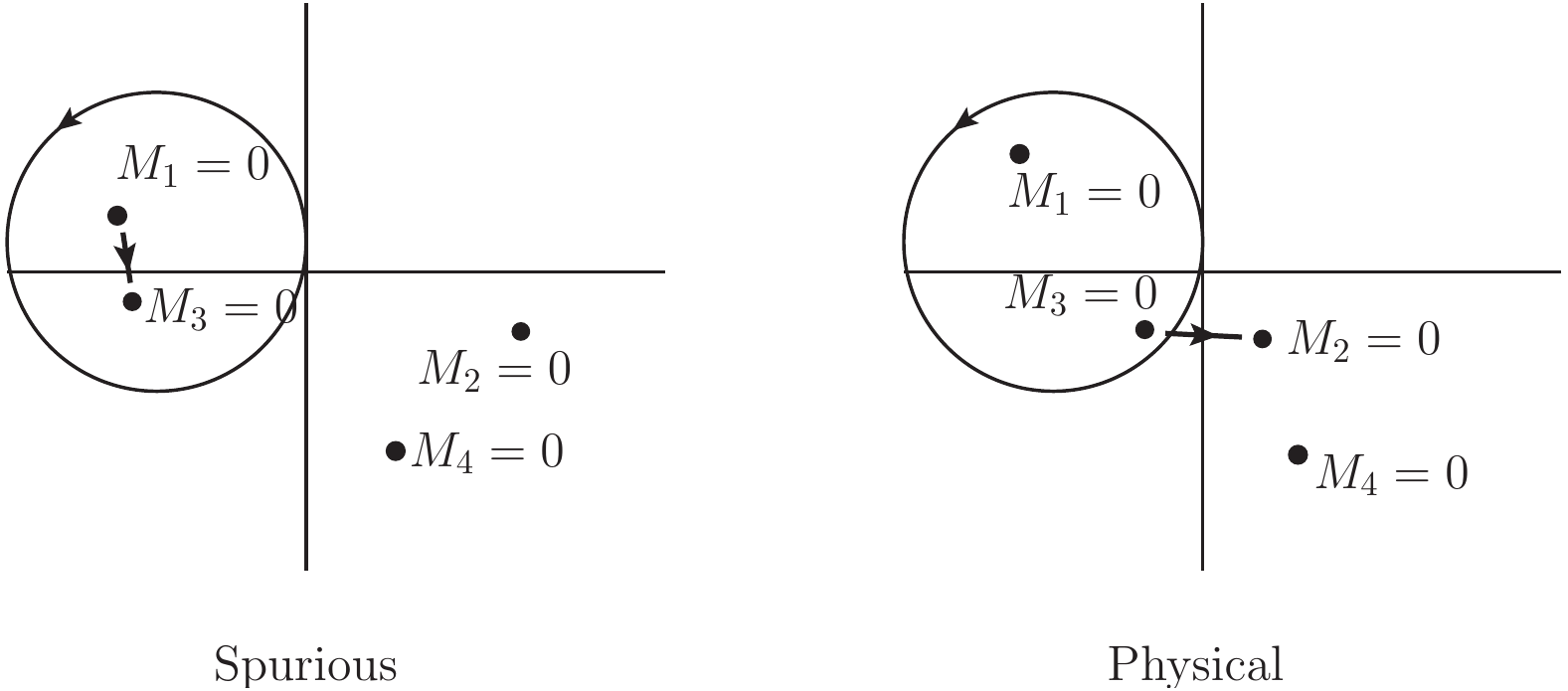} }} \, .
\eqe
This makes it manifest that a spurious pole is not a true singularity of the BCFW result, they always cancel out in pair. That is because one can always deform the contour with $M_1 =0$ and $M_3=0$ inside to the other one, where ${M_{1}} = {M_{3}} =0$ singularity is absent. This is precisely the picture that was revealed in the pioneering paper~\cite{Grassmanniandual}, where BCFW terms were first identified as the residues of a Grassmannian integral.

With the detailed study of the eight-point example, let us move on to the ten-point tree-level amplitude. Let us emphasize here that the tree-level contour for the ten-point amplitude is actually not known in literature, as we will see in a moment that it is rather simple to obtain the integration contour with the help of permutation paths and on-shell diagrams. The amplitude is given by a sum of following five BCFW diagrams, 
\eq  
\vcenter{\hbox{ \includegraphics[scale=0.85]{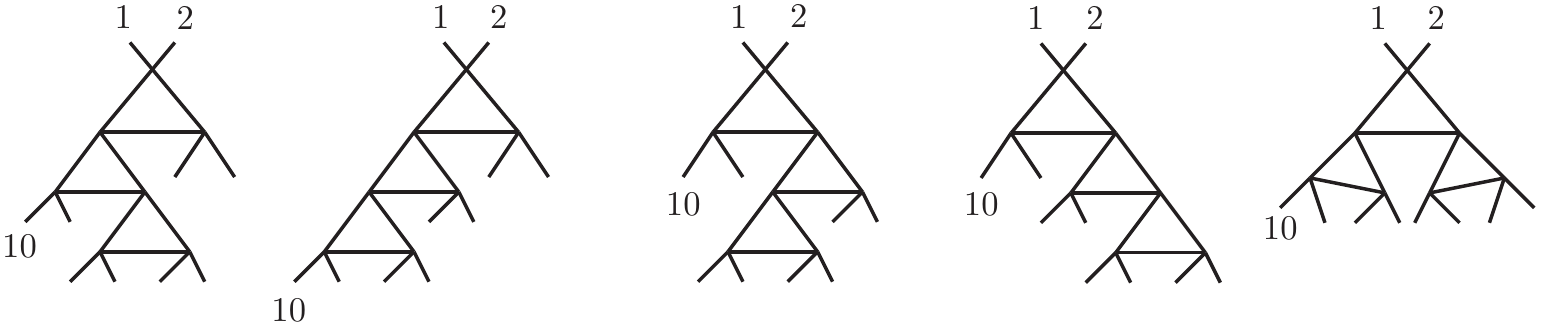} }} \, .
\eqe
Again one can easily read off vanishing minors from permutation paths for each diagram. For instance, for the first diagram, we have the permutation, 
\eqa
\sigma_{1} = [1,4][2,7][3,9][5,8][6, 10], 
\eqae
from which we can find the ranks of the consecutive minors. We find that the vanishing minors are $M_4, M_5$ and $M_1$. We will denote the corresponding contour as $\{4, 5, 1 \}$, namely the zero locus at $M_4 = M_5 = M_1 =0$. Similarly we can find out the integration contours for all other diagrams. List all of them in the order of five diagrams, we have,
\eqa
\{4, 5, 1 \}, \,\, \{ 5, 1, 2 \}, \,\, \{ 3, 4, 5 \}, \,\, \{ 2, 3,4 \} ,\,\, \{ 1,2, 3 \} \, ,
\eqae
So barely with any calculation we just obtain a totally new result regarding the integration contours of the ten-point tree-level amplitude, which can be nicely summarized as $\{i, i+1, i+2 \} $ for $i=1, \ldots, 5.$

Starting from twelve points, something totally new happens. We encounter diagrams that are identified with   ``composite residues"~\cite{Grassmanniandual}. To see this consider one of the fourteen BCFW diagrams for the twelve-point amplitude,
\eq   \label{twelveexample}
\vcenter{\hbox{ \includegraphics[scale=0.41]{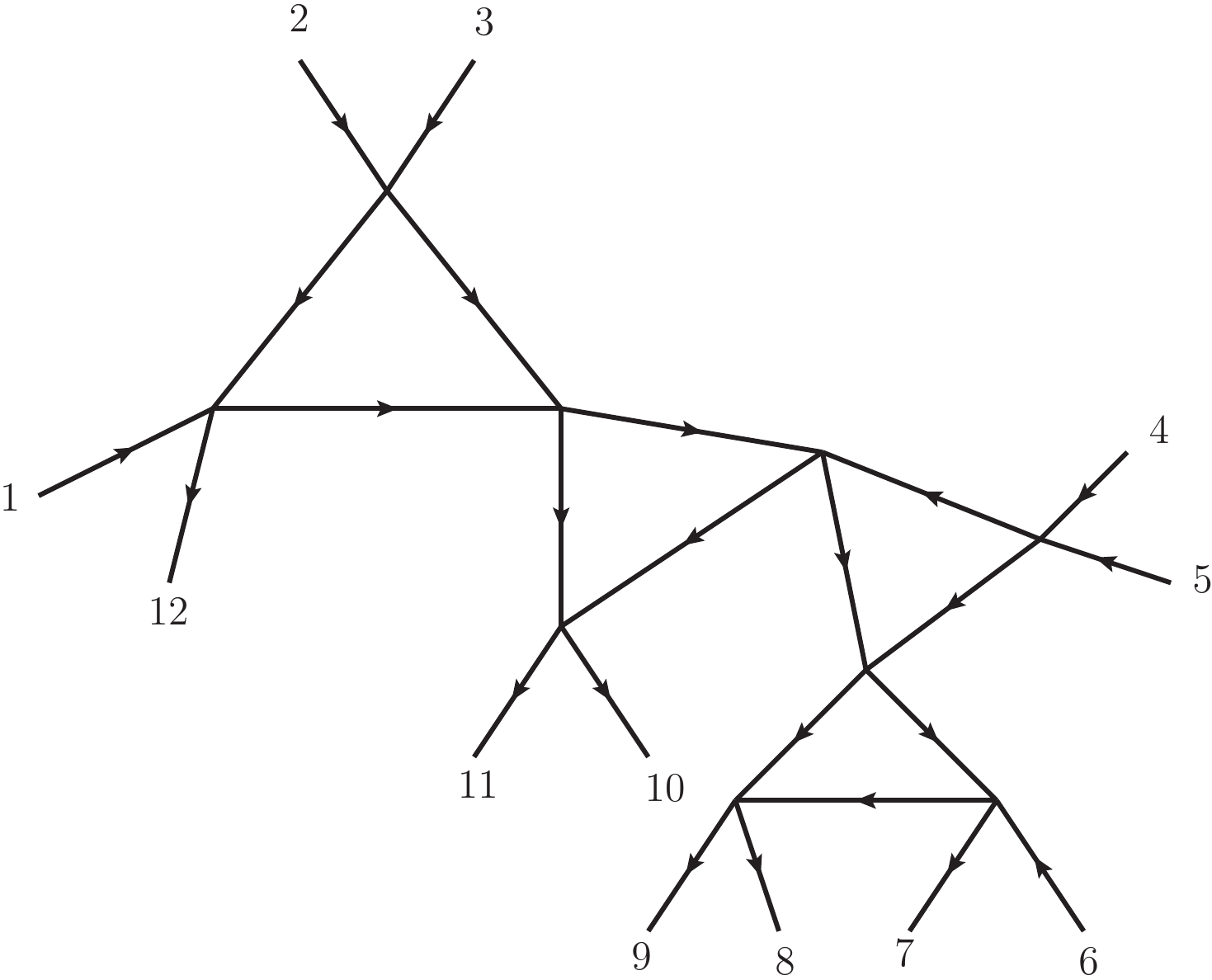} }} \, .
\eqe
To facilitate the analysis, we have chosen a particular gauge, as indicated by the arrows in the diagram. The permutation path is given by
\eqa
\sigma_{12} = [1,7][2,10][3,12][4,8][5,11][6,9] \, .
\eqae 
Denote the ranks of consecutive minors as $R_i \equiv{\rm rank} \left( \vec{c}_i, \vec{c}_{i+1}, \ldots, \vec{c}_{i+5} \right)$, we find, 
\eqa
R_1 =6, \, R_2 =6, \, R_3 =5, \, R_4 =4, \, R_5 =5, \, R_6 =5 \, ,
\eqae 
where $R_4 =4$ is because of the fact that we have both $[4,8]$ and $[6,9]$ in $\sigma_{12}$. So we have four vanishing minors, $M_3, M_4, M_5, M_6$. Note that while both $M_4$ and $M_5$ vanishes, the number of constraints involved are different as $M_5$ has rank one less than full rank, while $M_4$ is two less than full rank. We know that the Grassmannian integral in eq.~(\ref{GrassInt}) is a six-dimensional contour integral for $k=6$. As the total number of reduced ranks in each minor is $0+0+1+2+1+1=5$, one immediately see that this does not enough to account for all the integration valuables. This reflects the presence of composite residues, originally introduced in $\mathcal{N} =4$ SYM in~\cite{Grassmanniandual}. With the help of the on-shell diagram, we can study composite residue rather easily now. The relevant minors for our discussion are: 
\eqa
M_3 = \left |\begin{array}{cc} c_{1,7} & c_{1,8} \\ c_{2,7} & c_{2,8}   \end{array}\right |, \, 
M_4 = \left |\begin{array}{ccc} c_{4,10} & c_{4,11} & c_{4,12} \\ c_{5,10} & c_{5,11} & c_{5,12}
\\ c_{6,10} & c_{6,11} & c_{6,12}   \end{array}\right |, \,
M_5 = \left |\begin{array}{cc} c_{5,11} & c_{5,12} \\ c_{6,11} & c_{6,12}   \end{array}\right |, \, 
M_6 =  c_{6, 12}, 
\eqae
where the fact that we are in canonical gauge has been used, and for $M_4, M_5, M_6$ we have applied orthogonal relations to simplify the result. Without any calculation, one can read off from the on-shell diagram, eq.~(\ref{twelveexample}), that $c_{6, 12}=0$, that's simply because there is no decorated path going from leg $6$ to leg $12$. Under this condition, $M_5$ factorizes, namely, 
\eqa
M_6 =  c_{6, 12}=0 \quad \Rightarrow  \quad M_5 = - c_{5,12}\, c_{6,11} \, .
\eqae
One can further read off from the diagram directly that both $c_{5,12}$ and $ c_{6,11}$ in $M_5$ vanish. Stating the above analysis in another way, under the condition $M_6=0$, minor $M_5$ factorizes and reveals two poles instead of one. This is precisely the hallmark of a composite residue.  Now, given $c_{6,12} = c_{5, 12} = c_{6, 11}=0$, we further find that $M_4$ reduces to
\eqa
M_4 = c_{6, 10} \, c_{5, 11} \, c_{4, 12} \, .
\eqae
Again purely from decorated paths of the on-shell diagram, we find that $c_{6, 10}$ as well as $c_{4, 12}$ in above $M_4$ vanish. So, again, the residue of minor $M_4 =0$ becomes composite after we set $M_6$ and $M_5$ to be zero. As we have mentioned, this is actually already reflected in the fact that the rank of the corresponding matrix $R_4$ is $4$ instead of $5$. In conclusion, although only four minors vanish, we have identified all 6 conditions that went into localizing the six-dimensional integral. Use the same notation as that of ref.~\cite{Grassmanniandual}, for the BCFW term denoted by eq.~(\ref{twelveexample}), it can be identified with the following contour in the original Grassmannian integral eq.~(\ref{GrassInt}) for $k=6$, 
\eqa
\{ 3, 4^2, 5^2, 6 \} \, ,
\eqae
where the superscript $2$ denotes the position of the composite residues. Similar analysis can apply to other twelve-point as well as higher-point diagrams, we will not go into details here. 
\subsection{Permutation and representative} \label{section:representative}
For all reduced diagrams, we have a well-defined permutation which encodes the stratification of the corresponding configuration. Not surprisingly, this permutation also gives us a map of how to start from the trivial permutation, and successively apply adjacent transpositions to build up the final permutation. We illustrate this procedure in this subsection.

To begin we first define what is a trivial permutation.  For a given ordering, the trivial permutation is that whose two-cycles only involve adjacent points. For example:
\eq
\vcenter{\hbox{ \includegraphics[scale=0.7]{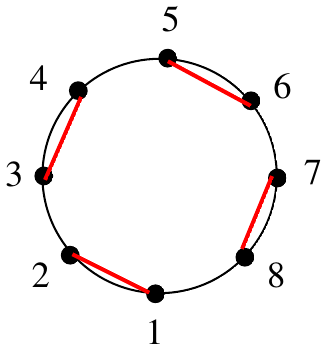}  }}\,.
\eqe 
where each redline connects two points, say $i_1,j_1$, that are in the same two cyclic $[i_1,j_1]$. For simplicity, we will order all two cycles such that $i_1<j_1$ for each $[i_1,j_1]$. Then the trivial Grassmannian would be defined as:
\eq
\vcenter{\hbox{\includegraphics[scale=0.6]{trivial}}}\rightarrow\quad C_{i\alpha}=\left(\begin{array}{cccccccc}1 & i & 0 & 0 & 0 & 0 & 0 & 0 \\0 & 0 & 1 & i & 0 & 0 & 0 & 0 \\0 & 0 & 0 & 0 & 1& i & 0 & 0 \\0 & 0 & 0 & 0 & 0 & 0 & 1 & i\end{array}\right)\,.
\eqe

We now claim that any non-trivial permutation can be decomposed into successive steps of adjacent transmutations, that brings it back to the trivial permutation. The procedure is as follows: \textit{Given any permutation, order all two-cycles vertically according to their first entry, i.e. if $i_1<i_2$ then $[i_1,j_1]$ comes on top of $[i_2,j_2]$. If the permutation is not a trivial permutation, then start from the top and for the first-pair of $i_1,i_2$ separated only by trivial permutation pairs, such that $j_1<j_2$, perform the following transposition: 
\eq
\begin{array}{c} [i_{1},j_{1}] \\ \left[i_{2},j_{2}\right]  \end{array} \rightarrow \begin{array}{c} [i_{1},j_{2}] \\ \left[i_{2},j_{1}\right]  \end{array} \,.
\eqe
Repeatedly apply such transposition until trivial permutation is reached. Note that if one reaches to the bottom of the two-cycle tower and trivial permutation is not reached, then exchange the two entries of the first top cycle and repeat the process. }

Let's illustrate the process with some examples. First consider the six-point permutation $[1,4]$, $[2,5]$, and $[3,6]$. Following the above we have 
\eqa
\nonumber&&\begin{array}{c} [1,4] \\ \left[2,5\right] \\  \left[3,6\right] \end{array}\;\; \vcenter{\hbox{\includegraphics[scale=0.8]{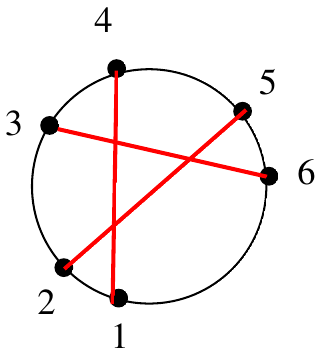}}}\;\;\underrightarrow{(1,2)}\;\;\begin{array}{c} [1,5] \\ \left[2,4\right] \\  \left[3,6\right] \end{array}\;\; \vcenter{\hbox{\includegraphics[scale=0.8]{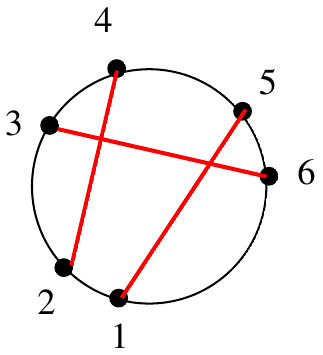}}}\;\;\underrightarrow{(2,3)}\;\;\begin{array}{c} [1,5] \\ \left[2,6\right] \\  \left[3,4\right] \end{array}\;\; \vcenter{\hbox{\includegraphics[scale=0.8]{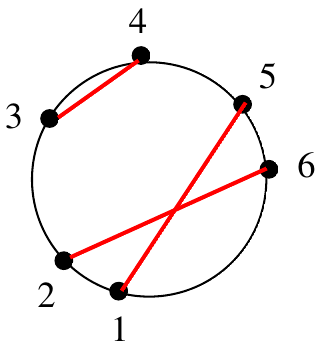}}}\;\;\\
&&\rightarrow \begin{array}{c} [5,1] \\ \left[2,6\right] \\  \left[3,4\right] \end{array} \quad \underrightarrow{(5,2)}\quad \;\;\begin{array}{c} [5,6] \\ \left[1,2\right] \\  \left[3,4\right] \end{array}\;\; \vcenter{\hbox{\includegraphics[scale=0.8]{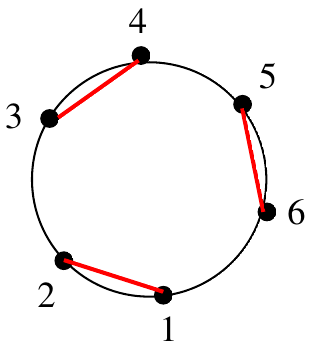}}}
\label{6ptdecomp}
\eqae
Note on the second line, we have reached the bottom of the tower and trivial permutation has not been reached yet, so we restart the process by exchanging the two entries in the top two-cycle and repeat the procedure. As further example, we consider the eight-point permutation $[1,4], [2,7], [3,6],[5,8]$:
\eqa
&&\nonumber\begin{array}{c} [1,4] \\ \left[2,6\right] \\  \left[3,7\right] \\ \left[5,8\right] \end{array}\;\; \vcenter{\hbox{\includegraphics[scale=0.8]{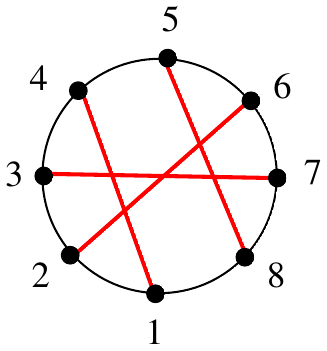}}}\;\;\underrightarrow{(1,2)}\;\;\begin{array}{c} [1,6] \\ \left[2,4\right] \\  \left[3,7\right] \\ \left[5,8\right] \end{array}\;\; \vcenter{\hbox{\includegraphics[scale=0.8]{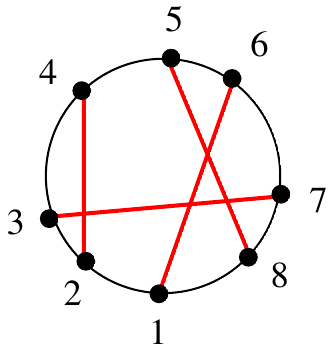}}}\;\;\underrightarrow{(2,3)}\;\;\begin{array}{c} [1,6] \\ \left[2,7\right] \\  \left[3,4\right] \\ \left[5,8\right] \end{array}\;\; \vcenter{\hbox{\includegraphics[scale=0.8]{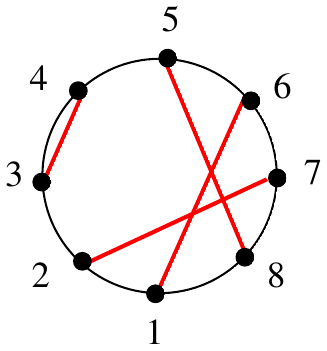}}}\\
\nonumber&&\underrightarrow{(1,2)}\;\;\begin{array}{c} [1,7] \\ \left[2,6\right] \\  \left[3,4\right] \\ \left[5,8\right] \end{array}\;\; \vcenter{\hbox{\includegraphics[scale=0.8]{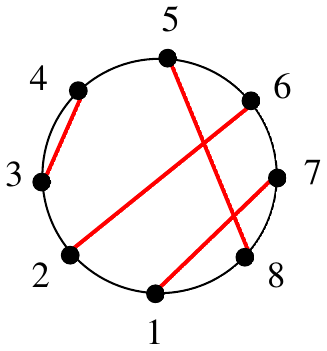}}}\;\; \underrightarrow{(2,5)}\;\;\begin{array}{c} [1,7] \\ \left[2,8\right] \\  \left[3,4\right] \\ \left[5,6\right] \end{array}\;\; \vcenter{\hbox{\includegraphics[scale=0.8]{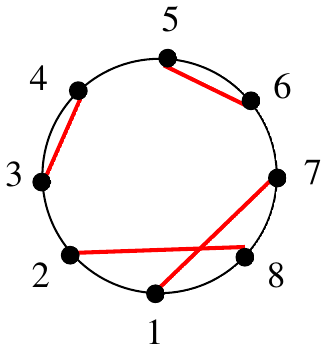}}}\;\; \underrightarrow{(1,2)}\;\;\begin{array}{c} [1,8] \\ \left[2,7\right] \\  \left[3,4\right] \\ \left[5,6\right] \end{array}\;\; \vcenter{\hbox{\includegraphics[scale=0.8]{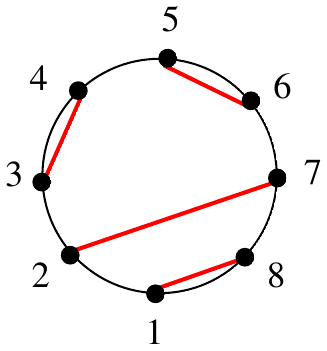}}}\,.\\
&&\rightarrow \begin{array}{c} [8,1] \\ \left[2,7\right] \\  \left[3,4\right] \\ \left[5,6\right] \end{array}\;\; \underrightarrow{(2,8)}\;\;\begin{array}{c} [8,7] \\ \left[2,1\right] \\  \left[3,4\right] \\ \left[5,6\right] \end{array}\;\; \vcenter{\hbox{\includegraphics[scale=0.8]{trivial}}}
\eqae

A fascinating fact is that we can reverse the process and build up a representative for a given permutation starting from the trivial permutation. For example beginning with the OG$_2$, $[1,3],[2,4]$, the previous rules tells us that it is built from applying a $(1,2)$ transmutation on the trivial permutation $[1,4],\; [2,3]$. The transmutation implies that we perform a SO(2) rotation on the two columns $1$ and $2$:
\eq
\left(\begin{array}{cccc}i & 0 & 0 & 1 \\0 & i & 1 & 0\end{array}\right)\;\;\rightarrow \;\;\left(\begin{array}{cccc}ic & is & 0 & 1 \\is & -ic & 1 & 0\end{array}\right)\, ,
\eqe  
for simplicity, here and in the following discussion, we only choose one branch, the other one can be equally considered. This is graphically represented as:
\eq
\includegraphics[scale=0.8]{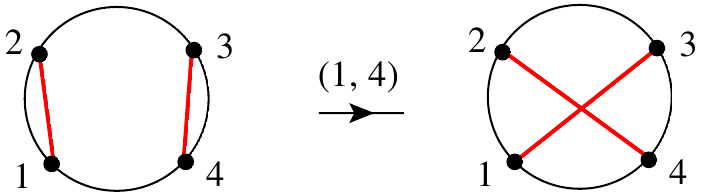}\,.
\eqe

It is straightforward to generalize the above to higher points: for each transmutation $(i_1,i_2)$, we rotate the two columns as:
\eq
(\cdots\vec{c}_{i_1} \cdots \vec{c}_{i_2}\cdots)\quad\quad \underrightarrow{(i_1,i_2)}\quad\quad (\cdots(c\vec{c}_{i_1}+ s\vec{c}_{i_2})\cdots (s\vec{c}_{i_1}-c\vec{c}_{i_2})\cdots)\,.
\eqe
It is easy to see that this process preserves orthogonality. Let us now consider one more non-trivial example, the build up of $[1,4],[2,5],[3,6]$ by reversing the process in eq.(\ref{6ptdecomp}). Starting with $[5,6],[2,1],[3,4]$
\eqa 
\nonumber &&\left(\begin{array}{cccccc}1 & i & 0 & 0 & 0 & 0 \\0 & 0 & i & 1 & 0 & 0 \\0 & 0 & 0 & 0 & i & 1\end{array}\right)\underrightarrow{(5,2)}\left(\begin{array}{cccccc}1 & ic_1 & 0 & 0 & is_1 & 0 \\0 & 0 & i & 1 & 0 & 0 \\0 & is_1 & 0 & 0 & -ic_1 & 1\end{array}\right)\underrightarrow{(2,3)}\left(\begin{array}{cccccc}1 & ic_1c_2 & ic_1s_2 & 0 & is_1 & 0 \\0 & is_2 & -ic_2 & 1 & 0 & 0 \\0 & is_1c_2 & is_1s_2 & 0 & -ic_1 & 1\end{array}\right)\\
&&\underrightarrow{(1,2)}\left(\begin{array}{cccccc}c_3+ ic_1c_2s_3  & s_3-ic_1c_2c_3 & ic_1s_2 & 0 & is_1 & 0 \\is_2s_3  & -is_2c_3 & -ic_2 & 1 & 0 & 0 \\ is_1c_2s_3 & -is_1c_2c_3 & is_1s_2 & 0 & -ic_1 & 1\end{array}\right) \, .
\eqae
In the above example some entries are complex rather than real or pure imaginary as we had previously, which may seem to lead some complexity for the minors. Surprisingly, all the consecutive minors are rather simple in this representation! Explicitly, we have: 
\eqa
(123) = s_1, \, (234) = -i s_1 s_2 s_3,  \, (345) = s_2 \, .
\eqae
We note that they are all simple products of permutation parameters $s_i$. As we will prove in the following section, this nice feature is actually a general fact of the on-shell diagrams (or permutations) constructed from BCFW recursion relations: all consecutive minors of such on-shell diagrams are simple products of $c_i$ and $s_i$'s. What is surprising is that following the same proof, one can show that representations built from permutations are always products of only $s_i$'s!

\subsection{Canonical coordinates}  \label{section:canonical}
Here we will proceed to prove that the non-vanishing consecutive minors of the OG$_k$ representations constructed by BCFW recursion relations will be given by a one term product of vertex parameters of $c_i$ and $s_i$. Since the vanishing of vertex parameters is corresponding to the singularities (or boundaries) of the on-shell diagram, this result implies that the boundaries of the on-shell diagrams always correspond to the vanishing of consecutive minors! This fact is closely related the notion of ``positivity" as we will discuss in the following section. 

We will prove this statement by induction. Firstly it is a trivial fact that consecutive minors of OG$_2$ with canonical or alternating gauge fixing are simple products of BCFW parameters $c$ and $s$. With the assumption that the consecutive minors of lower-point on-shell diagrams constructed by BCFW recursion relation are simple products of $c_i$ and $s_i$, in what follows we will prove that higher-point on-shell diagrams by BCFW recursion recursion have the same property. 

Let us start by considering the factorization diagram,
\eq   \label{factorization}
\vcenter{\hbox{ \includegraphics[scale=0.4]{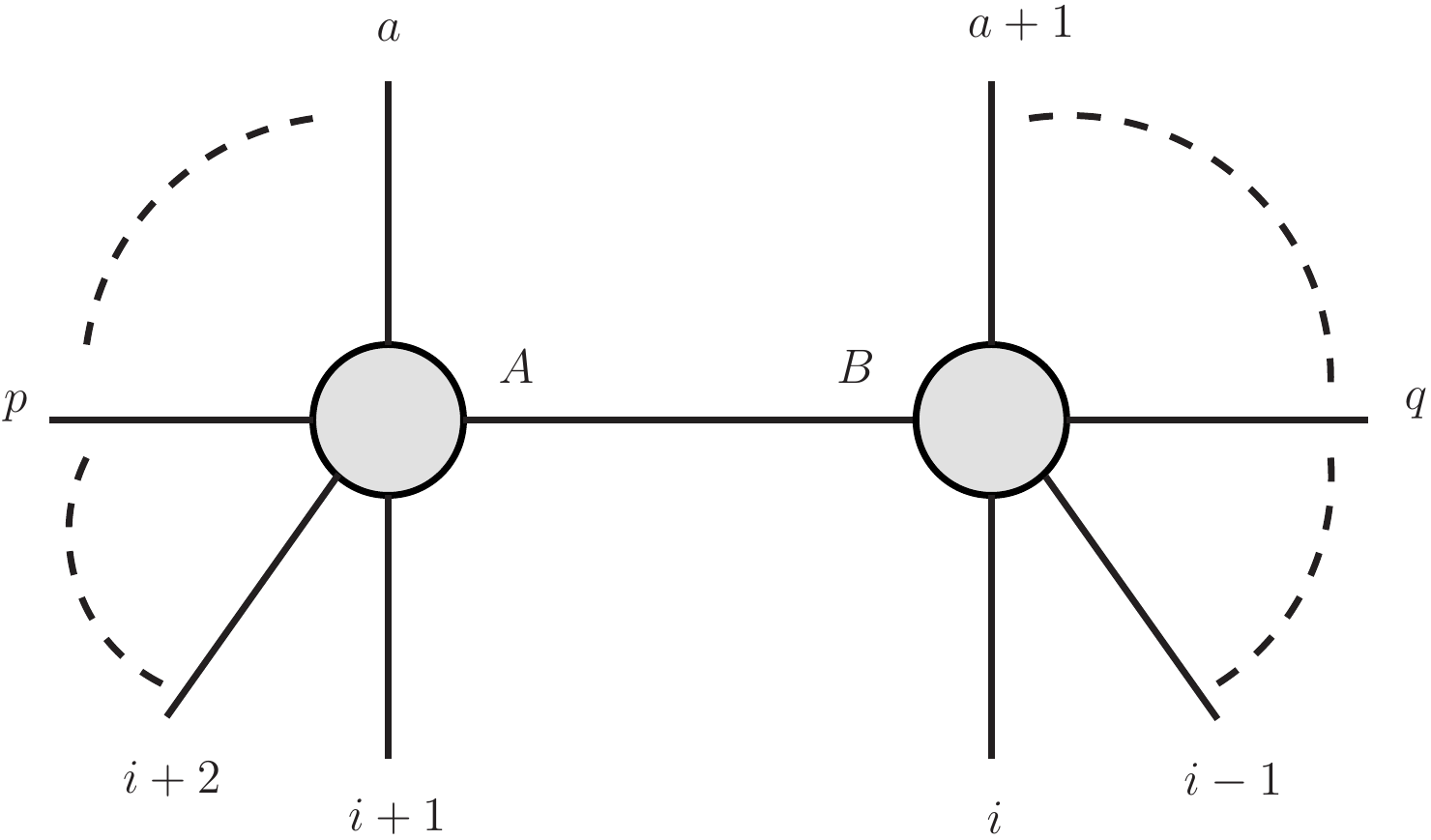} }} \, .
\eqe
Here we are interested in consecutive minors only. Firstly if the minor only involves the legs from one side of the factorization diagram, according to our assumption for lower-point on-shell diagrams, then it is a simple product of $c_i$ and $s_i$. Now, consider the case when legs of the minor $(p, \ldots, q)$ are spread on both sides of factorization diagram, let us denote it as $(p, \ldots,a, a+1, \ldots, q)$, as shown in eq.~(\ref{factorization}). According to amalgamation rules, we have\footnote{Here we will ignore possible factors such as $\pm i$ since they are irrelevant to our discussion.}
\eqa \label{amalgamation}
(p,\ldots,a, a+1, \ldots, q)|_{OG_k} = (A, p, \ldots, q)|_{OG_{k+1}} + (B, p, \ldots, q)|_{OG_{k+1}}. 
\eqae 
From permutation or the explicit form of the Grassmannian, 
\eq
C_{\alpha i}=\left(\begin{array}{cccccccc} \vec{c}_{i+1} & \cdots & \vec{c}_{a} & \vec{c}_A & \vec{0} & \vec{0} & \vec{0} & \vec{0} 
 \\ \vec{0} & \vec{0} & \vec{0} & \vec{0} & \vec{c}_{B} & \vec{c}_{a+1} & \cdots &  \vec{c}_{i}\end{array}\right) \, ,
\eqe
we see that, for $(p,\ldots,a, a+1, \ldots, q)|_{OG_k}$ to be non-vanishing, one of two terms in the sum of eq.~(\ref{amalgamation}) should vanish, and the other non-vanishing one is factorized into a simple product of two lower-point consecutive minors. This proves that the minors of all factorization diagrams are simple products of $c_i$ and $s_i$. 

Next, we add BCFW bridge to legs $i$ and $i+1$ of the factorization diagram, 
\eq   \label{factorizationBCFW}
\vcenter{\hbox{ \includegraphics[scale=0.35]{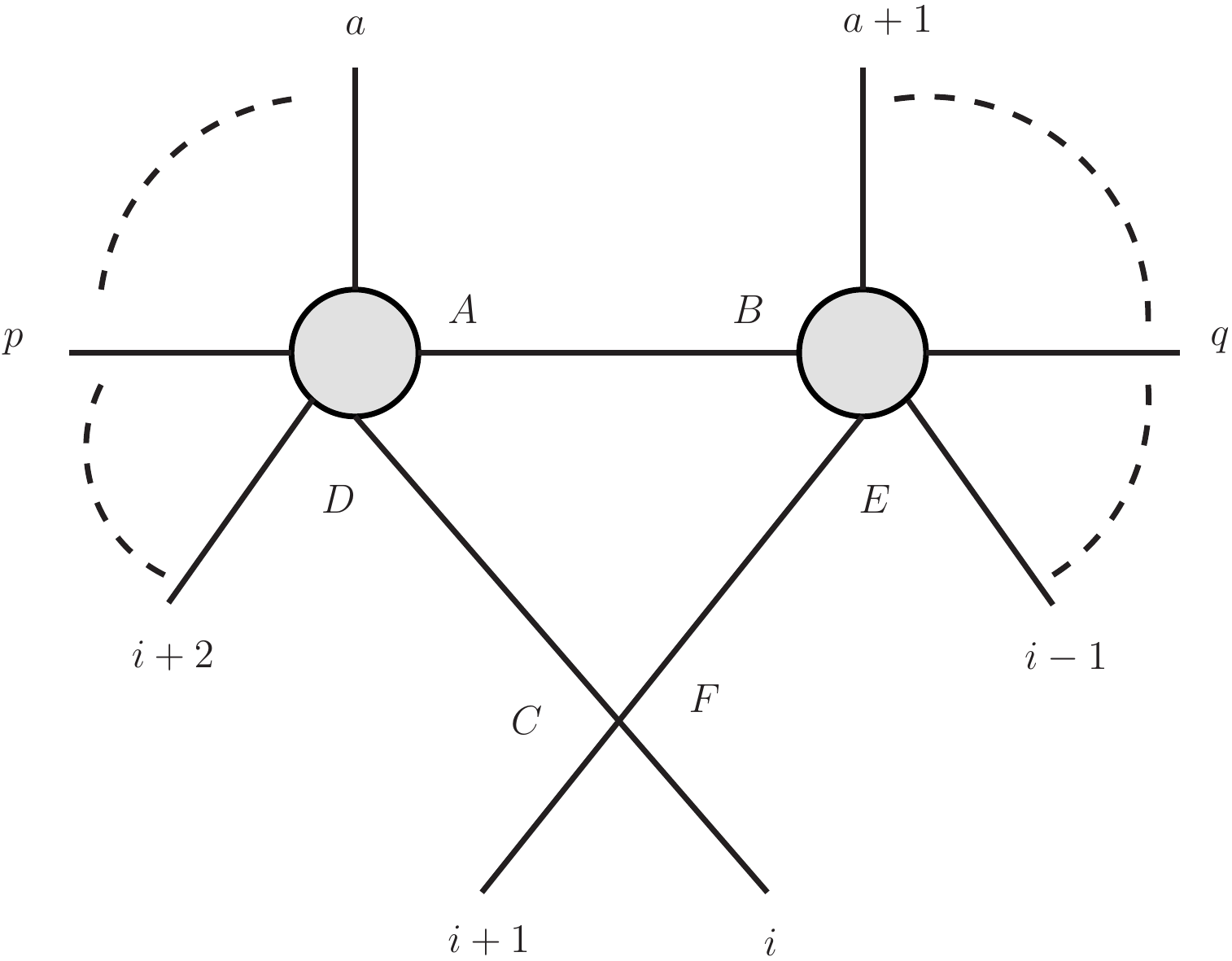} }} \, .
\eqe
Firstly, consecutive minors not involving $i$ and $i+1$ will of course not be changed after adding the BCFW bridge. The minors with both $i$ and $i+1$ is equivalent to the one without $i$ and $i+1$ by orthogonality condition, so they will not be changed either. So we only need to consider the case when the minor contains $i+1$ only, namely minor $(i+1,\ldots, i+k)|_{OG_k}$. Furthermore we can assume that this minor involves all the legs on the left-hand side of the factorization diagram, if this is not the case, we could relate $(i+1,\ldots, i+k)|_{OG_k}$ to $(k+i+1,\ldots, i)|_{OG_k}$ by orthogonal condition, which would involve all the legs on the right-hand side, and the following proof equally applies to that case.

With this set-up, we first like to prove that for $(i+1,\ldots, i+k)|_{OG_k}$ to be non-vanishing, $a$ in the diagram of eq.~(\ref{factorizationBCFW}) must be $i+3$, namely the amplitude on the left-hand side is four-point. To see this, note that if the permutation path connects leg $i+2$ into any legs among $i+3, \ldots, a$, then $(i+1,\ldots, i+k)|_{OG_k}$ vanishes due to reduced rank. It cannot go into $D$ either, as permutation paths in ABJM theory cannot connect a leg into its adjacent neighbour if there are no bubbles. So leg $i+2$ must be connected to $A$. Similarly, leg $i+3$ must permute into $D$ since we already have $i+2$ going to $A$. Then leg $i+4$ has nowhere to go for $(i+1,\ldots, i+k)|_{OG_k}$ being non-zero. So we indeed find $a = i+3$ for this case. Now, by applying amalgamation rule, we obtain
\eqa
(i+1,i+2, \ldots, i+k)|_{OG_k} &=& (C, i+1, \ldots, i+k)|_{OG_{k+1}}  + (D, i+1, \ldots, i+k )|_{OG_{k+1}} \nonumber \\
&=& (C, i+1)|_{OG_{2}} \times (i+2, \ldots, i+k )|_{OG_{k-1}}  + 0 \, ,
\eqae
where the second in the sum vanishes is precisely because of the fact that $a = i+3$, and the permutation path contains $[D, i+3]$. So we have proved that for this case $(i+1, \ldots, i+k)|_{OG_k}$ is a simple product of the consecutive minors of lower-point amplitudes, by induction this completes our proof.\footnote{Note attaching a BCFW bridge may introduce a closed loop, depending on
the gauge choice, and since the amalgamation rules are only correct upto an overall factor
( the Jacobian coming from the closed loop). So when there are closed loops in the on-shell diagrams, the
consecutive minors are simple products of BCFW parameters upto some overall factors.} 

Furthermore if the BCFW bridge is in the canonical gauge, and is simply corresponding to BCFW shifts on $\vec{c}_i$ and $\vec{c}_{i+1}$ as the case of the section \ref{section:representative}, then the effect of adding a BCFW bridge (or adjacent transmutation) can be simplified to
\eqa 
(i+1, i+2, \ldots, i+k)|_{OG_k} =
c (i+1,i+2 \ldots, i+k)|_{{OG_k}_F} - s (i,i+2 \ldots, i+k)|_{{OG_k}_F} \, , \nonumber 
\eqae
where OG${_k}_F$ denotes the Grassmannian of the factorization diagram, namely before adding the bridge. From eq.~(\ref{factorizationBCFW}) with $a=i+3$, we find the first term in the above equation vanishes, while the second term is given by the amalgamation rule
\eqa
s\, (i,i+2 \ldots, i+k)|_{{OG_k}_F} &=& s \left[ (A, i,i+2 \ldots, i+k)|_{{OG_{k+1}}}  +  (B, i,i+2 \ldots, i+k)|_{{OG_{k+1}}} \right] \nonumber \\
&=&
0+ s\, (i+2, i+3)|_{OG_2}\times (i, B, i+4 \ldots, i+k)|_{{OG_{k-1}}} \, .
\eqae
So for the diagram constructed from vertices with canonical gauge only, we see that the minors will be a simple product of $s_i$'s only as we observed previously. That's simply because of the fact that four-point vertex in the canonical gauge only has one singularity, namely at $s_i \rightarrow 0$.

\section{The Positive Orthogonal Grassmannian\label{OrthogSec}}
As we have discussed in the previous section, the on-shell diagram constructs a particular representation of positroid stratification in OG$_k$.  Looking at the explicit representation in eq.(\ref{OnShellRep}), one might wonder whether the $d\log$ singularities of the on-shell diagram involve the vanishing of non-consecutive minors. An important notion that was realized in the work of ref.~\cite{NimaBigBook}, is that if one assumes that the columns of the Grassmannian is real, then there is well-defined region of the Grassmannian where all ratios of all ordered minors satisfy:
\eq
\frac{(i_1,i_2,\cdots,i_k)}{(j_1,j_2,\cdots,j_k)}>0\,.
\label{Positivity}
\eqe 
This is called the ``\textit{positive}" Grassmannian, denoted by G$_+(k,n)$. As discussed in ref.~\cite{NimaBigBook}, positivity is a generalization of convexity of a polygon in RP$^2$, which ensures that its boundary only involves the vanishing of consecutive minors. Remarkably, the on-shell diagrams indeed always give representation in the positive Grassmannian, as one can show that amalgamation preserves positivity.  

As we proved generally that for on-shell diagrams constructed from the BCFW recursion relations, all non-vanishing consecutive minors are always given by a simple product of $s_i$ and $c_i$'s. Furthermore, for the representation constructed by the information of permutation paths, all consecutive minors are given by a simple product of $s_i$'s. Thus the $d\log$ singularities of eq.(\ref{OnShellRep}) always correspond to the vanishing of consecutive minors. Given this observation, one would expect that for ABJM theory, the on-shell diagrams also give representations of some ``positive" orthogonal Grassmannian.

An immediate obstruction to defining a positive part of the orthogonal Grassmannian is the fact that even for the fundamental OG$_2$, the consecutive minors alternate between purely real or purely imaginary.\footnote{Interestingly, one can show that in the canonical or cyclic gauge, by choosing all angles to be real, the minors always satisfy $|M_i|\leq1$. }  There is a simple reason why this is always the case: the orthogonal condition implies eq.(\ref{BranchDef}) and thus if one minor is real, then its complement can be imaginary. The extra factor of $i$ in eq.(\ref{BranchDef}) is present as a consequence of the fact that the ``signature" of the Grassmannian is defined to be all plus. Thus if we are allowed to analytically continue to split signature, then one can now have all real minors! In order for such continuation to be well-defined under the operation of amalgamation, it is natural to define the signature to be alternating along a given cyclic ordering, i.e. $\eta^{ij}=(+,-,+,\cdots,-)$. With this signature, the minors can all be real and positivity can be defined! The analytic continuation can be done by redefining all even columns as:
\eq
c_{\alpha,2i}\rightarrow i\tilde{c}_{\alpha,2i}\,.
\label{AnalyticCont}
\eqe
After this redefinition, the orthogonality condition $\tilde{C}\cdot \tilde{C}^T=0$, where $\tilde{C}$ is the Grassmannian whose even columns are simply $\tilde{c}$, now has alternating signature. Note that this is reminiscent to the four-dimensional spinor-helicity formalism where one defines the spinors to be in split signature such that the powerful tools of holomorphicity can be utilized. Another appealing aspect of split signature orthogonal Grassmannian is that the alternating signature exactly matches with the chirality of the legs. More importantly, on-shell diagrams which cannot be consistently assigned chirality will not have a representation in the split signature orthogonal Grassmannian. 

Now that we have real minors, before positivity can be defined there are several subtleties to take into account. First, with alternating split-signature, each minor is equivalent with their complement up to a sign:
\eq
\frac{M_{\sigma}}{M_{\bar{\sigma}}}=\pm\,.
\eqe 
Note that up to now, for a given $\sigma$ we have not defined the ordering of its complement. Here we claim that for alternating split signature, focusing on ordered minors, the ratio between $M_{\sigma}$ and its ordered complement  $M_{\bar{\sigma}}$ has uniform sign in a given branch. In other words, in SO$_+(k)$ all ordered minors are equivalent to their complement. Second, we have to make sure that positivity of all minors is consistent with relations implied by the orthogonality constraint. For example, for all consecutive minors, orthogonality implies~\cite{SangminGrass}\,,
\eq
M_{i}M_{i+1}=M_{i+k}M_{i+1+k}(-1)^{k-1}\,.
\eqe
Remarkably positivity of ordered minor is consistent with the above identity:
\eq
k=2:\; (12)(23)=-(34)(41)=(34)(14),\; k=4:\; (1234)(2345)=(5678)(1678)\,.
\eqe
For non-consecutive minors, we have the identities in eq.(\ref{OrthoMinor}). Taking into account the redefinition in eq.(\ref{AnalyticCont}), we have for example 
\eqa
\nonumber&&k=3:\; -(123)(356)+(124)(456)=0,\; \\
&&k=4:\; -(2345)(5781)+(2346)(6781)=-(2345)(1578)+(2346)(1678)=0\,. 
\eqae
Again positivity for all ordered minors is consistent with the above identities. A general case can be proved as what follows. 

We are interested in the case when the identity of eq.(\ref{OrthoMinor}) involves a sum of two terms only, so let us denote those two integers appeared in the sum as $a$ and $b$, also assuming $a<b$. So we have a list of ordered integers 
\eqa \label{2klist}
\ell := \{ 1, 2, \ldots, a, \ldots, b, \ldots, 2k \} \, ,
\eqae
which can further be separated into three smaller lists: $\ell_1 := \{ 2, \ldots, a-1 \}$, $\ell_2 :=\{a+1, \ldots, b-1 \}$ and $\ell_3:=\{ b+1, \ldots, 2k-1 \}$, whose lengths are denoted $n_a, n_{ab}$ and $n_b$ respectively. Here we consider the case when $a$ and $b$ are both even or odd, namely $n_{ab}$ is odd. The other possibility can be proved in the same way. Let us quote the two-term identity here 
\eqa \label{twotermidentity}
(1, \ldots, a)(a, \ldots, 2k) + (1, \ldots, b)(b, \ldots, 2k)=0 \, , 
\eqae
where ``$\ldots$" in the minor $(1, \ldots, a)$ are ordered integers with $n'_a$ of them from $\ell_1$, $n'_{ab}$ from $\ell_2$, and $n'_b$ from $\ell_3$. So to make $(1, \ldots, a)$ being ordered, we need move $a$ to the left by $(n'_{ab}+n'_b)$ steps. Similarly, as for the minor $(a, \ldots, 2k)$, we need move $a$ to the right by $(n_a- n'_a)$ steps to make it be in ordered. So in total we find to make the minors $(1, \ldots, a)$ and $(a, \ldots, 2k)$ to be in ordered, we introduce a factor of 
\eqa
(-1)^{(n'_{ab}+n'_b) + (n_a- n'_a)}.
\eqae
By a similar analysis we find that to make the minors $(1, \ldots, b)$ and $(b, \ldots, 2k)$ in order, we introduce a factor of $(-1)^{n'_{b}+(n_a-n'_a)+ (n_{ab}- n'_{ab})}$. Because $n_{ab}$ is odd, we have
\eqa
(-1)^{(n'_{ab}+n'_b) + (n_a- n'_a)} = - (-1)^{n'_{b}+(n_a-n'_a)+ (n_{ab}- n'_{ab})} \, ,
\eqae
namely an extra minus sign appears in eq.~(\ref{twotermidentity}) to rearrange all the minors in order. This insures the consistency of positivity and split-signature orthogonality.

Now we can proceed to show that orthogonality is preserved by amalgamation. For general amalgamation of OG$_k$ and OG$_{k'}$ to OG$_{k+k'}$, the proof is trivial since the ordered minor of  OG$_{k+k'}$ is simply the product of that of OG$_k$ and OG$_{k'}$. For amalgamation that reduces OG$_{k}$ to OG$_{k-1}$, as in eq.(\ref{AmalgaRule}), due to the analytic continuation, the minor of OG$_{k-1}$ is given as:
\eq
(i_1,\cdots, i_{k-1})=(i_1,\cdots,i_{k-1},A)+(i_1,\cdots,i_{k-1},B)\,.
\label{AmalgaRule2}
\eqe
where $A,B$ are the columns in OG$_k$ whose corresponding spinors are identified and integrated away. For simplicity, we've gauge fixed all odd columns to be unity and thus there are no Jacobian factors.\footnote{These factors are irrelevant since positivity is strictly defined for ratios of minors as indicated in eq.(\ref{Positivity})} Thus if the OG$_k$ is positive, so will the amalgamated OG$_{k-1}$. 

In summary, we've shown that for split signature orthogonal Grassmannian, starting with the positive OG$_2$, all OG$_k$ that are obtained through amalgamation will also be positive.

\section{Conclusion and a peek at loop level} \label{section:conclusion}

In this paper, we've studied in detail the relation between on-shell diagrams and residues of the orthogonal Grassmannian integral originally proposed in ref.~\cite{SangminGrass}. More precisely, the on-shell diagrams represents a $n_v$-dimensional sub manifold of the orthogonal Grassmannian subject to $2k-3$ constraints, where $n_v$ is the number of vertices and $2k-3$ is the number of bosonic delta functions enforcing constraints beyond that of momentum conservation. Much like in $\mathcal{N}=4$ SYM, the linear-dependency of consecutive minors, is the invariant data that is encoded in the on-shell diagrams. This linear-dependency forms a stratification of the orthogonal Grassmannian: starting from the top-cell in OG$_k$, successive linear dependencies among consecutive columns of the Grassmannian are the boundaries of the top-cell. As the top-cell is $k(k-1)/2$ dimensional, generic on-shell diagrams have dimensions fewer than the top-cell and can be shown to be co-dimension $k(k-1)/2-n_v$ boundaries of the top-cell. Furthermore, the stratification is encoded in the permutation paths that is associated with each on-shell diagrams. 

Remarkably, using the permutation of an on-shell diagram, one can reconstruct a representation of OG$_k$ that not only reflects the stratification, but most importantly, gives non-vanishing consecutive minors that are always of the form:
\eq
M_i=\prod_{n\in n_v}\sin\theta_{n}
\eqe
i.e. the consecutive minors are always given as a simple product of the vertex variables. As the canonical measure is simply $d\theta/\sin\theta$, this immediately leads to the conclusion that the singularities of the on-shell diagrams again correspond to the vanishing of consecutive minors, i.e. the singularity is precisely the boundary of the stratification. 

Such a property was ensured for the on-shell diagrams in $\mathcal{N}=4$ SYM by the fact that on-shell diagrams are related to the positive region of the Grassmannian, where all ordered minors are positive. Here we demonstrated that positivity can be defined for orthogonal Grassmannian by analytically continuing the Grassmannian into split signature. More importantly, the relations among ordered minors that are implied by the orthogonal constraint respects positivity. It is then straightforward to show that analytically continued Grassmannians for the corresponding on-shell diagrams are always positive.  

Armed with the stratification of the on-shell diagrams in the BCFW recursion relation, we can easily determine the tree-level contours in the original Grassmannian integral. It would be interesting to study the relation between this contour and the contour that localizes the Grassmannian to an integral over punctures in the Riemann sphere~\cite{ABJMString} (or equivalently ~\cite{Cachazo:2013iaa}) which reproduces the tree-level amplitude. 

A natural extension of our current work is to consider loop-level amplitudes. A remarkable property of the on-shell diagram representation of scattering amplitudes is that it can be written as an integral with canonical $d\log$ integration measure, subject to $2k-3$ bosonic constraints. We believe that similar representation can be achieved for loop amplitudes of ABJM as well. Here we like to demonstrate explicitly that some of the loop integrands can be written in a $d\log$ form. For example, let's consider one-loop four-point amplitude constructed in ref.~\cite{ABJMTwoL41}:
\eq
\mathcal{A}^{\rm 1-loop}_4=\mathcal{A}^{\rm tree}_4\int d^3X_0\frac{\epsilon(01234)}{(0\cdot 1)(0\cdot 2)(0\cdot 3)(0\cdot 4)}
\label{Old}
\eqe
where $(i\cdot j)=X_i\cdot X_j$ and region momenta $X_i$ are the five-dimensional embedding coordinates. To achieve the $d\log$ form we parametrize the loop integration region as
\eq
X_0= X_1+a_2X_2+a_3X_3+a_4X_4+a_\epsilon \epsilon(*1234) \, ,
\label{LoopP}
\eqe 
where we've used the projective invariance of the integrand to scale the coefficient of $X_1$ to be $1$. After expressing the projective measure in this parametrization, we then obtain the integrand in a $d\log$ form,\footnote{In momentum-space a similar $d\log$ form can be obtained, given as 
$$\mathcal{A}^{\rm 1-loop}_4 = \mathcal{A}^{\rm tree}_4 \int d \log({ (\ell -p_1)^2 \over  \ell^2} ) \wedge d \log({ (\ell -p_1 - p_2)^2  \over  \ell^2} ) \wedge d \log({ (\ell + p_4)^2 \over \ell^2} ),$$
which immediately integrates to zero, as four-point one-loop amplitude vanishes at $3d$.}
\eqa
\nonumber \mathcal{A}^{\rm 1-loop}_4&=&\mathcal{A}^{\rm tree}_4\int\frac{(1\cdot 3)^2(2\cdot 4)^2da_1\wedge da_2 \wedge da_4 \wedge da_\epsilon}{(a_3(1\cdot3)+a_2a_4(2\cdot4)+a_\epsilon^2(1\cdot3)^2(2\cdot4)^2)}\frac{a_\epsilon}{a_2a_3a_4}\\
&=&\mathcal{A}^{\rm tree}_4\int \prod_{i=2}^4d\log a_i\,,
\eqae
where we've localized the $da_\epsilon$ integral. Note however, this is only valid in a local patch. To see this recall that we have set the coefficient in front of $X_1$ to be 1. This implies that $(0\cdot 3)\neq0$. Thus for configuration where $(0\cdot 3)=0$, the parametrization in eq.(\ref{LoopP}) is invalid. In fact, there appears to be no universal patches for which the integrand in eq.(\ref{Old}) can be written as a product of three $d\log$s. 

We now turn to the on-shell recursion relation for obtaining all planar loop amplitudes in ABJM theory as was proposed in ref.~\cite{NimaAllLoop}:
 \eq
\mathcal{A}_n^{\ell}=\vcenter{\hbox{\includegraphics[scale=0.6]{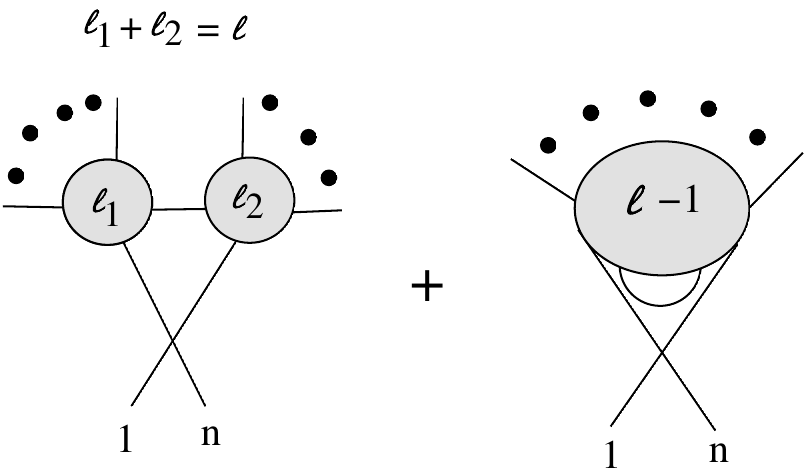}}} \, .
\label{recurssion}
\eqe 
For the four-point amplitude, the recursion relation simplifies since there is no factorization diagram. We have the following on-shell diagram representation for the recursive result of four-point one-loop amplitude:
\eq
\vcenter{\hbox{\includegraphics[scale=0.7]{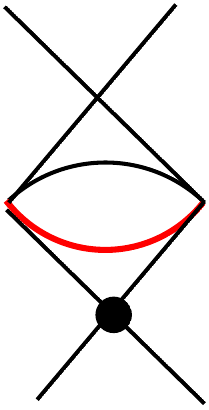}}}\,,
\label{RecurrseResult}
\eqe
where the black vertex represents the addition of a BCFW bridge, while the red line indicates taking the forward limit of six-point tree-level amplitude. It is straightforward to show that, by applying three steps of bubble reduction eq.(\ref{bubblereductionB}), the on-shell diagram of four-point one-loop amplitude is also given by three $d\log$ multiplying the tree-level amplitude. However it has been a difficulty to prove directly two $d\log$ forms match with each other. The difficulty may perhaps be caused by the local chart issue as we have discussed. As this is outside the scope of this paper, we will address this issue in a future work.

\section{Acknowledgements}
We thank Song He for collaboration at the early stages of this work. We would also like to thank Nima Arkani-Hamed,  Jacob Bourjaily, Andreas Brandhuber, Simon Caron-Huot, Sangmin Lee, Gabriele Travaglini and Jaroslav Trnka for helpful discussions on these topics. Y-t.H would especially like to thank Henriette Elvang, Cynthia Keeler, Tomas Lam, Timothy Olson, David Speyer and Sam Roland for the many discussions in the weekly meetings at university of Michigan. Y-t.H. is thankful for the hospitality of Simons Center for Geometry and Physics and the Perimeter Institute, where part of this work was done. Research at the Perimeter Institute is supported in part by the Government of Canada through NSERC and by
the Province of Ontario through MRI. The work of C.W. is supported by the STFC Grant ST/J000469/1, ``String theory, gauge theory and duality"

\appendix
\section{The orthogonal constraint\label{AppendixA}}
Here we show that the orthogonal constraint, eq.~(\ref{orthogonal}), is equivalent to the other relation appeared in the context,
\eq \label{orthogonalApp}
\sum_{a}(i_1,\cdots,i_{k-1},a)(j_1,\cdots, j_{k-1},a)=0\,.
\eqe 
Let's consider gauge fixing the $k\times 2k$ matrix to be, 
\eq
\left(\begin{array}{ccccccc}1 & 0 & 0 & 0 & c_{1, k+1} & \cdots & c_{1,2k} \\0 & 1 & 0 & 0 & c_{2, k+1} & \cdots & c_{2,2k} \\0 & 0 & \cdots & 0 & \vdots & \vdots & \vdots \\0 & 0 & 0 & 1 & c_{k, k+1} & \cdots & c_{k,2k}\end{array}\right)
\label{GaugeApp}
\eqe

First consider the case where $\{i_1,\cdots,i_{k-1}\}=\{j_1,\cdots,j_{k-1}\}$. Without loss of generality, from now on one choose $\{i_1,\cdots,i_{k-1}\}$ to be the first $k-1$ columns in eq.(\ref{GaugeApp}). In such case it is easy to see that 
\eq \label{diagonallApp}
\sum_{a}(i_1,\cdots,i_{k-1},a)(i_1,\cdots,i_{k-1},a)= 1+\sum_{j=k+1}^{2k}c_{k, j}^2=0
\eqe
where the second equality is exactly the orthogonal constraint. 

Next, let's consider $\{i_1,\cdots,i_{k-2}\}=\{j_1,\cdots,j_{k-2}\}$ while $i_{k-1}\neq j_{k-1}$. With out lost of generality, we can set $j_{k-1}$ to be column $k$ in eq.(\ref{GaugeApp}). One can see that for this choice,  
\eq \label{offdiagonallApp}
\sum_{a}(i_1,\cdots,i_{k-1},a)(i_1,\cdots,i_{k-2},j_{k-1},a)=\sum_{j=k+1}^{2k} c_{k, j}c_{k-1,j}=0
\eqe
which is simply the off-diagonal part of the orthogonal constraint. Note eq.~(\ref{diagonallApp}) and (\ref{offdiagonallApp}) prove the fact that one can derive the orthogonal constraint, eq.~(\ref{orthogonal}), from eq.~(\ref{orthogonalApp}).  

Now if $\{i_1,\cdots,i_{k-3}\}=\{j_1,\cdots,j_{k-3}\}$ while $\{i_{k-1},i_{k-2}\}\neq \{j_{k-1},j_{k-2}\}$. We again set $j_{k-2}$ to be column $k$ and $j_{k-1}$ to be column $k+1$ in eq.(\ref{GaugeApp}) and one finds: 
\eqa
\nonumber&& \sum_{a}(i_1,\cdots,i_{k-1},a)(i_1,\cdots,i_{k-3},j_{k-2},j_{k-1},a)=\sum_{j=k+2}^{2k} c_{k, j}(c_{k-1,j}c_{k-2,k+1}-c_{k-2,j}c_{k-1,k+1})\\
&&=-c_{k, k+1}(c_{k-1,k+1}c_{k-2,k+1}-c_{k-2,k+1}c_{k-1,k+1})=0\,.
\eqae
where the off-diagonal part of the orthogonal constraint is used again to arrive at the last result. 

The above analysis generalizes straightforwardly as $\{i_1,\cdots,i_{k-1}\}$ and $\{j_1,\cdots,j_{k-1}\}$ differ beyond three entries. 
\section{The Soft Exchange \label{SoftGlue}}
Here we show that the soft-gluon singularity of the six-point tree-level amplitude is indeed proportional to the four-point tree-level amplitude. In terms of Feynman diagrams, there are two types of diagrams contributing: 
\eq
\vcenter{\hbox{ \includegraphics[scale=0.7]{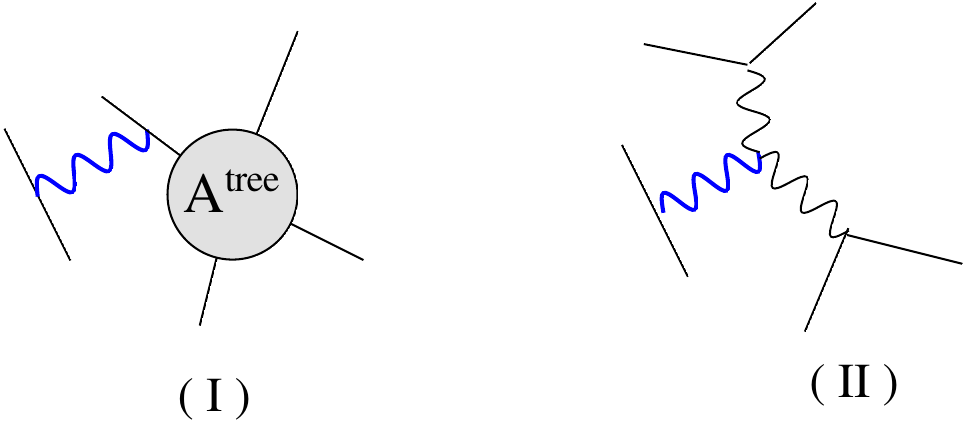} }}
\eqe
Diagram (I) are ones which the soft gluon connects to one of the matter lines of the four-point tree-diagram. It's contribution is given as:
\eq
diagram\,(I): \sim 2\frac{\epsilon(1\,q\, 5)}{q^2(q\cdot p_5)}A^{\rm tree}\,.
\eqe
This is to be compared with that of diagram (II) whose contribution is given as:
\eq
diagram\,(II): \sim 4\frac{\epsilon(4\, 5\, 6)\epsilon(q\, 3\, 2)-\epsilon(4\, 5\, q)\epsilon(6 \,3 \,2)}{q^2(p_4\cdot p_5)(p_2\cdot p_3)}\,.
\eqe
As one can see, as $q\rightarrow0$, the residue of diagram (II) vanishes, where as that of diagram (I) does not. Thus we see that on the soft pole, the residue is simply the four-point amplitude. Furthermore, the non-vanishing residue is invariant under the rescaling $q\rightarrow a q$. Thus the integral one dimensional integral of the residue in eq.(\ref{sixptfactorization2}), which can be separated via a change of variables, is precisely this extra scale factor.

\section{Double soft limit} \label{inversesoftAppendix}
%
%
%
%
\begin{figure}[h]
\scalebox{1}{
\centerline{\includegraphics[height=3.8cm]{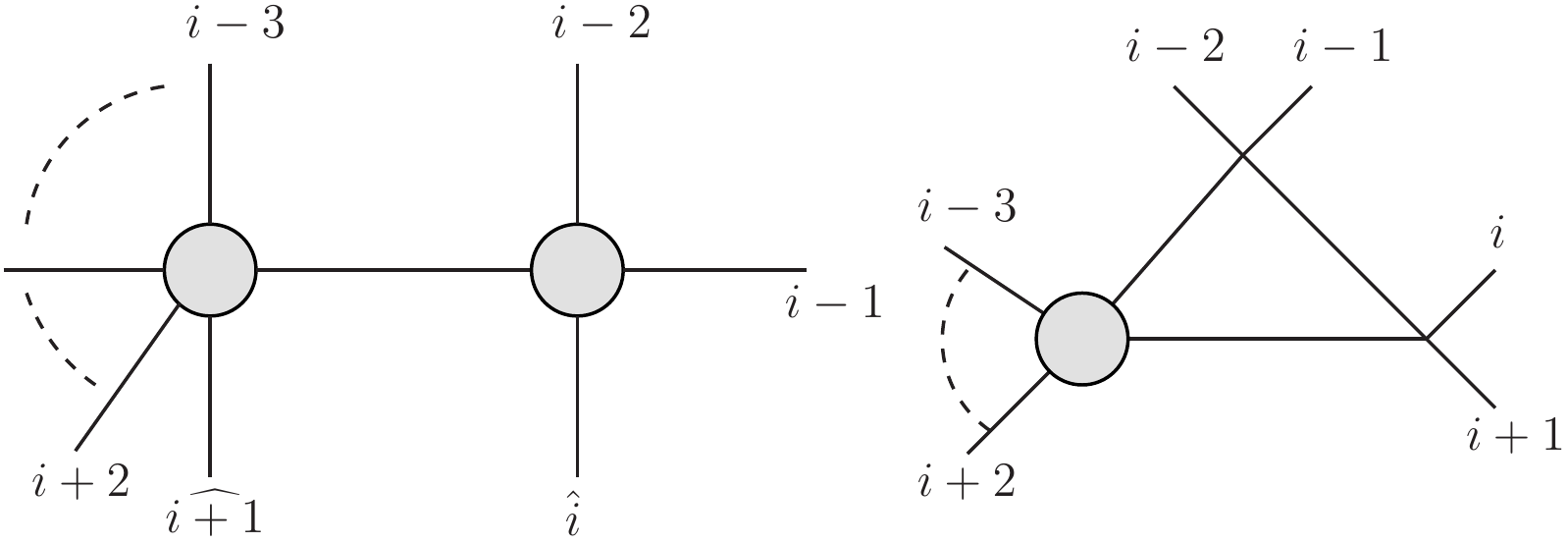}}
}
\caption{\it
The special BCFW diagram which contributes to the double soft limit of $p_{i-1}, p_i \rightarrow 0$, on the right is the corresponding on-shell diagram, which will be called as inverse-soft constructible.}
 \label{fig:softlimit}
 \end{figure}
%
%
%
%
Here we consider double soft-limit of tree-level amplitudes. To be precise, let us consider double soft limit of $p_{i-1}, p_i \rightarrow 0$. It is convenient to represent the tree-level amplitude by BCFW diagrams with legs $p_i$ and $p_{i+1}$ shifted. The dominated BCFW diagram at this limit is the one with four-point amplitude $A(-\hat{P}, i-2, i-1, \hat{i})$ on one-side as shown in Fig.~\ref{fig:softlimit}, that's because four-point amplitude has soft singularity as we discussed previously. The result of the BCFW diagram is given as
\eqa
&{}& \int d^3 \eta_{\hat{P}}  A(-\hat{P}, i-2, i-1, \hat{i}) {1 \over (p_{i-2} + p_{i-1} +p_i)^2}   A(\widehat{i+1}, \ldots, i-3, \hat{P}) \\
&=&
{ \delta^{(3)}(\langle i-2,\hat{P} \rangle \eta_{i-2}+ \langle i-1,\hat{P} \rangle  \eta_{i-1}+ \langle i,\hat{P} \rangle \eta_{i} )
\over \langle \hat{i}, i-1\rangle \langle i-1, i-2 \rangle (p_{i-2} + p_{i-1} +p_i)^2 }    A(\widehat{i+1}, \ldots, i-3, \hat{P}), \nonumber
\eqae
where the BCFW shifts are
\eqa
\lambda_{\hat{i}} = c \lambda_i + s \lambda_{i+1}, \quad  \lambda_{\widehat{i+1}} = c \lambda_{i+1} - s \lambda_{i}.
\eqae
The orthogonal BCFW parameters $c$ (and $s$) may be determined by on-shell condition,
\eqa
\hat{P}^2 =  \langle \hat{i}, i-2 \rangle ^2 + \langle i-1, \hat{i} \rangle ^2 + \langle i-2, i-1 \rangle ^2 =0.
\eqae
When we have $p_{i-1}, p_i \rightarrow 0$, the on-shell condition simplifies greatly, 
\eqa
\hat{P}^2 \rightarrow s^2 \langle i+1, i-2 \rangle ^2  =0,
\eqae
which means $s=0$ and $c^2=1$, and consequently 
\eqa
\lambda_{\hat{i}} \rightarrow \lambda_i, \quad \lambda_{\widehat{i+1}} \rightarrow \lambda_{i+1}, \quad
\hat{P} \rightarrow p_{i-2}.
\eqae
So under the double soft-limit, this particular BCFW diagram simplifies dramatically and reduces to 
\eqa
{ \delta^{(3)}( \langle i-1, i-2 \rangle  \eta_{i-1}+ \langle i, i-2 \rangle \eta_{i} )
\over \langle i, i-1\rangle \langle i-1, i-2 \rangle }  {1 \over 2 p_{i-2} \cdot( p_{i-1} +p_i) }   A(i+1, \ldots, i-3, i-2),
\eqae
from which we deduce the supersymmetric double soft factor of ABJM theory,
\eqa
S_d(i-1, i) = { \delta^{(3)}( \langle i-1, i-2 \rangle  \eta_{i-1}+ \langle i, i-2 \rangle \eta_{i} )
\over \langle i, i-1\rangle \langle i-1, i-2 \rangle }  {1 \over 2 p_{i-2} \cdot( p_{i-1} +p_i) }.
\eqae
For the special case when leg $i-1$ is scalar $\Phi_{i-1}$, and leg $i$ is anti-scalar $\bar{\Phi}_i$, the double soft factor reduces to
\eqa
{ \langle i-1, i-2 \rangle  \langle i, i-2 \rangle
\over \langle i, i-1\rangle  }  {1 \over 2 p_{i-2} \cdot( p_{i-1} +p_i) } .
\eqae
We have confirmed the this result from an explicit Feynman diagram calculation. 

From above discussion, we see that this particular BCFW diagram plays exactly the same role as its analogue, so-called inverse-soft diagrams, in $\mathcal{N}=4$ SYM.\footnote{For detailed discussion on inverse-soft diagrams and their applications in $\mathcal{N}=4$ SYM, see \cite{Bourjaily:2010kw, Nandan:2012rk, NimaBigBook} } So we will use the same terminology here by referring the on-shell diagram Fig.~\ref{fig:softlimit} as inverse-soft diagram in ABJM theory. In $\mathcal{N}=4$ sYM one has both $k$-increasing and $k$-preserving inverse-soft diagrams, here of course we only have $k$-increasing case. Note all tree-level amplitudes in ABJM theory are inverse-soft constructible, since the tree-level on-shell diagrams can always be represented as triangles only. For instance, six-point tree-level amplitude may be viewed as adding two legs to a four-point amplitude, 
\eq \label{inversesoft}
\vcenter{\hbox{ \includegraphics[scale=0.4]{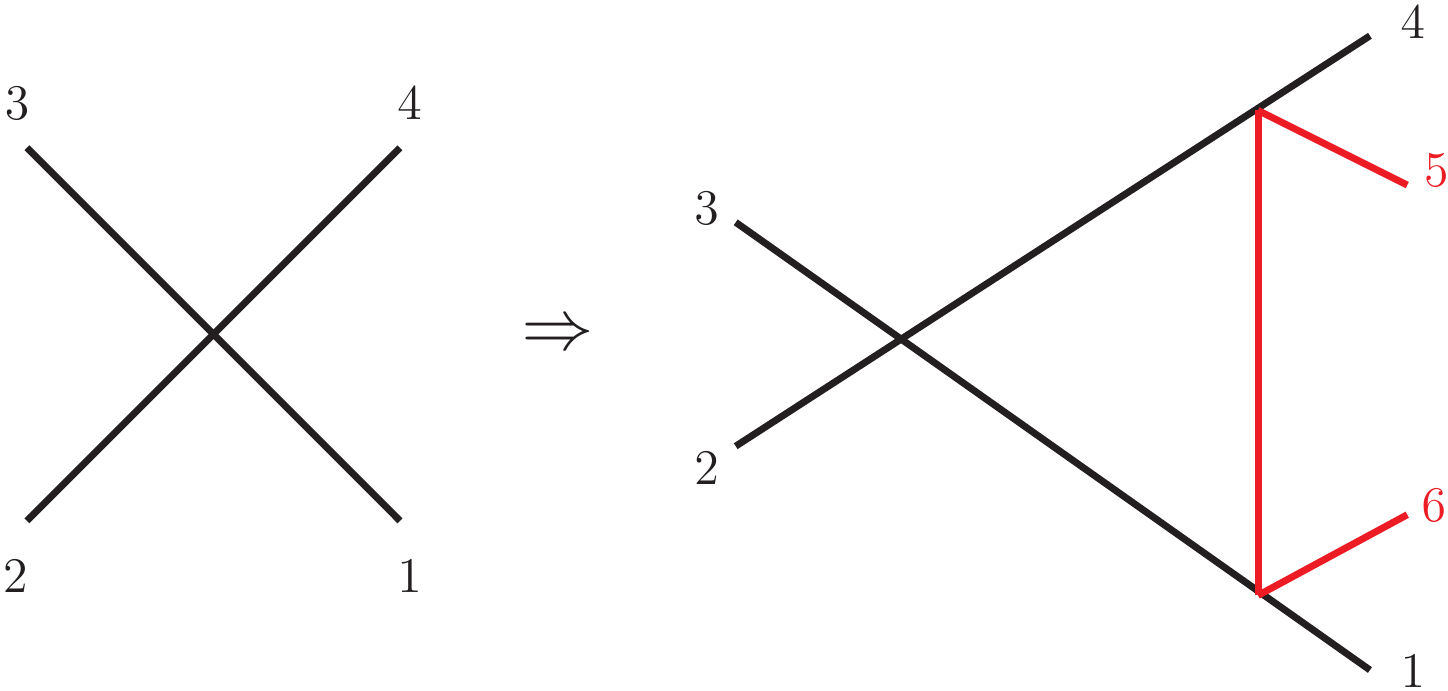} }} \, .
\eqe

\end{document}